\documentclass[a4paper,12pt]{article}
\usepackage[francais, english]{babel}
\usepackage[T1]{fontenc}
\usepackage[babel]{csquotes}
\MakeAutoQuote{«}{»}
\input xy
\xyoption{all}
\usepackage{amssymb,amsmath}
\usepackage{graphicx}
\usepackage{geometry}
\usepackage{stmaryrd}
\usepackage[sectionbib]{chapterbib}
\usepackage{mathrsfs}
\usepackage[autolanguage]{numprint}
\usepackage{mathtools}
\usepackage{booktabs}
\usepackage{multirow}
\usepackage{afterpage}
\usepackage{pdflscape}
\usepackage{bbold}
\usepackage{caption}
\usepackage{siunitx}
\usepackage{eso-pic}
\usepackage{authblk}

\makeatletter
\DeclareFontFamily{OMX}{MnSymbolE}{}
\DeclareSymbolFont{MnLargeSymbols}{OMX}{MnSymbolE}{m}{n}
\SetSymbolFont{MnLargeSymbols}{bold}{OMX}{MnSymbolE}{b}{n}
\DeclareFontShape{OMX}{MnSymbolE}{m}{n}{
    <-6>  MnSymbolE5
   <6-7>  MnSymbolE6
   <7-8>  MnSymbolE7
   <8-9>  MnSymbolE8
   <9-10> MnSymbolE9
  <10-12> MnSymbolE10
  <12->   MnSymbolE12
}{}
\DeclareFontShape{OMX}{MnSymbolE}{b}{n}{
    <-6>  MnSymbolE-Bold5
   <6-7>  MnSymbolE-Bold6
   <7-8>  MnSymbolE-Bold7
   <8-9>  MnSymbolE-Bold8
   <9-10> MnSymbolE-Bold9
  <10-12> MnSymbolE-Bold10
  <12->   MnSymbolE-Bold12
}{}

\DeclareMathAlphabet{\mathscrbf}{OMS}{mdugm}{b}{n}

\let\llangle\@undefined
\let\rrangle\@undefined
\DeclareMathDelimiter{\llangle}{\mathopen}%
                     {MnLargeSymbols}{'164}{MnLargeSymbols}{'164}
\DeclareMathDelimiter{\rrangle}{\mathclose}%
                     {MnLargeSymbols}{'171}{MnLargeSymbols}{'171}
\makeatother

\allowdisplaybreaks[1]
\makeatletter

\@addtoreset{equation}{section}
\makeatother

\makeatletter
\def\hlinewd#1{%
\noalign{\ifnum0=`}\fi\hrule \@height #1 %
\futurelet\reserved@a\@xhline}
\makeatother

\newcommand{\Avogadro}{\mathcal{N}_{\textsc{a}}}
\newcommand{\Backward}{\mathfrak{B}}

\newcommand{\Chemistry}{\mathcal{C}}
\newcommand{\ChemK}{\mathcal{K}}

\newcommand{\CrossSection}{\sigma}
\newcommand{\Current}{J}
\newcommand{\current}{j}

\newcommand{\density}{n}
\newcommand{\Elements}{\mathfrak{A}}
\newcommand{\element}{\mathfrak{a}}
\newcommand{\Energy}{\mathcal{E}}

\newcommand{\Force}{F}
\newcommand{\Forward}{\mathfrak{F}}

\newcommand{\HeatFlux}{\mathcal{Q}}
\newcommand{\Heavy}{\mathfrak{H}}
\newcommand{\heavy}{h}

\newcommand{\Invariants}{\mathcal{I}}

\newcommand{\Knudsen}{\textsc{k}_{n}}

\newcommand{\PseudoMach}{M_{h}}
\newcommand{\Mass}{\mathfrak{m}}
\newcommand{\MeanV}{\mathcal{V}}
\newcommand{\Meanv}{v}
\newcommand{\Molecule}{\mathfrak{M}}
\newcommand{\MoleMass}{m}

\newcommand{\pressure}{p}
\newcommand{\QuantumSpace}{\mathfrak{Q}}
	
\newcommand{\ran}{\rm{Span}}
\newcommand{\Reaction}{\mathscr{R}}
\newcommand{\RU}{R}
\newcommand{\Scattering}{\mathcal{S}}

\newcommand{\Species}{\mathfrak{S}}
\newcommand{\Streaming}{\mathcal{D}}

\newcommand{\Temperature}{T}
\newcommand{\TransitionProbabilities}{\mathcal{W}}

\title{\textbf{Kinetic theory of two-temperature polyatomic plasmas}}

\author[,1]{Jean-Maxime Orlac'h \thanks{Corresponding author: \texttt{jean-maxime.orlach@polytechnique.edu}}}
\author[2]{Vincent Giovangigli}
\author[1]{Tatiana Novikova}
\author[1]{Pere Roca i Cabarrocas}
\date{{October, 10th 2017}}

\affil[1]{Laboratoire de Physique des Interfaces et des Couches Minces (LPICM), CNRS, Ecole Polytechnique, 91128 Palaiseau, France.}
\affil[2]{Centre de Mathématiques Appliquées (CMAP), CNRS, Ecole Polytechnique, 91128 Palaiseau, France.}

\begin{document}

\thispagestyle{empty}

\maketitle
%\pagebreak[4]

\newenvironment{maliste}%d
{ \begin{list}%
	{$\bullet$}%
	{\setlength{\labelwidth}{50pt}%
	 \setlength{\leftmargin}{50pt}%
	 \setlength{\rightmargin}{35pt}%
	 \setlength{\itemsep}{\parsep}}}%
{ \end{list} }

\section*{Abstract}

We investigate the kinetic theory of two-temperature plasmas for reactive \textit{polyatomic} gas mixtures. The Knudsen number is taken proportional to the square root of the mass ratio between electrons and heavy-species, and thermal non-equilibrium between electrons and heavy species is allowed. The kinetic non-equilibrium framework also requires a weak coupling between electrons and internal energy modes of heavy species. The zeroth-order and first-order fluid equations are derived by using a generalized Chapman-Enskog method. Expressions for transport fluxes are obtained in terms of macroscopic variable gradients and the corresponding transport coefficients are expressed as bracket products of species perturbed distribution functions. The theory derived in this paper provides a consistent fluid model for non-thermal multicomponent plasmas. \\
{} \\
\textbf{Keywords:} kinetic theory; non-thermal plasmas; polyatomic gas mixtures; Chapman-Enskog method; transport coefficients.

\section{Introduction}

The kinetic theory of plasmas has been an important subject of research over the past decades. The sound derivation of a multicomponent fluid plasma model is indeed of crucial interest for a wide range of practical applications, ranging from laboratory plasmas, space plasmas, to re-entry vehicles and atmospheric phenomena.

Application of the Chapman-Enskog theory to ionized gas mixtures was first discussed by Chapman and Cowling for monoatomic binary mixtures \cite{ChapmanCowling}, and by Ferziger and Kaper for multicomponent mixtures of monoatomic gases \cite{FerzigerKaper}, in a regime where all species have the same temperature, assuming that the electric field is not intense. More recently, Giovangigli and Graille extended the work of Ferziger and Kaper to the case of reactive polyatomic gas mixtures \cite{GiovangigliGraille2003} \cite{GiovangigliGraille2009}. Interactions between particles at distances greater than the Debye length are considered to be mediated by the electric and magnetic fields while those at shorter distance are considered to be true collisions. Thus, the collision operator generally used is the Boltzmann collision operator with shielded potentials. The Fokker-Planck operator can also be used, but the latter yields identical results within a few percent accuracy \cite{FerzigerKaper}. Besides, the number of particles in a Debye sphere must be large, the cyclotron radius and the wavelength of any electromagnetic wave must be larger than the Debye length.

Higher order evaluations of transport coefficients have been performed by Kaneko and coworkers for binary neutral mixtures in uniform magnetic fields in a simplified steady kinetic framework \cite{Kaneko1960,Kaneko1980}. Convergence properties of the Chapman-Enskog expansion for transport coefficients of magnetized argon plasmas have been investigated by Bruno and coworkers \cite{BrunoCapitelliDangola2003,BrunoLaricchiutaCapitelliCatalfamoChikhaouiKustovaGiordano2004,BrunoCatalfamoLaricchiutaGiordanoCapitelli2006}. The degree of anisotropy of various transport coefficients induced by the magnetic field has been studied in terms of the electron Hall parameter. Bruno and coworkers have established in particular the important influence of the Ramsauer minimum in the electron-argon cross sections on the transport coefficients of magnetized argon plasmas \cite{BrunoLaricchiutaCapitelliCatalfamoChikhaouiKustovaGiordano2004,BrunoCatalfamoLaricchiutaGiordanoCapitelli2006}. 

The first kinetic model for a binary gas mixture composed of light and heavy species was derived by Lorentz \cite{Lorentz} \cite{ChapmanCowling}. This model, also referred to as the \og Lorentzian gas \fg{}, assumed that the heavy molecules were not altered by their collisions with light particles, and that mutual encounters between light particles were of negligible influence compared to encounters with heavy molecules. However, when considering a multiscale analysis of the Boltzmann equations for ionized gases in the fluid regime, both the Knudsen number $\Knudsen$ \textit{and} the ratio of the electron mass over the heavy-species characteristic mass $\varepsilon = \sqrt{\Mass_{e}^{0}/\Mass_{\heavy}^{0}}$ tend to zero. Mixtures of monoatomic gases that are not at thermodynamic equilibrium with multitemperature transport arising from small electron-to-ion mass ratio asymptotics have been investigated by Chmieleski and Ferziger \cite{ChmieleskiFerziger1967}, and in the fully ionized case by Braginskii using the Landau equation \cite{Braginskii1958,Braginskii1965}. Multitemperature models also naturally arise with the presence of strong electric fields, in particular in swarm physics \cite{MasonMcDaniel1988}. As shown by Petit and Darrozes \cite{PetitDarrozes}, the Boltzmann equations exhibit a singularity in the limit $\varepsilon~\to~0, \Knudsen~\to~0$, which might be solved upon assuming that the Knudsen number is proportional to the small parameter $\varepsilon$
\begin{equation}
\Knudsen \propto \varepsilon.
\end{equation}
Such a scaling is associated with the thermal non-equilibrium between light and heavy species. This scaling was first applied by Degond and Lucquin \cite{DegondLucquin} \cite{Degond} to the derivation of a two-temperature macroscopic fluid model for a binary mixture made of electrons and positive ions. In the meantime, Magin and Degrez \cite{MaginDegrez} developed a model for multicomponent non-thermal plasmas, where they introduced the scaling of Petit and Darrozes \cite{PetitDarrozes} in order to account for thermal non-equilibrium. Their model was improved by Graille, Magin and Massot, who further investigated the strongly magnetized case for a multicomponent mixture of monoatomic gases \cite{GrailleMaginMassot2009}. The macroscopic equations in the zeroth-order and first-order regimes, together with expressions for the transport fluxes and the transport coefficients, have been obtained. New bracket expressions have been established for the perpendicular and transverse diffusion, thermal diffusion and partial thermal conductivity coefficients as well as for the shear viscosity coefficients. Positivity properties of multicomponent diffusion matrices have been investigated and the mathematical structure of transport linear systems has also been addressed.

However, in many applications gas mixtures are made of polyatomic molecules, whose internal energy structure may have a significant influence on the transport coefficients of the mixture \cite{ErnGiovangigli1995Thermal} \cite{TantosGhiroldiValourgeorgisFrezzotti2016}. Additionally, the presence of excited atoms or molecules is common in a so-called \og low-temperature \fg{} plasma.

In this paper, we generalize the results of Graille, Magin and Massot to multicomponent mixtures of \textit{polyatomic} gases. We assume that there is only one velocity in the mixture and discard multifluid models where each species has its own velocity \cite{Braginskii1965}. It is worth mentioning that macroscopic multifluid conservation equations lead to very serious mathematical pathologies \cite{Cordier1995}. The mixture of polyatomic species is described in a semi-classical framework by using Wang Chang-Uhlenbeck-de Boer equations. These equations preaverage the collision cross sections over degeneracies \cite{Raizer1987,ErnGiovangigli}. The chemical source terms appearing in the Boltzmann equations are taken essentially from Ludwig and Heil \cite{LudwigHeil}, Alexeev et al. \cite{AlexeevChikhaouiGrushin}, Ern and Giovangigli \cite{ErnGiovangigli1998} and Gr\"{u}nfeld \cite{Grunfeld1993}. These chemical source terms are valid for arbitrary chemical mechanisms. We also assume that the distribution functions do not depend on any of the angular momenta \cite{MasonMcDaniel1988}.

We study the Enskog expansion and obtain macroscopic equations in the zeroth- and first-order regimes, together with transport fluxes and transport coefficients. We further investigate the positivity properties of the resulting heavy-species multicomponent diffusion matrices. These properties, established here for the exact matrices arising from the kinetic theory of gases, are a key point in numerical approximations of multicomponent diffusion, where these properties must be enforced by the computational algorithms used to evaluate the transport coefficients.

Our paper is organized as follows. In section~\ref{SecKTKinFrame}, we introduce the kinetic framework. In section~\ref{SecKTAsymptEx}, we set the scaling hypotheses and derive an asymptotic expansion of the Boltzmann equations. In section~\ref{SecKTChapEnsk}, the Chapman-Enskog procedure is applied on the basis of the proposed scaling. The transport fluxes are expressed in terms of macroscopic variable gradients and transport coefficients in section~\ref{SecKTTransCoef}. Finally, section~\ref{SecKTConsEq} synthesizes the macroscopic equations obtained.

\section{Kinetic Framework}
\label{SecKTKinFrame}

In this section, we describe a theoretical framework for the kinetic theory of polyatomic non-thermal plasmas. We first introduce a generalized Boltzmann equation, which is then reformulated in the heavy-species reference frame. We next present collisional invariants of the collision operators.

\subsection{Generalized Boltzmann equation}

The starting point is the Boltzmann equation for reactive polyatomic ionized gas mixtures derived from Ref. \cite{GiovangigliGraille2003} in a semiclassical framework. It preaverages the collision cross sections over the degeneracies and can be derived from the Waldmann-Snider quantum mechanical Boltzmann equation \cite{Waldmann}.

The plasma is described as a multicomponent gas mixture of electrons, neutrals, and ions. The electron has only one internal degree of freedom, namely its ground state, and its distribution function reads $f_{e} \left( t, \boldsymbol{x}, \boldsymbol{c}_{e} \right)$, where $t$ is the time, $\boldsymbol{x}$ is the three-dimensional spatial coordinate, and $\boldsymbol{c}_{e}$ the velocity of the electron. We denote by $\Heavy$ the indexing set of heavy species, which can be ionized or not. For each $i \in \Heavy$, we denote by $\QuantumSpace_{i}$ the set of internal degrees of freedom associated with the $i^{\text{th}}$ heavy species. The distribution function for the $i^{\text{th}}$ heavy species then reads $f_{i} \left(t, \boldsymbol{x}, \boldsymbol{c}_{i}, \textsc{i} \right)$, where $\boldsymbol{c}_{i}$ denotes the velocity of the molecule, and $\textsc{i} \in \QuantumSpace_{i}$ its quantum state. Finally, we denote by $\Species = \Heavy \cup \left\lbrace e \right\rbrace$ the indexing set of the plasma species. The species distribution functions are then governed by a generalized Boltzmann equation that takes into account the reactive aspect of the mixture. We denote by $f_{\heavy} = \left( f_{i} \right)_{i \in \Heavy}$ the family of heavy-species distribution functions, and by $f = \left( f_{k} \right)_{k \in \Species} = \left( f_{e}, f_{\heavy} \right)$ the complete family of species distribution functions. The subscript \og $\heavy$ \fg{} refers to the set of heavy species. Finally, we denote by $q_{k}$ the charge carried by the $k^{\text{th}}$ species, while $\boldsymbol{E}$ and $\boldsymbol{B}$ refer to the electric and magnetic fields, respectively.

The Boltzmann equation governing the species distribution functions reads in an inertial reference frame \cite{ChapmanCowling} \cite{FerzigerKaper} \cite{Giovangigli} \cite{GrailleMaginMassot2009}
\begin{equation}
\Streaming_{k}(f_{k}) = \Scattering_{k}(f) + \Chemistry_{k}(f), \quad k \in \Species,
\label{KTBoltzmann}
\end{equation}
where $\Streaming_{k}$ denotes the usual streaming differential operator
\begin{equation}
\Streaming_{k}(f_{k}) = \partial_{t}f_{k} + \boldsymbol{c}_{k} \cdot \boldsymbol{\partial_{x}}f_{k} + \frac{q_{k}}{\Mass_{k}}\left[ \boldsymbol{E} + \boldsymbol{c}_{k} \wedge \boldsymbol{B} \right] \cdot \boldsymbol{\partial}_{\boldsymbol{c}_{k}}f_{k}, \quad k \in \Species,
\end{equation}
while $\Scattering_{k}\left(f\right)$ and $\Chemistry_{k} \left( f \right)$ are the scattering or nonreactive source term, and the chemically reactive source term, respectively.

Under the assumption that the system is dilute, the scattering source term can be written in the form
\begin{equation}
\Scattering_{k}\left( f \right) = \sum_{l \in \Species} \Scattering_{k l}\left( f_{k}, f_{l} \right),
\label{KTScattering}
\end{equation}
where $\Scattering_{k l}$ denotes the scattering source term for the $k^{\text{th}}$ species due to collisions with molecules of the $l^{\text{th}}$ species
\begin{equation}
\Scattering_{k l}\left( f \right) = \sum\limits_{\substack{ \textsc{k}' \in \QuantumSpace_{k} \\ \textsc{l}, \textsc{l}' \in \QuantumSpace_{l} }} \int \big( f_{k}'f_{l}' \frac{a_{k \textsc{k}} a_{l \textsc{l}}}{a_{k \textsc{k}'} a_{l \textsc{l}'}} - f_{k}f_{l} \big) |\boldsymbol{c}_{k}-\boldsymbol{c}_{l}| \CrossSection_{kl}^{\textsc{k}\textsc{l}\textsc{k}'\textsc{l}'} \, \mathrm{d} \boldsymbol{\omega}_{kl}' \mathrm{d} \boldsymbol{c}_{l}.
\label{KTCrossScattering}
\end{equation}
We have denoted by $a_{k \textsc{k}}$ the degeneracy of the $\textsc{k}^{\text{th}}$ quantum energy shell of the $k^{\text{th}}$ species, $\boldsymbol{\omega}_{kl} = (\boldsymbol{c}_{k}- \boldsymbol{c}_{l})/|\boldsymbol{c}_{k}- \boldsymbol{c}_{l}|$ and $\boldsymbol{\omega}_{kl}' = (\boldsymbol{c}_{k}'- \boldsymbol{c}_{l}')/|\boldsymbol{c}_{k}'- \boldsymbol{c}_{l}'|$ the directions of the relative velocities, respectively before and after collision, and $\CrossSection_{kl}^{\textsc{k}\textsc{l}\textsc{k}'\textsc{l}'}$ the differential cross-section associated with a binary collision between a molecule of the $k^{\text{th}}$ species in internal energy state $\textsc{k}$ and a molecule of the $l^{\text{th}}$ species in internal  energy state $\textsc{l}$.

The differential cross-section is taken in the classical form \cite{Waldmann} \cite{ChapmanCowling} \cite{FerzigerKaper} \cite{GrailleMaginMassot2009}
\begin{equation}
\CrossSection_{kl}^{\textsc{k}\textsc{l}\textsc{k}'\textsc{l}'} = \CrossSection_{kl}^{\textsc{k}\textsc{l}\textsc{k}'\textsc{l}'} \left( \frac{\mu_{kl} g_{kl}^{2}}{k_{\textsc{b}} T^{0}}, \boldsymbol{\omega}_{kl} \cdot \boldsymbol{\omega}_{kl}' \right),
\end{equation}
where $\mu_{kl} = \Mass_{k}\Mass_{l}/(\Mass_{k}+\Mass_{l})$ is the reduced mass of the pair of particles, $g_{kl} = |\boldsymbol{c}_{k}-\boldsymbol{c}_{l}|$ is their relative velocity, $T^{0}$ is a reference temperature which is common to all species, and $k_{\textsc{b}}$ is the Boltzmann constant. One could also work with transition probabilities $\TransitionProbabilities_{kl}^{\textsc{k} \textsc{l} \textsc{k}' \textsc{l}'}$ rather than with classical collision cross-sections $\CrossSection_{kl}^{\textsc{k}\textsc{l}\textsc{k}'\textsc{l}'}$. Transition probabilities are notably interesting with reactive collisions, since the term $\Chemistry_{k}(f)$ is then much easier to write. For binary collisions, transition probabilities and differential cross-sections are related through the following identity \cite{LudwigHeil} \cite{AlexeevChikhaouiGrushin} \cite{GrunfeldGeorgescu}
\begin{equation}
|\boldsymbol{c}_{k} - \boldsymbol{c}_{l}| \CrossSection_{kl}^{\textsc{k}\textsc{l}\textsc{k}'\textsc{l}'} \, \mathrm{d} \boldsymbol{\omega}_{kl}' = \TransitionProbabilities_{kl}^{\textsc{k} \textsc{l} \textsc{k}' \textsc{l}'} \, d \boldsymbol{c}_{k}' d \boldsymbol{c}_{l}'.
\end{equation}
Classically, the forward and reverse collision cross-sections exhibit reciprocity relations in the form \cite{Waldmann} \cite{WaldmannTrubenbacher} \cite{ChapmanCowling} \cite{FerzigerKaper}
\begin{equation}
a_{k \textsc{k}} a_{l \textsc{l}} ~ g_{kl} ~ \CrossSection_{kl}^{\textsc{k}\textsc{l}\textsc{k}'\textsc{l}'} \, \mathrm{d} \boldsymbol{C}_{k} \mathrm{d} \boldsymbol{C}_{l} \mathrm{d} \boldsymbol{\omega}_{kl}' = a_{k \textsc{k}'} a_{l \textsc{l}'} ~ g_{kl}' ~ \CrossSection_{kl}^{\textsc{k}'\textsc{l}'\textsc{k}\textsc{l}} \, \mathrm{d} \boldsymbol{C}_{k}' \mathrm{d} \boldsymbol{C}_{l}' \mathrm{d} \boldsymbol{\omega}_{kl}.
\label{KTScattering Reciprocity Relation} 
\end{equation}

For the reactive, or chemistry, source term $\Chemistry_{k}(f)$, we consider a chemical reaction mechanism composed of an arbitrary number of elementary reactions. Unlike for the scattering process, we take into account multiple reactive collisions, including triple reactive collisions since recombination reactions cannot often proceed otherwise \cite{Kuscer} \cite{AlexeevChikhaouiGrushin} \cite{GrunfeldGeorgescu}. If we denote by $\Reaction$ the set of reactions, each chemical reaction $r \in \Reaction$ can be written in the form
\begin{equation}
\sum_{k \in \Forward^{r}} \Molecule_{k} \rightleftharpoons \sum_{k \in \Backward^{r}} \Molecule_{k}, \quad r \in \Reaction,
\end{equation}
where $\Molecule_{k}$ denotes the chemical symbol of the $k^{\text{th}}$ species, and where $\Forward^{r}$ and $\Backward^{r}$ are, respectively, the indices for the reactant and product species in the $r^{\text{th}}$ elementary reaction, counted with their order of multiplicity. The letters $\Forward^{r}$ and $\Backward^{r}$ are mnemonic for the forward and backward directions, respectively. We denote by $\nu_{kr}^{\mathrm{f}}$ and $\nu_{kr}^{\mathrm{b}}$ the stoichiometric coefficients of the $k^{\text{th}}$ species among reactants and products, respectively, and we also denote by $\textsc{f}^{r}$ and $\textsc{b}^{r}$ the indices of internal energy states for reactants and products, respectively. In other words, the forward and backward coefficients $\nu_{kr}^{\mathrm{f}}$ and $\nu_{kr}^{\mathrm{b}}$ are the order of multiplicity of the $k^{\text{th}}$ species in $\Forward^{r}$ and $\Backward^{r}$, respectively. For a given $k \in \Species$, $\Forward_{k}^{r}$ denotes the set of reactant indices where the index for the $k^{\text{th}}$ species has been removed only once and we introduce a similar notation for $\Backward_{k}^{r}$, $\textsc{f}_{\textsc{k}}^{r}$ and $\textsc{b}_{\textsc{k}}^{r}$.

The reactive source term, $\Chemistry_{k}(f)$, then reads \cite{ErnGiovangigli1998}
\begin{equation}
\Chemistry_{k}(f) = \sum_{r \in \Reaction} \Chemistry_{k}^{r}(f),
\label{KTChemistrySource}
\end{equation}
where $\Chemistry_{k}^{r}(f)$ is the source term for the $k^{\text{th}}$ species due to the $r^{\text{th}}$ elementary reaction
\begin{align}
\Chemistry_{k}^{r}(f) & = \nu_{k}^{r\mathrm{f}} \sum\limits_{ \textsc{f}_{\textsc{k}}^{r}, \textsc{b}^{r}} \int \Big( \prod_{j \in \Backward^{r}} f_{j} ~ \frac{\prod\limits_{j \in \Backward^{r}} \beta_{j \textsc{j}}}{\prod\limits_{i \in \Forward^{r}} \beta_{i \textsc{i}}} ~ - ~ \prod\limits_{i \in \Forward^{r}} f_{i} \Big)  \TransitionProbabilities_{\Forward^{r} \Backward^{r}}^{\textsc{f}^{r} \textsc{b}^{r}} \, \prod_{i \in \Forward_{k}^{r}} \mathrm{d} \boldsymbol{c}_{i} \prod_{j \in \Backward^{r}} \mathrm{d} \boldsymbol{c}_{j}
\label{KTChemistrySpeciesperReaction} \\
{} & + \nu_{k}^{r\mathrm{b}} \sum\limits_{ \textsc{f}^{r}, \textsc{b}_{\textsc{k}}^{r}} \int \Big( \prod\limits_{i \in \Forward^{r}} f_{i} ~ - ~ \frac{\prod\limits_{j \in \Backward^{r}} \beta_{j \textsc{j}}}{\prod\limits_{i \in \Forward^{r}} \beta_{i \textsc{i}}} ~ \prod_{j \in \Backward^{r}} f_{j} \Big) \TransitionProbabilities_{\Forward^{r} \Backward^{r}}^{\textsc{f}^{r} \textsc{b}^{r}} \, \prod_{i \in \Forward^{r}} \mathrm{d} \boldsymbol{c}_{i} \prod_{j \in \Backward_{k}^{r}} \mathrm{d} \boldsymbol{c}_{j}, \nonumber
\end{align}
where $\beta_{k \textsc{k}} = h_{\textsc{p}}^{3}/(a_{k \textsc{k}} \Mass_{k}^{3})$, and $h_{\textsc{p}}$ is the Planck constant. The quantity $\TransitionProbabilities_{\Forward^{r} \Backward^{r}}^{\textsc{f}^{r} \textsc{b}^{r}}$ is the transition probability for a reactive collision in which the reactants $\Forward^{r}$ with internal energy states $\textsc{f}^{r}$ are transformed into products $\Backward^{r}$ with internal energy states $\textsc{b}^{r}$. The sums over $\textsc{f}^{r}$, respectively $\textsc{f}_{\textsc{k}}^{r}$, represent the sums over $\textsc{i} \in \QuantumSpace_{i}$ for all $i \in \Forward^{r}$, respectively $i \in \Forward_{k}^{r}$. Similarly, the sums over $\textsc{b}^{r}$, respectively $\textsc{b}_{\textsc{k}}^{r}$, represent the sums over $\textsc{j} \in \QuantumSpace_{j}$ for all $j \in \Backward^{r}$, respectively $j \in \Backward_{k}^{r}$. Several examples for different types of reactions are given in \cite{Giovangigli}. Finally, the reactive transition probabilities exhibit the following symmetry properties \cite{LudwigHeil} \cite{GrunfeldGeorgescu} \cite{AlexeevChikhaouiGrushin} \cite{Giovangigli}
\begin{equation}
\TransitionProbabilities_{\Forward^{r} \Backward^{r}}^{\textsc{f}^{r} \textsc{b}^{r}} \prod_{j \in \Backward^{r}} \beta_{j \textsc{j}} = \TransitionProbabilities_{\Backward^{r} \Forward^{r}}^{\textsc{b}^{r} \textsc{f}^{r}} \prod_{i \in \Forward^{r}} \beta_{i \textsc{i}}.
\label{KTChemistryReciprocityRelation} 
\end{equation}

\subsection{Heavy-species reference frame}

Given the strong disparity of masses between electrons and heavy species, it is natural to choose a reference frame associated with the motion of the heavy species \cite{GrailleMaginMassot2009}. We thus introduce the mean electron and mean heavy-species velocities, given by
\begin{eqnarray}
\rho_{e} \boldsymbol{\Meanv}_{e} & = & \int \Mass_{e} \boldsymbol{c}_{e} f_{e} \, \mathrm{d} \boldsymbol{c}_{e}, \\
\rho_{\heavy} \boldsymbol{\Meanv}_{\heavy} & = & \sum\limits_{j \in \Heavy} \sum\limits_{\textsc{j} \in \QuantumSpace_{j}} \int \Mass_{j} \boldsymbol{c}_{j} f_{j} \, \mathrm{d} \boldsymbol{c}_{j},
\end{eqnarray}
where the subscript \og $\heavy$ \fg{} refers to the set of heavy species, and $\rho_{\heavy} = \sum_{j \in \Heavy} \rho_{j}$ is the heavy-species mass density, so that $\rho = \rho_{e} + \rho_{\heavy}$. The hydrodynamic velocity of the fluid is then given by
\begin{equation}
\rho \boldsymbol{\Meanv} = \rho_{e} \boldsymbol{\Meanv}_{e} + \rho_{\heavy} \boldsymbol{\Meanv}_{\heavy}. \label{KTMeanv}
\end{equation}
We now introduce the peculiar velocity of the $k^{\text{th}}$ species with respect to the heavy-species reference frame
\begin{equation}
\boldsymbol{C}_{k} = \boldsymbol{c}_{k}-\boldsymbol{\Meanv}_{\heavy}, \quad k \in \Species,
\end{equation}
and denote by $f_{k}(t,\boldsymbol{x},\boldsymbol{C}_{k},\textsc{k})$ the distribution function of the $k^{\text{th}}$ species in the new reference frame.

In the heavy-species reference frame, the streaming operator $\Streaming_{k}$ reads \cite{GrailleMaginMassot2009}
\begin{align}
\Streaming_{k}(f_{k}) = {} & \partial_{t}f_{k} + \left(\boldsymbol{C}_{k}+ \boldsymbol{\Meanv}_{\heavy} \right) \cdot \boldsymbol{\partial_{x}}f_{k} + \frac{q_{k}}{\Mass_{k}} \left[ \boldsymbol{E} + \left(\boldsymbol{C}_{k}+ \boldsymbol{\Meanv}_{\heavy}\right) \wedge \boldsymbol{B} \right] \cdot \boldsymbol{\partial}_{\boldsymbol{C}_{k}} f_{k} \\
{} & - \frac{D \boldsymbol{\Meanv}_{\heavy}}{Dt} \cdot \boldsymbol{\partial}_{\boldsymbol{C}_{k}}f_{k} - \left( \boldsymbol{\partial}_{\boldsymbol{C}_{k}}f_{k} \otimes \boldsymbol{C}_{k} \right) : \boldsymbol{\partial_{x}} \boldsymbol{\Meanv}_{\heavy}, \nonumber
\end{align}
where $\frac{D}{Dt}$ is the time derivative following the heavy-species velocity reference frame
\begin{equation}
\frac{D}{Dt} = \partial_{t} + \boldsymbol{\Meanv}_{\heavy} \cdot \boldsymbol{\partial_{x}}.
\end{equation}
The scattering source term \eqref{KTCrossScattering} may be rewritten using the new velocities $\boldsymbol{C}_{k}$, $k \in \Species$, in the form
\begin{equation}
\Scattering_{k l}\left( f_{k}, f_{l} \right) = \sum_{l \in \Species \vphantom{\textsc{l}'}} \sum\limits_{\substack{ \textsc{k}' \in \QuantumSpace_{k} \\ \textsc{l}, \textsc{l}' \in \QuantumSpace_{l} }} \int \big( f_{k}'f_{l}' \frac{a_{k \textsc{k}} a_{l \textsc{l}}}{a_{k \textsc{k}'} a_{l \textsc{l}'}} - f_{k}f_{l} \big) |\boldsymbol{C}_{k}-\boldsymbol{C}_{l}| \CrossSection_{kl}^{\textsc{k}\textsc{l}\textsc{k}'\textsc{l}'} \, \mathrm{d} \boldsymbol{\omega}_{kl}' \mathrm{d} \boldsymbol{C}_{l},
\label{KTKineticScatHframe}
\end{equation}
and as well the reactive source term \eqref{KTChemistrySpeciesperReaction} now reads in the new reference frame
\begin{align}
\Chemistry_{k}^{r}(f) & = \nu_{k}^{r\mathrm{f}} \sum\limits_{ \textsc{f}_{\textsc{k}}^{r}, \textsc{b}^{r}} \int \big( \prod_{j \in \Backward^{r}} f_{j} ~ \frac{\prod\limits_{j \in \Backward^{r}} \beta_{j \textsc{j}}}{\prod\limits_{i \in \Forward^{r}} \beta_{i \textsc{i}}} ~ - ~ \prod\limits_{i \in \Forward^{r}} f_{i} \big)  \TransitionProbabilities_{\Forward^{r} \Backward^{r}}^{\textsc{f}^{r} \textsc{b}^{r}} \, \prod_{i \in \Forward_{k}^{r}} \mathrm{d} \boldsymbol{C}_{i} \prod_{j \in \Backward^{r}} \mathrm{d} \boldsymbol{C}_{j} \label{KTChemistrySourcePerReaction} \\
{} & + \nu_{k}^{r\mathrm{b}} \sum\limits_{ \textsc{f}^{r}, \textsc{b}_{\textsc{k}}^{r}} \int \big( \prod\limits_{i \in \Forward^{r}} f_{i} ~ - ~ \frac{\prod\limits_{j \in \Backward^{r}} \beta_{j \textsc{j}}}{\prod\limits_{i \in \Forward^{r}} \beta_{i \textsc{i}}} ~ \prod_{j \in \Backward^{r}} f_{j} \big) \TransitionProbabilities_{\Forward^{r} \Backward^{r}}^{\textsc{f}^{r} \textsc{b}^{r}} \, \prod_{i \in \Forward^{r}} \mathrm{d} \boldsymbol{C}_{i} \prod_{j \in \Backward_{k}^{r}} \mathrm{d} \boldsymbol{C}_{j}. \nonumber
\end{align}

\subsection{Collisional invariants}
\label{KTCollisional Invariants}

Collisional invariants are associated with macroscopic conservation equations and are therefore of fundamental importance. Collisional invariants of the scattering operator $\Scattering$ are functionals $\psi = \left( \psi_{l} \right)_{l \in \Species}$, where $\psi_{l} = \psi_{l}(t,\boldsymbol{x},\boldsymbol{C}_{l},\textsc{l})$ is a scalar or tensor function of $t$, $\boldsymbol{x}$, $\boldsymbol{C}_{l}$, and $\textsc{l}$,  whose values summed over the particles involved in a nonreactive collision do not change during the collision
\begin{equation}
\psi_{k} + \psi_{l} = \psi_{k}' + \psi_{l}', \qquad \qquad k,l \in \Species,
\end{equation}
where $\psi_{k}' = \psi_{k}(t,\boldsymbol{x},\boldsymbol{C}_{k}',\textsc{k}')$ and $\psi_{l}' = \psi_{l}(t,\boldsymbol{x},\boldsymbol{C}_{l}',\textsc{l}')$.

There are $n^{s}+4$ linearly independent scalar collisional invariants, which can be taken in the form \cite{WaldmannTrubenbacher}
\begin{equation}
\left\lbrace \begin{array}{lll}
\displaystyle \psi^{k} = \left( \delta_{kl} \right)_{l \in \Species}, & \displaystyle k \in \Species, \vphantom{\frac{1}{2}} \\
\displaystyle \psi^{n^{s}+\nu} =\left(  \Mass_{l} C_{l \nu} \right)_{l \in \Species}, & \displaystyle \nu \in \left\lbrace 1,2,3 \right\rbrace, \vphantom{\frac{1}{2}} \\
\displaystyle \psi^{n^{s}+4} = \Big( \frac{1}{2} \Mass_{l} \boldsymbol{C}_{l} \cdot \boldsymbol{C}_{l} + E_{l\textsc{l}}  \Big)_{l \in \Species}, & \displaystyle {}
\end{array}
\right.
\end{equation}
where $n^{s}$ is the number of species in $\Species$, $C_{l \nu}$ is the component of $\boldsymbol{C}_{l}$ in the $\nu^{\text{th}}$ spatial coordinate, and $E_{l \textsc{l}}$ is the internal energy of the $l^{\text{th}}$ species in the $\textsc{l}^{\text{th}}$ quantum state. The three kinds of collisional invariants thus defined correspond respectively to the conservation of chemical species, the conservation of momentum, and the conservation of energy during nonreactive collisions. For micropolar fluids there is an additional linearly independent summational invariant, accounting for the conservation of angular momentum \cite{GraillePhD} \cite{KaganAfanasev}, but we consider in this study isotropic mixtures only, so that there are no micropolar effects. We name $\Invariants$ the space of collisional invariants with respect to the scattering operator $\Scattering$, i.e., the space spanned by the family $\left( \psi^{p}\right)_{ 1 \leq p \leq n^{s}+4 }$.

We further introduce a tensorial product defined for scalar functions $\xi = \left( \xi_{l} \right)_{l \in \Species}$ and $\zeta = \left( \zeta_{l} \right)_{l \in \Species}$ as
\begin{equation}
\llangle \xi, \zeta \rrangle = \sum\limits_{k \in \Species} \sum\limits_{\textsc{k} \in \QuantumSpace_{k}} \int \xi_{k} \zeta_{k} \, \mathrm{d} \boldsymbol{C}_{k},
\end{equation}
and more generally for tensors $\boldsymbol{\xi} = \left( \boldsymbol{\xi}_{l} \right)_{l \in \Species}$ and $\boldsymbol{\zeta} = \left( \boldsymbol{\zeta}_{l} \right)_{l \in \Species}$ as
\begin{equation}
\llangle \boldsymbol{\xi}, \boldsymbol{\zeta} \rrangle = \sum\limits_{k \in \Species} \sum\limits_{\textsc{k} \in \QuantumSpace_{k}} \int \boldsymbol{\xi}_{k} \odot \boldsymbol{\zeta}_{k} \, \mathrm{d} \boldsymbol{C}_{k},
\end{equation}
where $\odot$ stands for the fully contracted product in space \cite{ErnGiovangigli} \cite{GraillePhD}.

The scattering operator $\Scattering$ and the corresponding collisional invariants then satisfy the classical relations
\begin{equation}
\llangle \psi^{p}, \Scattering(f) \rrangle = 0, \qquad p \in \left\lbrace 1, \ldots, n^{s}+4 \right\rbrace.
\label{KTInvariantsConservation}
\end{equation}
Indeed, using reciprocity relation \eqref{KTScattering Reciprocity Relation} and symmetrizing between $k$ and $l$, one can establish that for any $\psi = \left( \psi_{l} \right)_{l \in \Species}$
\begin{eqnarray}
4 \llangle \psi, \Scattering(f) \rrangle & = & \sum_{k,l \in \Species \vphantom{\textsc{l}'}} ~ \sum\limits_{\textsc{k},\textsc{k}' \in \QuantumSpace_{k}} ~ \sum\limits_{\textsc{l}, \textsc{l}' \in \QuantumSpace_{l}} ~ \int \big( \psi_{k} + \psi_{l} - \psi_{k}' - \psi_{l}' \big) \\
{} & {} & \times \big( f_{k}'f_{l}' \frac{a_{k \textsc{k}} a_{l \textsc{l}}}{a_{k \textsc{k}'} a_{l \textsc{l}'}} - f_{k}f_{l} \big) ~ g_{kl} ~ \CrossSection_{kl}^{\textsc{k}\textsc{l}\textsc{k}'\textsc{l}'} \, \mathrm{d} \boldsymbol{\omega}_{kl}' \, \mathrm{d} \boldsymbol{C}_{k} \mathrm{d} \boldsymbol{C}_{l}, \nonumber
\end{eqnarray}
which is zero as soon as $\psi$ is a collisional invariant.

For our case of interest, it turns out that we have additional orthogonality relations, by considering pairwise interaction terms separately. Indeed, we can decompose the scalar product $\llangle \cdot \rrangle$ as
\begin{equation}
\llangle \xi, \zeta \rrangle = \llangle \xi_{e}, \zeta_{e} \rrangle_{e} + \llangle \xi_{\heavy}, \zeta_{\heavy} \rrangle_{\heavy},
\label{KTScalarProduct}
\end{equation}
where
\begin{eqnarray}
\llangle \xi_{e}, \zeta_{e} \rrangle_{e} = \int \xi_{e} \odot \zeta_{e} \, \mathrm{d} \boldsymbol{C}_{e}, \\
\llangle \xi_{\heavy}, \zeta_{\heavy} \rrangle_{\heavy} = \sum\limits_{j \in \Heavy} \sum\limits_{\textsc{j} \in \QuantumSpace_{j}} \int \xi_{j} \odot \zeta_{j} \, \mathrm{d} \boldsymbol{C}_{j},
\end{eqnarray}
and obtain, as in \cite{GrailleMaginMassot2009}, the following orthogonality property for pairwise interactions:
\begin{eqnarray}
\llangle \psi_{e}^{p}, \Scattering_{ee} \rrangle_{e} = 0, \vphantom{\sum_{j \in \Heavy}}
\label{KTPairwiseInvariantsConservationElectron} \\
\llangle \psi_{\heavy}^{p}, \Scattering_{\heavy e} \rrangle_{\heavy} + \sum_{j \in \Heavy} \llangle \psi_{e}^{p}, \Scattering_{ej} \rrangle_{e} = 0,
\label{KTPairwiseInvariantsConservationHeavyElectron} \\
\sum_{j \in \Heavy} \llangle \psi_{\heavy}^{p}, \Scattering_{\heavy j} \rrangle_{\heavy} = 0,
\label{KTPairwiseInvariantsConservationHeavy}
\end{eqnarray}
for any $p \in \left\lbrace 1, \ldots, n^{s}+4 \right\rbrace$.

Unlike for the nonreactive source term, the species are not conserved during reactive collisions, and only chemical elements are conserved \cite{Giovangigli}
\begin{equation}
\sum_{k \in \Species} \nu_{k}^{r\mathrm{f}} \element_{kl} = \sum_{k \in \Species} \nu_{k}^{r\mathrm{b}} \element_{kl}, \quad r \in \Reaction, \, l \in \Elements,
\label{KTElementsStoichio}
\end{equation}
where $\element_{kl}$ is the number of $l^{\text{th}}$ atom in the $k^{\text{th}}$ species, and $\Elements$ denotes the indexing set for the atoms present in the mixture. The conservation of total mass during reactive collisions then follows from equation \eqref{KTElementsStoichio}
\begin{equation}
\sum_{k \in \Species} \nu_{k}^{r\mathrm{f}} \Mass_{k} = \sum_{k \in \Species} \nu_{k}^{r\mathrm{b}} \Mass_{k}.
\label{KTMassStoichio}
\end{equation}

\section{Asymptotic Expansion of the Boltzmann Equations}
\label{SecKTAsymptEx}

Dimensional analysis is a necessary preliminary to the Chapman-Enskog procedure. In this regard, we follow the scaling first introduced by Petit and Darrozes \cite{PetitDarrozes}. We take as small parameter $\varepsilon$ the square root of the ratio between the characteristic masses. As shown by Petit and Darrozes, when both the Knudsen number $\Knudsen$ and the mass ratio $\varepsilon$ tend to zero, $\Knudsen$ must be chosen proportional to $\varepsilon$. The scaling introduced here will serve as a basis for the derivation of a scaled Boltzmann equation, in which the different terms will depend on the small parameter $\varepsilon$ \cite{ChapmanCowling} \cite{FerzigerKaper} \cite{Giovangigli} \cite{GrailleMaginMassot2009}.

\subsection{Choice of scaling}

The reference quantities used in the scaling are denoted by the superscript "0". Most of the reference quantities are common to all species, though it is necessary to distinguish between electron and heavy-species respective characteristic masses, velocities, and kinetic time scales. Also, the characteristic cross-section for inelastic scattering of electrons by heavy species is denoted by $\CrossSection_{e \heavy}^{\text{in},0} = \CrossSection_{\heavy e}^{\text{in},0}$, while the characteristic cross-section for other scattering processes is denoted by $\CrossSection^{0}$.

\renewcommand{\arraystretch}{1.25}

\begin{table}[h!]
\begin{center}
\caption{Reference quantities \cite{GrailleMaginMassot2009}.}
\label{TableReferencequantities}
\begin{tabular}{lcc}
  \midrule
  \textbf{Physical entity}                                          & \multicolumn{2}{c}{\textbf{Common to all species}} \\
  \midrule
  Temperature                                 & \multicolumn{2}{c}{$\Temperature^{0}$} \\
  Number density                              & \multicolumn{2}{c}{$\density^{0}$} \\
  Charge                                      & \multicolumn{2}{c}{$q^{0}$} \\
  Scattering cross-section                  & \multicolumn{2}{c}{$\CrossSection^{0}$} \\
  Mean free path                              & \multicolumn{2}{c}{$l^{0} = \frac{1}{\density^{0} \CrossSection^{0}}$} \\
  Macroscopic time scale                      & \multicolumn{2}{c}{$t^{0}$} \\
  Hydrodynamic velocity                       & \multicolumn{2}{c}{$\Meanv^{0}$} \\
  Macroscopic length                          & \multicolumn{2}{c}{$L^{0} = \Meanv^{0} t^{0}$} \\
  Electric field                              & \multicolumn{2}{c}{$E^{0}$} \\
  Magnetic field                              & \multicolumn{2}{c}{$B^{0}$} \\
  Reactive source term                        & \multicolumn{2}{c}{$\Chemistry_{k}^{0}$, $k \in \Species$} \\
  \midrule
  {}                                          & \textbf{Electrons} & \textbf{Heavy species} \\
  \midrule
  Mass                                        & $\Mass_{e}^{0} = \Mass_{e}$           & $\Mass_{\heavy}^{0}$ \\
  Thermal speed                               & $C_{e}^{0}$                           & $C_{\heavy}^{0}$ \\
  Kinetic timescale                           & $t_{e}^{0} = \frac{l^{0}}{C_{e}^{0}}$ & $t_{\heavy}^{0} = \frac{l^{0}}{C_{\heavy}^{0}}$ \\
  \midrule
  {}                                          & \multicolumn{2}{c}{\textbf{Hybrid}} \\
  \midrule
  Inelastic scattering cross-sections & \multicolumn{2}{c}{$\CrossSection^{\text{in},0}_{he}$} \\
  \midrule
\end{tabular}
\end{center}
\end{table}

\renewcommand{\arraystretch}{1}

\paragraph{Mass ratio}
The ratio of the electron mass $\Mass_{e}^{0} = \Mass_{e}$ to the characteristic heavy-species mass $\Mass_{\heavy}^{0}$ is such that
\begin{equation}
\sqrt{\frac{\Mass_{e}^{0}}{\Mass_{\heavy}^{0}}} = \varepsilon \ll 1.
\label{KTMassRatio}
\end{equation}
The non-dimensional number $\varepsilon$ will be the key parameter driving the asymptotic analysis of the plasma.

\paragraph{Temperatures}
The reference temperature is the same for electrons and heavy species \cite{GrailleMaginMassot2009}:
\begin{equation}
\Temperature_{e}^{0} = \Temperature_{\heavy}^{0} = \Temperature^{0}.
\label{KTReferenceTemperature}
\end{equation}
This means that the electron temperature $\Temperature_{e}$ and the heavy-species temperature $\Temperature_{\heavy}$  will remain of the same order of magnitude in the model.

\paragraph{Velocities}
As a consequence of assumptions \eqref{KTMassRatio}-\eqref{KTReferenceTemperature}, electrons exhibit a larger thermal speed than heavy species
\begin{align}
 & C_{\heavy}^{0} = \sqrt{\frac{k_{\textsc{b}} \Temperature^{0}}{\Mass_{\heavy}^{0}}},
 \label{KTCh0} \\
 & C_{e}^{0} = \sqrt{\frac{k_{\textsc{b}} \Temperature^{0}}{\Mass_{e}^{0}}} = \frac{1}{\varepsilon} C_{\heavy}^{0}.
 \label{KTCe0}
\end{align}
Besides, the pseudo-Mach number, defined as the reference hydrodynamic velocity $\Meanv^{0}$ divided by the heavy-species thermal speed $C_{\heavy}^{0}$, is of order one \cite{GrailleMaginMassot2009}
\begin{equation}
\PseudoMach = \frac{\Meanv^{0}}{C_{\heavy}^{0}} \propto 1.
\label{KTMach}
\end{equation}
In other words, there is only one reference velocity for the heavy species.

\paragraph{Densities}
As stated in \cite{PetitDarrozes}, the \og weakly ionized \fg{} limit is not singular with respect to the limits $\Knudsen \to 0$ and $\sqrt{\Mass_{e}^{0} / \Mass_{\heavy}^{0}} \to 0$. Therefore, we adopt the same scaling for both electron and heavy-species densities:
\begin{equation}
\density_{e}^{0} = \density_{\heavy}^{0} = \density^{0}.
\end{equation}
The results for a weakly ionized plasma will then follow by taking the limit $\density_{e}^{0} / \density_{\heavy}^{0} \to 0$.

\paragraph{Mean free path}
The characteristic mean free path \cite{GrailleMaginMassot2009}
\begin{equation}
l^{0} = \frac{1}{\density^{0}\CrossSection^{0}}.
\label{KTMeanFreePath}
\end{equation}
is imposed by the carrier gas density $\density^{0}$ and the reference elastic scattering cross-section $\CrossSection^{0}$ \cite{ChapmanCowling} \cite{FerzigerKaper}, and is thus common to all species \cite{GrailleMaginMassot2009}.

\paragraph{Time scales}
From \eqref{KTCh0} and \eqref{KTCe0}, the kinetic timescales, or the relaxation times of the distribution functions towards their respective quasi-equilibrium states, are given by
\begin{align}
 & t_{e}^{0} = \frac{l^{0}}{C_{e}^{0}}, \\
 & t_{\heavy}^{0} = \frac{l^{0}}{C_{\heavy}^{0}},
\end{align}
and therefore \cite{GrailleMaginMassot2009} $t_{e}^{0} = \varepsilon t_{\heavy}^{0}$. The macroscopic timescale $t^{0}$ is one order of magnitude larger than the heavy-species kinetic timescale $t_{\heavy}^{0}$, so that there are three distinct relevant time scales $t_{e}^{0}$, $t_{\heavy}^{0}$, and $t^{0}$ \cite{GrailleMaginMassot2009}:
\begin{equation}
t_{e}^{0} = \varepsilon t_{\heavy}^{0} = \varepsilon^{2} t^{0}.
\label{KTTimeScales}
\end{equation}

\paragraph{Inelastic scattering cross-sections}

As can be seen from \eqref{KTTimeScales}, electron thermalization is the fastest process corresponding to the kinetic scale $t_{e}^{0}$. This thermalization is ensured by scattering collisions between electrons, and elastic collisions between electrons and heavy species, which given the strong mass disparity do not involve any energy exchange at the lowest order, but allow for isotropization of the electron distribution function in the heavy-species reference frame \cite{GrailleMaginMassot2009}. Since the heavy species' internal degrees of freedom thermalize at $\Temperature_{\heavy}$, further allowing for energy exchange between these internal degrees of freedom and electrons at the lowest order would actually require $\Temperature_{e} = \Temperature_{\heavy}$. Thus the reference differential cross-section associated with inelastic scattering between heavy species and electrons $\CrossSection^{\text{in},0}_{\heavy e}$ must be negligible compared to the reference cross-section $\CrossSection^{0}$ for other scattering collisions.

The second fastest kinetic process is the thermalization of heavy species \cite{GrailleMaginMassot2009}, corresponding to the time scale $t_{\heavy}^{0} = \varepsilon^{-1} t_{e}^{0}$ as described in \eqref{KTTimeScales}. This thermalization arises from elastic and inelastic collisions between heavy species, while elastic collisions with electrons are of negligible influence at this order due to the strong mass disparity. Again, since the heavy species' internal degrees of freedom thermalize at $\Temperature_{\heavy}$, inelastic collisions between electrons and heavy species must be neglibible at the lowest order of the Chapman-Enskog expansion for heavy species, otherwise one would have $\Temperature_{\heavy} = \Temperature_{e}$. In other words, $\CrossSection^{\text{in},0}_{\heavy e}$ must be negligible compared to $\varepsilon\CrossSection^{0}$.

Therefore, the inelastic scattering collisions between electrons and heavy species are assumed to be two orders of magnitude slower than the corresponding elastic collisions
\begin{equation}
\CrossSection^{\text{in},0}_{\heavy e} = \varepsilon^{2} \CrossSection^{0}.
\end{equation}
The latter requirement will be discussed in more details in subsection \ref{SubSecKTScalingInelastic}, and other regimes will be addressed in the conclusion.

%In a future study, we intend to consider a model where some of the heavy species internal energy levels do not thermalize at $\Temperature_{\heavy}$ \cite{MaginGrailleMassot2012} \cite{Munafo} but rather at some internal temperature $\Temperature_{\heavy}^{\text{int}}$, and where energy transfer between electrons and the hottest internal energy levels of the heavy species would be possible, so that $\Temperature_{\heavy}^{\text{int}} = \Temperature_{e}$.

\paragraph{Knudsen number}
The macroscopic length scale is based on a reference convective length
\begin{equation}
L^{0} = \Meanv^{0} t^{0}.
\end{equation}
As a consequence of the proposed scaling, the Knudsen number
\begin{equation}
\Knudsen = \frac{l^{0}}{L^{0}} = \frac{\varepsilon}{\PseudoMach}
\end{equation}
is small compared to $1$, justifying the choice of a continuum description of the gas.

\paragraph{Electric field}
The reference electrical and thermal energies are of the same order of magnitude, namely
\begin{equation}
q^{0} E^{0} L^{0} = k_{\textsc{b}} \Temperature^{0}.
\label{KTElectricFieldScaling}
\end{equation}

\paragraph{Magnetic field}

The intensity of the magnetic field is related to the Hall numbers of electron and heavy species, defined as the Larmor frequencies, respectively $q^{0}B^{0}/\Mass_{e}$ and $q^{0}B^{0}/\Mass_{\heavy}^{0}$, multiplied by the corresponding kinetic timescales. The magnetic field is assumed to be proportional to a power of $\varepsilon$ by means of an integer $0 \leq b \leq 1$:
\begin{align}
\beta_{e} & = \frac{q^{0} B^{0}}{\Mass_{e}^{0}} t_{e}^{0} = \varepsilon^{1-b},
\label{KTHallElectron} \\
\beta_{\heavy} & = \frac{q^{0} B^{0}}{\Mass_{\heavy}^{0}} t_{\heavy}^{0} = \varepsilon \beta_{e}.
\label{KTHallHeavy}
\end{align}
The case $b=1$ corresponds to strongly magnetized plasmas, the case $b = 0$ to weakly magnetized plasmas.

\renewcommand{\arraystretch}{1.5}

\begin{table}[h!]
\begin{center}
\caption{Relative scales for the main plasma physical properties.}
\label{TableScaling}
\begin{tabular}{cc}
  \midrule
  \textbf{Reference quantity} & \textbf{Scaling relationships} \\
  \midrule
  Characteristic masses & $\Mass_{e}^{0} = \varepsilon^{2} \, \Mass_{\heavy}^{0}$ \\
  Time scales & $t_{e}^{0} = \varepsilon \, t_{\heavy}^{0} = \varepsilon^{2} \, t^{0}$ \\
  Length scales & $l^{0} = \frac{\varepsilon}{\PseudoMach} \, L^{0}$ \\
  Velocities & $\Meanv^{0} = \PseudoMach \, C_{\heavy}^{0} = \varepsilon \PseudoMach \, C_{e}^{0}$ \\
  Energies & $\Mass_{e}^{0}\big(C_{e}^{0}\big)^{2} = \Mass_{\heavy}^{0}\big(C_{\heavy}^{0}\big)^{2} = q^{0} E^{0} L^{0} = k_{\textsc{b}} \Temperature^{0} $ \\
  Larmor frequencies & $\frac{q^{0}B^{0}}{\Mass_{\heavy}^{0}} t_{\heavy}^{0} = \varepsilon \, \frac{q^{0}B^{0}}{\Mass_{e}^{0}} t_{e}^{0} = \varepsilon^{2-b}$ \\
  Differential cross-sections & $\CrossSection^{\text{in},0}_{\heavy e} = \CrossSection^{\text{in},0}_{e \heavy} = \varepsilon^{2} \CrossSection^{0}$ \\
  Reactive source term &  $\Chemistry_{k}^{0} = \varepsilon \frac{\density^{0}}{t^{0}(C_{k}^{0})^{3}}$, $k \in \Species$ \\
  \midrule
\end{tabular}
\end{center}
\end{table}

\renewcommand{\arraystretch}{1}

\paragraph{Chemistry}

The chemical reactions are slow compared to other plasma phenomena, and the reactive source term for the $k^{\text{th}}$ species $\Chemistry_{k}(f)$ is of order $1$ in $\varepsilon$, irrespective of the detailed scaling properties of each chemical reaction $r$:
\begin{equation}
\Chemistry_{k}^{0} = \varepsilon \frac{\density^{0}}{t^{0}(C_{k}^{0})^{3}}, \quad k \in \Species,
\end{equation}
where $C_{k}^{0}$ is the order of magnitude of the $k^{\text{th}}$ species peculiar velocity. The reference quantities and the scaling adopted are summarized in Table~\ref{TableReferencequantities} and Table~\ref{TableScaling}, respectively.

\paragraph{Remark}

The range of applicability of the fluid model derived here is subject to the assumption of Maxwellian equilibrium distributions that will be obtained in the following section. Apart from the thermal non-equilibrium between electrons and heavy species, two kinds of deviation from the local thermal equilibrium might occur. First, when the ratio of the electric field over pressure $\frac{E}{\pressure}$ is \og too high \fg{}, namely when the assumption $q^{0} E^{0} l^{0} = \varepsilon k_{\textsc{b}} \Temperature_{e}^{0} = \varepsilon k_{\textsc{b}} \Temperature^{0}$ is not valid, or when $E/\pressure \gtrsim (\sigma^{0}/q) (\Temperature_{e}^{0} / \Temperature^{0})$ \cite{Rax}, the electron distribution function can depart strongly from a Maxwellian distribution. Second, in the case of high frequency oscillations, namely when the collision frequencies $1/t_{e}^{0}$ and $1/t_{\heavy}^{0} = \varepsilon /t_{e}^{0}$ are comparable with the electric field frequency $\nu_{RF}$, the species distribution function may also depart strongly from the Maxwellian equilibrium \cite{Rax}. In both cases, one would need a kinetic model rather than a fluid model.

\subsection{Scaled Boltzmann equations}
\label{SubSecKTScaledBoltzmann}

The scaling is applied to the Boltzmann equation \eqref{KTBoltzmann} written in the heavy-species reference frame. For each variable $\phi$, we denote by
\begin{equation}
\hat{\phi} = \frac{\phi}{\phi^{0}}
\end{equation}
the corresponding adimensionalized quantity. The adimensionalized Boltzmann equations then read
\begin{align}
\partial_{\hat{t}}\hat{f}_{e} & + \frac{1}{\varepsilon} \, \big(\widehat{\boldsymbol{C}}_{e}+ \varepsilon \, \widehat{\boldsymbol{\Meanv}}_{\heavy} \big) \cdot \boldsymbol{\partial}_{\widehat{\boldsymbol{x}}}\hat{f}_{e} + \varepsilon^{-(1+b)} \, \frac{\hat{q}_{e}}{\widehat{\Mass}_{e}} \big[ \big(\widehat{\boldsymbol{C}}_{e}+ \varepsilon \, \widehat{\boldsymbol{\Meanv}}_{\heavy} \big) \wedge \widehat{\boldsymbol{B}} \big] \cdot \boldsymbol{\partial}_{\widehat{\boldsymbol{C}}_{e}} \hat{f}_{e} \nonumber \\*
{} & + \big( \frac{1}{\varepsilon} \, \frac{\hat{q}_{e}}{\widehat{\Mass}_{e}} \widehat{\boldsymbol{E}} - \varepsilon \, \frac{D \widehat{\boldsymbol{\Meanv}}_{\heavy}}{D\hat{t}} \big) \cdot \boldsymbol{\partial}_{\widehat{\boldsymbol{C}}_{e}}\hat{f}_{e} - \big( \boldsymbol{\partial}_{\widehat{\boldsymbol{C}}_{e}}\hat{f}_{e} \otimes \widehat{\boldsymbol{C}}_{e} \big) : \boldsymbol{\partial}_{\widehat{\boldsymbol{x}}} \widehat{\boldsymbol{\Meanv}}_{\heavy} 
\label{KTAdimensionalized Boltzmann electron equation} \\*
{} & = \frac{1}{\varepsilon^{2}} \, \widehat{\Scattering}_{ee} (\hat{f}_{e},\hat{f}_{e} ) + \frac{1}{\varepsilon^{2}} \, \sum_{j \in \Heavy} \widehat{\Scattering}_{ej} (\hat{f}_{e},\hat{f}_{j} ) + \varepsilon \, \widehat{\Chemistry}_{e}(\hat{f}), \nonumber
\\
\partial_{\hat{t}}\hat{f}_{i} & + \big(\widehat{\boldsymbol{C}}_{i}+ \, \widehat{\boldsymbol{\Meanv}}_{\heavy} \big) \cdot \boldsymbol{\partial}_{\widehat{\boldsymbol{x}}} \hat{f}_{i} + \varepsilon^{1-b} \, \frac{\hat{q}_{i}}{\widehat{\Mass}_{i}} \big[ \big(\widehat{\boldsymbol{C}}_{i}+ \, \widehat{\boldsymbol{\Meanv}}_{\heavy} \big) \wedge \widehat{\boldsymbol{B}} \big] \cdot \boldsymbol{\partial}_{\widehat{\boldsymbol{C}}_{i}} \hat{f}_{i} \nonumber \\*
{} & + \big( \frac{\hat{q}_{i}}{\widehat{\Mass}_{i}} \widehat{\boldsymbol{E}} - \frac{D \widehat{\boldsymbol{\Meanv}}_{\heavy}}{D\hat{t}} \big) \cdot \boldsymbol{\partial}_{\widehat{\boldsymbol{C}}_{i}}\hat{f}_{i} - \big( \boldsymbol{\partial}_{\widehat{\boldsymbol{C}}_{i}}\hat{f}_{i} \otimes \widehat{\boldsymbol{C}}_{i} \big) : \boldsymbol{\partial}_{\widehat{\boldsymbol{x}}} \widehat{\boldsymbol{\Meanv}}_{\heavy}
\label{KTAdimensionalized Boltzmann heavy-species equations} \\*
{} & = \frac{1}{\varepsilon^{2}} \, \widehat{\Scattering}_{ie} (\hat{f}_{i},\hat{f}_{e} ) + \frac{1}{\varepsilon} \sum_{j \in \Heavy} \widehat{\Scattering}_{ij} (\hat{f}_{i},\hat{f}_{j} ) + \varepsilon \, \widehat{\Chemistry}_{i} (\hat{f} ), \qquad i \in \Heavy. \nonumber
\end{align}
The Chapman-Enskog method is then applied to the adimensionalized equations \eqref{KTAdimensionalized Boltzmann electron equation}-\eqref{KTAdimensionalized Boltzmann heavy-species equations} \cite{GrailleMaginMassot2009}. For the sake of simplicity the Mach number, which is of order $1$ from \eqref{KTMach}, can be taken equal to $1$, and the \og hat \fg {} symbol can be dropped, without affecting the fluid equations and transport fluxes derived. Alternatively, equations \eqref{KTAdimensionalized Boltzmann electron equation}-\eqref{KTAdimensionalized Boltzmann heavy-species equations} can be redimensionalized before applying the Chapman-Enskog method, keeping $\varepsilon$ as a formal parameter driving the asymptotic expansion. In this case, $\varepsilon$ is equal to $1$, eventually \cite{Kremer}. Either way, the scaled Boltzmann equations may be written as
\begin{align}
\partial_{t}f_{e} & + \frac{1}{\varepsilon} \, \left(\boldsymbol{C}_{e}+ \varepsilon \, \boldsymbol{\Meanv}_{\heavy} \right) \cdot \boldsymbol{\partial_{x}}f_{e} + \varepsilon^{-(1+b)} \, \frac{q_{e}}{\Mass_{e}} \left[ \left(\boldsymbol{C}_{e}+ \varepsilon \, \boldsymbol{\Meanv}_{\heavy} \right) \wedge \boldsymbol{B} \right] \cdot \boldsymbol{\partial}_{\boldsymbol{C}_{e}} f_{e} \nonumber \\*
{} & + \big( \frac{1}{\varepsilon} \, \frac{q_{e}}{\Mass_{e}} \boldsymbol{E} - \varepsilon \, \frac{D \boldsymbol{\Meanv}_{\heavy}}{Dt} \big) \cdot \boldsymbol{\partial}_{\boldsymbol{C}_{e}}f_{e} - \left( \boldsymbol{\partial}_{\boldsymbol{C}_{e}}f_{e} \otimes \boldsymbol{C}_{e} \right) : \boldsymbol{\partial_{x}} \boldsymbol{\Meanv}_{\heavy} 
\label{KTScaled Boltzmann electron equation} \\*
{} & = \frac{1}{\varepsilon^{2}} \, \Scattering_{ee}\left(f_{e},f_{e}\right) + \frac{1}{\varepsilon^{2}} \, \sum_{j \in \Heavy} \Scattering_{ej} \left(f_{e},f_{j} \right) + \varepsilon \, \Chemistry_{e}\left(f\right), \nonumber
\\
\partial_{t}f_{i} & + \left(\boldsymbol{C}_{i}+ \, \boldsymbol{\Meanv}_{\heavy} \right) \cdot \boldsymbol{\partial_{x}} f_{i} + \varepsilon^{1-b} \, \frac{q_{i}}{\Mass_{i}} \left[ \left(\boldsymbol{C}_{i}+ \, \boldsymbol{\Meanv}_{\heavy} \right) \wedge \boldsymbol{B} \right] \cdot \boldsymbol{\partial}_{\boldsymbol{C}_{i}} f_{i} \nonumber \\*
{} & + \big( \frac{q_{i}}{\Mass_{i}} \boldsymbol{E} - \frac{D \boldsymbol{\Meanv}_{\heavy}}{Dt} \big) \cdot \boldsymbol{\partial}_{\boldsymbol{C}_{i}}f_{i} - \left( \boldsymbol{\partial}_{\boldsymbol{C}_{i}}f_{i} \otimes \boldsymbol{C}_{i} \right) : \boldsymbol{\partial_{x}} \boldsymbol{\Meanv}_{\heavy}
\label{KTScaled Boltzmann heavy-species equations} \\*
{} & = \frac{1}{\varepsilon^{2}} \, \Scattering_{ie}\left(f_{i},f_{e}\right) + \frac{1}{\varepsilon} \sum_{j \in \Heavy} \Scattering_{ij} \left(f_{i},f_{j} \right) + \varepsilon \, \Chemistry_{i}\left(f\right), \qquad i \in \Heavy, \nonumber
\end{align}
where the partial scattering operators $\Scattering_{kl}$, $k, l \in \Species$, depend on $\varepsilon$, and are analyzed as follows.

For electron electron collisions the scaled scattering source term reads
\begin{equation}
\Scattering_{ee} (f_{e},\tilde{f}_{e}) \left( \boldsymbol{C}_{e} \right) = \int \CrossSection_{e\tilde{e}} g_{e\tilde{e}} \big( f_{e}' \tilde{f}_{e}' - f_{e} \tilde{f}_{e} \big) \, \mathrm{d} \boldsymbol{\omega}_{e\tilde{e}}' \mathrm{d} \widetilde{\boldsymbol{C}}_{e},
\label{KTScatteringElectronElectron}
\end{equation}
where $\widetilde{\boldsymbol{C}}_{e}$ and $\widetilde{\boldsymbol{C}}_{e}'$ represent the velocity of the electron collision partner, respectively before and after collision.

The formula for electron heavy-species scattering source term is similar, although we have to distinguish between elastic and inelastic collisions:
\begin{align}
\Scattering_{ei}\left(f_{e},f_{i}\right) \left( \boldsymbol{C}_{e} \right) = {} & \sum\limits_{\substack{\textsc{i} \in \QuantumSpace_{i}}} \int \CrossSection_{ei}^{\textsc{i} \textsc{i}}  g_{ei} \big( f_{e}' f_{i}' - f_{e} f_{i} \big) \, \mathrm{d} \boldsymbol{\omega}_{ei}' \mathrm{d} \boldsymbol{C}_{i}  \label{KTScatteringElectronHeavy} \\
{} & + \varepsilon^{2} ~ \sum\limits_{\substack{\textsc{i}, \textsc{i}' \in \QuantumSpace_{i} \\ \textsc{i}' \neq \textsc{i}}} \int \CrossSection_{ei}^{\textsc{i} \textsc{i}'} g_{ei} \big( f_{e}' f_{i}' \frac{a_{i \textsc{i}}}{a_{i \textsc{i}'}} - f_{e} f_{i} \big) \, \mathrm{d} \boldsymbol{\omega}_{ei}' \mathrm{d} \boldsymbol{C}_{i}, \nonumber
\end{align}
where $ \boldsymbol{\omega}_{ei} = (\boldsymbol{C}_{e}- \varepsilon \boldsymbol{C}_{i})/|\boldsymbol{C}_{e}- \varepsilon \boldsymbol{C}_{i}|$, $\boldsymbol{\omega}_{ei}' = (\boldsymbol{C}_{e}'- \varepsilon \boldsymbol{C}_{i}')/|\boldsymbol{C}_{e}'- \varepsilon \boldsymbol{C}_{i}'|$, $g_{ei} = |\boldsymbol{C}_{e}- \varepsilon \boldsymbol{C}_{i}|$, $g_{ei}' = |\boldsymbol{C}_{e}'- \varepsilon \boldsymbol{C}_{i}'|$, and $\mu_{ie} = \mu_{ei} = \Mass_{e}\Mass_{i}/(\Mass_{i} + \varepsilon^{2} \Mass_{e})$.

We obtain as well the source term corresponding to collisions of heavy species against electrons
\begin{align}
\Scattering_{ie}\left(f_{i},f_{e}\right) \left( \boldsymbol{C}_{i}, \textsc{i} \right) = {} & \int \CrossSection_{ie}^{\textsc{i} \textsc{i}} g_{ie} \big( f_{i}' f_{e}' - f_{i} f_{e} \big) \, \mathrm{d} \boldsymbol{\omega}_{ie}' \mathrm{d} \boldsymbol{C}_{e} \label{KTScatteringHeavyElectron} \\
{} & + \varepsilon^{2} ~ \sum\limits_{\substack{\textsc{i}' \in \QuantumSpace_{i} \\ \textsc{i}' \neq \textsc{i}}} \int \CrossSection_{ie}^{\textsc{i} \textsc{i}'} g_{ie} \big( f_{i}' f_{e}' \frac{a_{i \textsc{i}}}{a_{i \textsc{i}'}} - f_{i} f_{e} \big) \, \mathrm{d} \boldsymbol{\omega}_{ie}' \mathrm{d} \boldsymbol{C}_{e}, \nonumber
\end{align}
where $\boldsymbol{\omega}_{ie} = (\varepsilon \boldsymbol{C}_{i} - \boldsymbol{C}_{e})/|\varepsilon \boldsymbol{C}_{i} - \boldsymbol{C}_{e}|$, $\boldsymbol{\omega}_{ie}' = (\varepsilon \boldsymbol{C}_{i}' -\boldsymbol{C}_{e}')/|\varepsilon \boldsymbol{C}_{i}' - \boldsymbol{C}_{e}'|$, $g_{ie} = |\varepsilon \boldsymbol{C}_{i} - \boldsymbol{C}_{e}|$, and $g_{ie}' = |\varepsilon \boldsymbol{C}_{i}' - \boldsymbol{C}_{e}'|$.

Finally, we obtain for collisions between two heavy species
\begin{equation}
\Scattering_{ij}\left(f_{i},f_{j}\right) \left( \boldsymbol{C}_{i}, \textsc{i} \right) = \sum\limits_{\textsc{i}' \in \QuantumSpace_{i}} \sum\limits_{\textsc{j}, \textsc{j}' \in \QuantumSpace_{j}} \int \CrossSection_{ij}^{\textsc{i} \textsc{j} \textsc{i}' \textsc{j}'} g_{ij} \big( f_{i}' f_{j}' \frac{a_{i \textsc{i}} a_{j \textsc{j}}}{a_{i \textsc{i}'} a_{j \textsc{j}'}} - f_{i} f_{j} \big) \, \mathrm{d} \boldsymbol{\omega}_{ij}' \mathrm{d} \boldsymbol{C}_{j}.
\label{KTScatteringHeavyHeavy}
\end{equation}

Unlike for the nonreactive source terms just stated, we do not consider the different orders of magnitude in $\varepsilon$ associated with the motion of heavy species and electrons when computing the chemistry source terms. We only consider as a first approximation that chemical reactions occur \og slowly \fg{}, namely at order $\varepsilon$, and thus retain expression \eqref{KTChemistrySourcePerReaction} for $\Chemistry_{e}(f)$ and $\Chemistry_{i}(f)$, $i \in \Heavy$.

\subsection{Scaled collisional invariants}

We also apply the latter scaling to the space $\Invariants$ of collisional invariants of the scattering operator. The space of collisional invariants after scaling, denoted by $\Invariants^{\varepsilon}$, is spanned by the family $\psi^{l} = (\psi_{e}^{l}, \psi_{\heavy}^{l})$, $l \in \lbrace 1, \ldots, n^{s}+4 \rbrace$, defined as
\begin{equation}
\left\lbrace \begin{array}{lll}
\displaystyle \psi_{e}^{k} = \delta_{ke}, & \displaystyle \psi_{\heavy}^{k} = \left( \delta_{kj} \right)_{j \in \Heavy}, & \displaystyle k \in \Species, \vphantom{\frac{1}{2}} \\
\displaystyle \psi_{e}^{n^{s}+\nu} = \varepsilon ~ \Mass_{e} C_{e \nu}, & \displaystyle \psi_{\heavy}^{n^{s}+\nu} = \left( \Mass_{j} C_{j \nu}\right)_{j \in \Heavy}, & \displaystyle \nu \in \left\lbrace 1,2,3 \right\rbrace, \vphantom{\frac{1}{2}} \\
\displaystyle \psi_{e}^{n^{s}+4} = \frac{1}{2} \Mass_{e} \boldsymbol{C}_{e} \cdot \boldsymbol{C}_{e}, & \displaystyle \psi_{\heavy}^{n^{s}+4} = \Big( \frac{1}{2} \Mass_{j} \boldsymbol{C}_{j} \cdot \boldsymbol{C}_{j} + E_{j \textsc{j}} \Big)_{j \in \Heavy}. & {}
\end{array}
\right.
\end{equation}
From now on, $\psi^{l}$, $l \in \lbrace 1, \ldots, n^{s}+4 \rbrace$ will refer to the collisional invariants after scaling. We have seen in the previous section that the nonreactive collision operator can be written as
\begin{equation}
\Scattering = \left( \frac{1}{\varepsilon^{2}} \Scattering_{e}, \frac{1}{\varepsilon} \Scattering_{\heavy} \right),
\end{equation}
where $\Scattering_{e} = \Scattering_{ee} + \sum_{j \in \Heavy}\Scattering_{ej}$ and $\Scattering_{i} = \frac{1}{\varepsilon} \Scattering_{ie} + \sum_{j \in \Heavy} \Scattering_{ij}$, $i \in \Heavy$, are the scattering source terms for electron and $i^{\text{th}}$ heavy-species, respectively.

The orthogonality relations \eqref{KTPairwiseInvariantsConservationElectron}, \eqref{KTPairwiseInvariantsConservationHeavyElectron}, \eqref{KTPairwiseInvariantsConservationHeavy} remain valid after scaling. In particular, the cross-collision identities \eqref{KTPairwiseInvariantsConservationHeavyElectron} now read
\begin{eqnarray}
\sum_{j \in \Heavy} \llangle \psi_{e}^{e}, \Scattering_{ej} \rrangle_{e} = 0, \qquad \qquad \qquad  \llangle \psi_{\heavy}^{i}, \Scattering_{\heavy e} \rrangle_{\heavy} = 0, \quad i \in \Heavy,
\label{KTCrossInvariantsConservationNumber} \\
\llangle  \Mass_{\heavy} C_{\heavy \nu}, \Scattering_{\heavy e} \rrangle_{\heavy} + \varepsilon \sum_{j \in \Heavy} \llangle \Mass_{e} C_{e \nu}, \Scattering_{ej} \rrangle_{e} = 0, \quad \nu \in \left\lbrace 1,2,3 \right\rbrace,
\label{KTCrossInvariantsConservationMomentum} \\
\llangle \psi_{\heavy}^{n^{s}+4}, \Scattering_{\heavy e} \rrangle_{\heavy} + \sum_{j \in \Heavy} \llangle \psi_{e}^{n^{s}+4}, \Scattering_{ej} \rrangle_{e} = 0.
\label{KTCrossInvariantsConservationEnergy}
\end{eqnarray}

We also introduce two vector spaces obtained from projection of the space of collisional invariants $\Invariants^{\varepsilon}$. For all $\varepsilon$, $\Invariants_{e}^{\varepsilon}$ is the space spanned by
\begin{equation}
\left\lbrace \begin{array}{ll}
\displaystyle \psi_{e}^{e} = 1, & \displaystyle {} \vphantom{\frac{1}{2}} \\
\displaystyle \psi_{e}^{n^{s}+\nu} = \varepsilon ~ \Mass_{e} C_{e \nu}, & \displaystyle \nu \in \left\lbrace 1,2,3 \right\rbrace, \vphantom{\frac{1}{2}} \\
\displaystyle \psi_{e}^{n^{s}+4} = \frac{1}{2} \Mass_{e} \boldsymbol{C}_{e} \cdot \boldsymbol{C}_{e}, & \displaystyle {}
\end{array}
\right.
\end{equation}
and $\Invariants_{\heavy}$ is the space spanned by
\begin{equation}
\left\lbrace \begin{array}{ll}
\displaystyle \psi_{\heavy}^{i} = \left( \delta_{ij} \right)_{j \in \Heavy}, & \displaystyle {} \vphantom{\frac{1}{2}} \\
\displaystyle \psi_{\heavy}^{n^{s}+\nu} = \left( \Mass_{j} C_{j \nu}\right)_{j \in \Heavy}, & \displaystyle \nu \in \left\lbrace 1,2,3 \right\rbrace, \vphantom{\frac{1}{2}} \\
\displaystyle \psi_{\heavy}^{n^{s}+4} = \Big( \frac{1}{2} \Mass_{j} \boldsymbol{C}_{j} \cdot \boldsymbol{C}_{j} + E_{j \textsc{j}} \Big)_{j \in \Heavy}. & {}
\end{array}
\right.
\end{equation}

The macrosopic properties of the fluid mixture can then be expressed as partial scalar products of the partial distribution functions $f_{e}$ and $f_{\heavy}$ with the electron and heavy-species collisional invariants, respectively. When taking into account the scaling, one obtains
\begin{equation}
\left\lbrace \begin{array}{l}
\displaystyle \vphantom{\frac{1}{\varepsilon}} \llangle f_{e}, \psi_{e}^{e} \rrangle_{e} = \density_{e} \vphantom{\frac{1}{\varepsilon}} \\
\displaystyle \frac{1}{\varepsilon} ~ \llangle f_{e}, \Mass_{e} C_{e \nu} \rrangle_{e} = \rho_{e} \left( \Meanv_{e \nu} - \Meanv_{\heavy \nu} \right), \quad \nu \in \left\lbrace 1,2,3 \right\rbrace  \\
\displaystyle \llangle f_{e}, \psi_{e}^{n^{s}+4} \rrangle_{e} = \Energy_{e} + \varepsilon^{2} ~ \frac{1}{2} \rho_{e} \left(\boldsymbol{\Meanv}_{e}-\boldsymbol{\Meanv}_{\heavy} \right) \cdot \left( \boldsymbol{\Meanv}_{e}-\boldsymbol{\Meanv}_{\heavy}\right) \\
\end{array}
\right.
\label{KTMacroscopicPropertiesElectron}
\end{equation}
for electrons, and
\begin{equation}
\left\lbrace \begin{array}{ll}
\displaystyle \vphantom{\frac{1}{\varepsilon}} \llangle f_{\heavy}, \psi_{\heavy}^{j} \rrangle_{\heavy} = \density_{j}, & j \in \Heavy \vphantom{\psi_{\heavy}^{n^{s}+\nu}} \\
\displaystyle \vphantom{\frac{1}{\varepsilon}} \llangle f_{\heavy}, \psi_{\heavy}^{n^{s}+\nu} \rrangle_{\heavy} = 0, & \nu \in \left\lbrace 1,2,3 \right\rbrace \\
\displaystyle \vphantom{\frac{1}{\varepsilon}} \llangle f_{\heavy}, \psi_{\heavy}^{n^{s}+4} \rrangle_{\heavy} = \Energy_{\heavy} \\
\end{array}
\right.
\label{KTMacroscopicPropertiesHeavy}
\end{equation}
for the heavy species, where $\Energy_{e}$ and $\Energy_{\heavy}$ denote the respective internal energies per unit volume.

Note in particular that, in the limit $\varepsilon \rightarrow 0$, the space of electron collisional invariants $\Invariants_{e}^{\varepsilon}$, is reduced to the space of isotropic invariants, denoted by $\Invariants_{e}^{0}$, which is spanned by
\begin{equation}
\left\lbrace \begin{array}{l}
\displaystyle \psi_{e}^{e} = 1, \vphantom{\frac{1}{2}} \\
\displaystyle \psi_{e}^{n^{s}+4} = \frac{1}{2} \Mass_{e} \boldsymbol{C}_{e} \cdot \boldsymbol{C}_{e}.
\end{array}
\right.
\end{equation}
In other words, the electron momentum collisional invariant vanishes in the limit $\varepsilon \to 0$. This is because electron momentum is negligible before heavy-species momentum, and is related to the isotropization of the zeroth-order electron distribution function, as shall be seen later. The relevant sets of collisional invariants for our purpose are thus $\Invariants_{e}^{0}$ and $\Invariants_{\heavy}$.

\subsection{Asymptotic expansion of collision operators}
We now derive asymptotic expansions in powers of $\varepsilon$ for the scattering operators $\Scattering_{ei}$ and $\Scattering_{ie}$, $i \in \Heavy$. Conservation of momentum and energy during a binary collision between an electron and a molecule of the $i^{\text{th}}$ heavy species read, when taking into account the scaling with respect to $\varepsilon$
\begin{equation}
\left\lbrace \begin{array}{l}
\displaystyle \Mass_{i} \boldsymbol{C}_{i} + \varepsilon ~ \Mass_{e} \boldsymbol{C}_{e} = \Mass_{i} \boldsymbol{C}_{i}' + \varepsilon ~ \Mass_{e} \boldsymbol{C}_{e}', \vphantom{\frac{1}{\big( 1+ \varepsilon^{2} \frac{\Mass_{e}}{\Mass_{i}} \big)}} \\
\displaystyle \frac{1}{2}\Mass_{e} \frac{1}{\big( 1+ \varepsilon^{2} \frac{\Mass_{e}}{\Mass_{i}} \big)} g^{2} + E_{i \textsc{i}} = \frac{1}{2}\Mass_{e} \frac{1}{\big( 1+ \varepsilon^{2} \frac{\Mass_{e}}{\Mass_{i}} \big)} g^{'2} + E_{i \textsc{i}'}, \\
\end{array}
\right.
\end{equation}
where we have denoted by $g = g_{ie} = g_{ei} = |\varepsilon \boldsymbol{C}_{i}-\boldsymbol{C}_{e}|$ and by $g' = g_{ie}' = g_{ei}' = |\varepsilon \boldsymbol{C}_{i}'-\boldsymbol{C}_{e}'|$ the relative velocities of colliding species respectively before and after collision.
Setting $\Delta E_{\textsc{i} \textsc{i}'} = E_{i \textsc{i}'}-E_{i \textsc{i}}$, we rewrite the latter expression in the form
\begin{equation}
\left\lbrace \begin{array}{l}
\displaystyle \Mass_{i} \boldsymbol{C}_{i} + \varepsilon ~ \Mass_{e} \boldsymbol{C}_{e} = \Mass_{i} \boldsymbol{C}_{i}' + \varepsilon ~ \Mass_{e} \boldsymbol{C}_{e}', \vphantom{\frac{1}{\big( 1+ \varepsilon^{2} \frac{\Mass_{e}}{\Mass_{i}} \big)}} \\
\displaystyle \Delta E_{\textsc{i} \textsc{i}'} + \frac{1}{2}\Mass_{e} \frac{1}{\big( 1+ \varepsilon^{2} \frac{\Mass_{e}}{\Mass_{i}} \big)} \big( g^{'2} - g^{2} \big) = 0. \\
\end{array}
\right.
\label{KTScaledMomentumEnergy}
\end{equation}

\subsubsection{Scattering of heavy species by electrons}

The scattering operator for $i^{\text{th}}$-heavy-species electron collisions $\Scattering_{ie}$ was stated in equation \eqref{KTScatteringHeavyElectron}. The change of variable $\boldsymbol{C}_{e}^{(0)}= \frac{\boldsymbol{C}_{e} - \varepsilon \boldsymbol{C}_{i}}{\left( 1 + \varepsilon^{2} \frac{\Mass_{e}}{\Mass_{i}} \right)^{\frac{1}{2}}}$, with
\[
\mathrm{d} \boldsymbol{C}_{e} = \Big( 1 + \varepsilon^{2} \frac{\Mass_{e}}{\Mass_{i}} \Big)^{\frac{3}{2}} \, \mathrm{d} \boldsymbol{C}_{e}^{(0)},
\]
allows us to eliminate the dependance in $\varepsilon$ of the differential cross-section \cite{GrailleMaginMassot2009}, and yields
\begin{align}
\Scattering_{ie}\left(f_{i},f_{e}\right) \left( C_{i}, \textsc{i} \right) = {} & \int \CrossSection_{ie}^{\textsc{i} \textsc{i}} |\boldsymbol{C}_{e}^{(0)}| \Big( 1 + \varepsilon^{2} \frac{\Mass_{e}}{\Mass_{i}} \Big)^{2} \Big( f_{i}' f_{e}' - f_{i} f_{e} \Big) \, \mathrm{d} \boldsymbol{\omega}_{ie}' \mathrm{d} \boldsymbol{C}_{e}^{(0)}
\label{KTSiegamma} \\
{} & + \varepsilon^{2} \sum\limits_{\substack{\textsc{i}' \in \QuantumSpace_{i} \\ \textsc{i}' \neq \textsc{i}}} \int \CrossSection_{ie}^{\textsc{i} \textsc{i}'} |\boldsymbol{C}_{e}^{(0)}| \Big( 1 + \varepsilon^{2} \frac{\Mass_{e}}{\Mass_{i}} \Big)^{2} \Big( f_{i}' f_{e}' \frac{a_{i \textsc{i}}}{a_{i \textsc{i}'}} - f_{i} f_{e} \Big) \, \mathrm{d} \boldsymbol{\omega}_{ie}' \mathrm{d} \boldsymbol{C}_{e}^{(0)}, \nonumber
\end{align}
where
\begin{equation}
\CrossSection_{ie}^{\textsc{i} \textsc{i}'} = \CrossSection_{ie}^{\textsc{i} \textsc{i}'} \left( \Mass_{e} |\boldsymbol{C}_{e}^{(0)}|^{2}, \frac{\boldsymbol{C}_{e}^{(0)}}{|\boldsymbol{C}_{e}^{(0)}|} \cdot \boldsymbol{\omega}_{ie}' \right).
\label{KTSigmaieii}
\end{equation}
The variable $\boldsymbol{C}_{e}^{(0)}$ is the zeroth-order electron velocity before collision.

\paragraph{Expansion of the species velocities}

Conservation equations \eqref{KTScaledMomentumEnergy} associated with collisions between molecules of the $i^{\text{th}}$ heavy species and electrons read after change of variable
\begin{equation}
\left\lbrace \begin{array}{l}
\displaystyle \boldsymbol{C}_{i}' = \boldsymbol{C}_{i} + \varepsilon \frac{\Mass_{e}}{\Mass_{i}} \frac{1}{\big(1+\varepsilon^{2}\frac{\Mass_{e}}{\Mass_{i}} \big)^{\frac{1}{2}}} \bigg( \boldsymbol{C}_{e}^{(0)} + |\boldsymbol{C}_{e}^{(0)}| \Big( 1-\frac{\Delta E_{\textsc{i} \textsc{i}'}}{\frac{1}{2} \Mass_{e} |\boldsymbol{C}_{e}^{(0)}|^{2}} \Big)^{\frac{1}{2}} \boldsymbol{\omega}_{ie}' \bigg), \\
\displaystyle \boldsymbol{C}_{e}' = \varepsilon \boldsymbol{C}_{i} - \frac{1}{\big(1+\varepsilon^{2}\frac{\Mass_{e}}{\Mass_{i}} \big)^{\frac{1}{2}}} \bigg( |\boldsymbol{C}_{e}^{(0)}| \Big( 1-\frac{\Delta E_{\textsc{i} \textsc{i}'}}{\frac{1}{2} \Mass_{e} |\boldsymbol{C}_{e}^{(0)}|^{2}} \Big)^{\frac{1}{2}} \boldsymbol{\omega}_{ie}' - \varepsilon^{2}\frac{\Mass_{e}}{\Mass_{i}} \boldsymbol{C}_{e}^{(0)} \bigg),
\end{array}
\right.
\end{equation}
yielding the following asymptotic expansion:
\begin{equation}
\left\lbrace \begin{array}{l}
\displaystyle \boldsymbol{C}_{i}' = \boldsymbol{C}_{i} - \varepsilon \, \frac{\Mass_{e}}{\Mass_{i}} (\boldsymbol{C}_{e}'^{(0)} - \boldsymbol{C}_{e}^{(0)}) + O(\varepsilon^{3}), \\
\displaystyle \boldsymbol{C}_{e}' = \boldsymbol{C}_{e}'^{(0)} + \varepsilon \, \boldsymbol{C}_{i} - \frac{\varepsilon^{2}}{2} \, \frac{\Mass_{e}}{\Mass_{i}} (\boldsymbol{C}_{e}'^{(0)} - 2 \boldsymbol{C}_{e}^{(0)}) + O(\varepsilon^{4}), \\
\displaystyle \boldsymbol{C}_{e} = \boldsymbol{C}_{e}^{(0)} + \varepsilon \, \boldsymbol{C}_{i} + \frac{\varepsilon^{2}}{2} \, \frac{\Mass_{e}}{\Mass_{i}} \boldsymbol{C}_{e}^{(0)} + O(\varepsilon^{4}),
\end{array}
\right.
\end{equation}
where $\boldsymbol{C}_{e}'^{(0)} $ is the zeroth-order electron velocity after collision
\begin{equation}
\boldsymbol{C}_{e}'^{(0)} = - |\boldsymbol{C}_{e}'^{(0)}| \, \boldsymbol{\omega}_{ie}' = - |\boldsymbol{C}_{e}^{(0)}| \, \Big( 1 - \frac{\Delta E_{\textsc{i} \textsc{i}'}}{\frac{1}{2}\Mass_{e} |\boldsymbol{C}_{e}^{(0)}|^{2}} \Big)^{\frac{1}{2}} \boldsymbol{\omega}_{ie}'.
\end{equation}

We thus obtain an expansion of the distribution functions
\begin{align}
f_{i} \left(\boldsymbol{C}_{i}',\textsc{i}' \right) = {} & f_{i} \left( \boldsymbol{C}_{i},\textsc{i}' \right) - \varepsilon \, \frac{\Mass_{e}}{\Mass_{i}} \, \boldsymbol{\partial}_{\boldsymbol{C}_{i}} f_{i} \left( \boldsymbol{C}_{i}, \textsc{i}' \right) \cdot (\boldsymbol{C}_{e}'^{(0)} - \boldsymbol{C}_{e}^{(0)})
\label{KTfiCiprimeIprime} \\
{} & + \frac{\varepsilon^{2}}{2} \, \frac{\Mass_{e}^{2}}{\Mass_{i}^{2}} \, \boldsymbol{\partial}_{\boldsymbol{C}_{i} \boldsymbol{C}_{i}} f_{i} \left( \boldsymbol{C}_{i}, \textsc{i}' \right) : (\boldsymbol{C}_{e}'^{(0)} - \boldsymbol{C}_{e}^{(0)}) \otimes (\boldsymbol{C}_{e}'^{(0)} - \boldsymbol{C}_{e}^{(0)}) + O(\varepsilon^{3}),
\nonumber \\
f_{e} \left(\boldsymbol{C}_{e}' \right) = {} & f_{e}( \boldsymbol{C}_{e}'^{(0)}) + \varepsilon \, \boldsymbol{\partial}_{\boldsymbol{C}_{e}} f_{e}(\boldsymbol{C}_{e}'^{(0)}) \cdot \boldsymbol{C}_{i} + \frac{\varepsilon^{2}}{2} \, \boldsymbol{\partial}_{\boldsymbol{C}_{e} \boldsymbol{C}_{e}} f_{e}(\boldsymbol{C}_{e}'^{(0)}) : \boldsymbol{C}_{i} \otimes \boldsymbol{C}_{i}
\label{KTfeCeprime} \\
{} & - \frac{\varepsilon^{2}}{2} \, \frac{\Mass_{e}}{\Mass_{i}} \, \boldsymbol{\partial}_{\boldsymbol{C}_{e}} f_{e}(\boldsymbol{C}_{e}'^{(0)}) \cdot (\boldsymbol{C}_{e}'^{(0)} - 2 \boldsymbol{C}_{e}^{(0)}) + O(\varepsilon^{3}),
\nonumber \\
f_{e}\left(\boldsymbol{C}_{e} \right) = {} & f_{e}(\boldsymbol{C}_{e}^{(0)}) + \varepsilon \, \boldsymbol{\partial}_{\boldsymbol{C}_{e}} f_{e}(\boldsymbol{C}_{e}^{(0)}) \cdot  \boldsymbol{C}_{i} + \frac{\varepsilon^{2}}{2} \, \boldsymbol{\partial}_{\boldsymbol{C}_{e} \boldsymbol{C}_{e}} f_{e}(\boldsymbol{C}_{e}^{(0)}) : \boldsymbol{C}_{i} \otimes \boldsymbol{C}_{i}
\label{KTfeCe} \\
{} & + \frac{\varepsilon^{2}}{2} \, \frac{\Mass_{e}}{\Mass_{i}} \, \boldsymbol{\partial}_{\boldsymbol{C}_{e}} f_{e}(\boldsymbol{C}_{e}^{(0)}) \cdot \boldsymbol{C}_{e}^{(0)} + O(\varepsilon^{3}).
\nonumber
\end{align}

\paragraph{Expansion of $\Scattering_{ie}$}

Upon introducing the asymptotic development \eqref{KTfiCiprimeIprime}-\eqref{KTfeCe} in expression \eqref{KTSiegamma}, the collision operator $\Scattering_{ie}$, $i \in \Heavy$, can be expanded in the form
\begin{equation}
\Scattering_{ie} = \varepsilon ~ \Scattering_{ie}^{1} + ~ \varepsilon^{2} ~ \Scattering_{ie}^{2} + ~ O(\varepsilon^{3}).
\end{equation}
The zeroth-order collision operator $\Scattering_{ie}^{0}$ cancels. Indeed, from \eqref{KTSiegamma}
\begin{align*}
\Scattering_{ie}^{0} \left( f_{i} , f_{e} \right) \left( C_{i}, \textsc{i} \right) & = \int \CrossSection_{ie}^{\textsc{i} \textsc{i}} |\boldsymbol{C}_{e}^{(0)}| \big( f_{i}(\boldsymbol{C}_{i},\textsc{i}) f_{e} (\boldsymbol{C}_{e}'^{(0)}) - f_{i}(\boldsymbol{C}_{i},\textsc{i}) f_{e} (\boldsymbol{C}_{e}^{(0)}) \big) \, \mathrm{d} \boldsymbol{\omega}_{ie}' \mathrm{d} \boldsymbol{C}_{e}^{(0)} \\
& = f_{i}(\boldsymbol{C}_{i},\textsc{i}) \int \CrossSection_{ie}^{\textsc{i} \textsc{i}} |\boldsymbol{C}_{e}^{(0)}| \big( f_{e}(-|\boldsymbol{C}_{e}^{(0)}| \boldsymbol{\omega}_{ie}') - f_{e}(\boldsymbol{C}_{e}^{(0)}) \big) \, \mathrm{d} \boldsymbol{\omega}_{ie}' \mathrm{d} \boldsymbol{C}_{e}^{(0)},
\end{align*}
where $\CrossSection_{ie}^{\textsc{i} \textsc{i}}$ was given in \eqref{KTSigmaieii}. The successive changes of variable $\boldsymbol{C}_{e}^{(0)} = - |\boldsymbol{C}_{e}^{(0)}| \boldsymbol{\omega}_{ie}$, with $\mathrm{d} \boldsymbol{C}_{e}^{(0)} = |\boldsymbol{C}_{e}^{(0)}|^{2} \, \mathrm{d} |\boldsymbol{C}_{e}^{(0)}| \mathrm{d} \boldsymbol{\omega}_{ie}$, and $(\boldsymbol{\omega_{ie}}',\boldsymbol{\omega_{ie}}) \leftrightarrow (\boldsymbol{\omega_{ie}},\boldsymbol{\omega_{ie}}')$, then yield
\begin{align*}
\Scattering_{ie}^{0} & \left( f_{i} , f_{e} \right) \left( C_{i}, \textsc{i} \right) \vphantom{\int} \\
& =  f_{i}(\boldsymbol{C}_{i},\textsc{i}) \int \CrossSection_{ie}^{\textsc{i} \textsc{i}} |\boldsymbol{C}_{e}^{(0)}|^{3} \big( f_{e}(-|\boldsymbol{C}_{e}^{(0)}| \boldsymbol{\omega}_{ie}') - f_{e}(- |\boldsymbol{C}_{e}^{(0)}|\boldsymbol{\omega}_{ie}) \big) \, \mathrm{d} |\boldsymbol{C}_{e}^{(0)}| \mathrm{d} \boldsymbol{\omega}_{ie} \mathrm{d} \boldsymbol{\omega}_{ie}' \\
& = f_{i}(\boldsymbol{C}_{i},\textsc{i}) \int \CrossSection_{ie}^{\textsc{i} \textsc{i}} |\boldsymbol{C}_{e}^{(0)}|^{3} \big( f_{e}(-|\boldsymbol{C}_{e}^{(0)}| \boldsymbol{\omega}_{ie}) - f_{e}(- |\boldsymbol{C}_{e}^{(0)}|\boldsymbol{\omega}_{ie}') \big) \, \mathrm{d} |\boldsymbol{C}_{e}^{(0)}| \mathrm{d} \boldsymbol{\omega}_{ie} \mathrm{d} \boldsymbol{\omega}_{ie}' \\
& = - \Scattering_{ie}^{0} \left( f_{i} , f_{e} \right) \left( C_{i}, \textsc{i} \right), \vphantom{\int}
\end{align*}
where $\CrossSection_{ie}^{\textsc{i} \textsc{i}} = \CrossSection_{ie}^{\textsc{i} \textsc{i}} ( \Mass_{e} |\boldsymbol{C}_{e}^{(0)}|^{2}, \boldsymbol{\omega}_{ie} \cdot \boldsymbol{\omega}_{ie}' )$, so that finally
\begin{equation}
 \Scattering_{ie}^{0} \left( f_{i} , f_{e} \right) \left( \boldsymbol{C}_{i}, \textsc{i} \right) = 0, \qquad i \in \Heavy, \textsc{i} \in \QuantumSpace_{i}. \label{KTSie0}
\end{equation}
From similar calculations, the first-order term $\Scattering_{ie}^{1} \left( f_{i} , f_{e} \right) \left( \boldsymbol{C}_{i}, \textsc{i} \right)$ reads, for $i \in \Heavy$, $\textsc{i} \in \QuantumSpace_{i}$
\begin{equation}
 \Scattering_{ie}^{1} \left( f_{i} , f_{e} \right) \left( \boldsymbol{C}_{i}, \textsc{i} \right) = - \frac{\Mass_{e}}{\Mass_{i}} \, \boldsymbol{\partial}_{\boldsymbol{C}_{i}} f_{i} (\boldsymbol{C}_{i}, \textsc{i}) \cdot \int \Sigma_{\textsc{i} \textsc{i}}^{(1)} (|\boldsymbol{C}_{e}|^{2}) \, |\boldsymbol{C}_{e}| \, f_{e} (\boldsymbol{C}_{e}) \, \boldsymbol{C}_{e} \, \mathrm{d} \boldsymbol{C}_{e}, \label{KTSie1}
\end{equation}
where we have dropped the upperscript $(0)$ on the integration variable $\boldsymbol{C}_{e}$ for the sake of simplicity. The generalized momentum cross-section in thermal non-equilibrium context $\Sigma_{\textsc{i} \textsc{i}'}^{(l)}$ is defined, for given $i \in \Heavy$, $\textsc{i} \in \QuantumSpace_{i}$ and $\textsc{i}' \in \QuantumSpace_{i}$, by
\begin{equation}
\Sigma_{\textsc{i} \textsc{i}'}^{(l)} (|\boldsymbol{C}_{e}|^{2}) = 2 \pi \left( \frac{\Mass_{e}}{\Mass_{i}} \right)^{l} \int_{0}^{\pi} \CrossSection_{ie}^{\textsc{i} \textsc{i}'} \big(\Mass_{e} |\boldsymbol{C}_{e}|^{2}, \cos \theta \big) \left( 1 - \cos^{l} \theta \right) \sin \theta \, \mathrm{d} \theta, \quad l \in \mathbb{N}^{*},
\end{equation}
where the symbol $\theta$ represents the angle between the vectors ${\boldsymbol{\omega}_{ie}' = \varepsilon \boldsymbol{C}_{i}' - \boldsymbol{C}_{e}'}$ and ${\boldsymbol{\omega}_{ie} = \varepsilon \boldsymbol{C}_{i} - \boldsymbol{C}_{e}}$. For $l=1$, this cross-section represents the average momentum transferred in encounters between electrons and molecules of the $i^{\text{th}}$ heavy species with initial quantum state $\textsc{i}$ and final quantum state $\textsc{i}'$, for a given initial value of the electron kinetic energy $\Mass_{e} |\boldsymbol{C}_{e}|^{2}$. We also define
\begin{equation}
\Sigma_{\textsc{i} \textsc{i}'}^{(0)} (|\boldsymbol{C}_{e}|^{2}) = 2 \pi \int_{0}^{\pi} \CrossSection_{ie}^{\textsc{i} \textsc{i}'} \big(\Mass_{e} |\boldsymbol{C}_{e}|^{2}, \cos \theta \big) \sin \theta \, \mathrm{d} \theta, \qquad i \in \Heavy, \quad \textsc{i}, \textsc{i}' \in \QuantumSpace_{i}.
\end{equation}

Finally, the second-order term can be decomposed into an elastic and an inelastic contributions:
\begin{equation}
\Scattering_{ie}^{2} = \Scattering_{ie}^{2,\text{el}} + \Scattering_{ie}^{2,\text{in}}.
\label{KTSie2}
\end{equation}
The elastic contribution $\Scattering_{ie}^{2,\text{el}} \left( f_{i} , f_{e} \right) \left( \boldsymbol{C}_{i}, \textsc{i} \right)$ reads, for $i \in \Heavy$, $\textsc{i} \in \QuantumSpace_{i}$
\begin{align}
{} & \Scattering_{ie}^{2,\text{el}} \left( f_{i} , f_{e} \right) \left( \boldsymbol{C}_{i}, \textsc{i} \right) = 
\label{KTSie2el} \\
{} \nonumber \\
{} & - \frac{\Mass_{e}}{\Mass_{i}} ~ \boldsymbol{\partial}_{\boldsymbol{C}_{i}} \left( f_{i} (\boldsymbol{C}_{i}, \textsc{i}) \boldsymbol{C}_{i} \right) : \int \Sigma_{\textsc{i} \textsc{i}}^{(1)} (|\boldsymbol{C}_{e}|^{2}) \, |\boldsymbol{C}_{e}| \, \big( \boldsymbol{C}_{e} \otimes \boldsymbol{\partial}_{\boldsymbol{C}_{e}} f_{e} (\boldsymbol{C}_{e}) \big) \, \mathrm{d} \boldsymbol{C}_{e} \nonumber \\
{} \nonumber \\
{} & + \frac{1}{4} \frac{\Mass_{e}^{2}}{\Mass_{i}^{2}} ~ \boldsymbol{\partial}_{\boldsymbol{C}_{i} \boldsymbol{C}_{i}} f_{i} (\boldsymbol{C}_{i}, \textsc{i}) : \int \Sigma_{\textsc{i} \textsc{i}}^{(2)} (|\boldsymbol{C}_{e}|^{2}) \, |\boldsymbol{C}_{e}| \, f_{e} (\boldsymbol{C}_{e}) \, \big( |\boldsymbol{C}_{e}|^{2} \, \mathbb{I} - 3 \boldsymbol{C}_{e} \otimes \boldsymbol{C}_{e} \big) \, \mathrm{d} \boldsymbol{C}_{e} \nonumber \\
{} \nonumber \\
{} & + \frac{\Mass_{e}^{2}}{\Mass_{i}^{2}} ~ \boldsymbol{\partial}_{\boldsymbol{C}_{i} \boldsymbol{C}_{i}} f_{i} (\boldsymbol{C}_{i}, \textsc{i}) : \int \Sigma_{\textsc{i} \textsc{i}}^{(1)} (|\boldsymbol{C}_{e}|^{2}) \, |\boldsymbol{C}_{e}| \, f_{e} (\boldsymbol{C}_{e}) \, \big( \boldsymbol{C}_{e} \otimes \boldsymbol{C}_{e} \big) \, \mathrm{d} \boldsymbol{C}_{e}, \nonumber
\end{align}
while the inelastic term can be written as
\begin{align}
{} & \Scattering_{ie}^{2,\text{in}} \left( f_{i} , f_{e} \right) \left( \boldsymbol{C}_{i}, \textsc{i} \right)
\label{KTSie2in} \\
{} \nonumber \\
{} & = \sum\limits_{\substack{\textsc{i}' \in \QuantumSpace_{i} \\ \textsc{i}' \neq \textsc{i}}} \int \CrossSection_{ie}^{\textsc{i} \textsc{i}'} |\boldsymbol{C}_{e}| \Big( f_{i} (\boldsymbol{C}_{i}, \textsc{i}') f_{e} (\boldsymbol{C}_{e}'^{(0)}) \frac{a_{i \textsc{i}}}{a_{i \textsc{i}'}} - f_{i} \left( \boldsymbol{C}_{i}, \textsc{i} \right) f_{e} ( \boldsymbol{C}_{e} ) \Big) \, \mathrm{d} \boldsymbol{\omega}_{ie}' \mathrm{d} \boldsymbol{C}_{e} \nonumber \\
{} \nonumber \\
{} & = \sum\limits_{\substack{\textsc{i}' \in \QuantumSpace_{i} \\ \textsc{i}' \neq \textsc{i}}} \int \left( \Sigma_{\textsc{i}' \textsc{i}}^{(0)} (|\boldsymbol{C}_{e}|^{2}) f_{i}(\boldsymbol{C}_{i},\textsc{i}') - \Sigma_{\textsc{i} \textsc{i}'}^{(0)} (|\boldsymbol{C}_{e}|^{2}) f_{i}(\boldsymbol{C}_{i},\textsc{i}) \right) |\boldsymbol{C}_{e}| f_{e} ( \boldsymbol{C}_{e}) \, \mathrm{d} \boldsymbol{C}_{e}. \nonumber \\
{} \nonumber
\end{align}

\subsubsection{Scattering of electrons by heavy species}

The electron $i^{\text{th}}$-heavy-species scattering term $\Scattering_{ei}$ was stated in \eqref{KTScatteringElectronHeavy}. Unlike for the heavy-species electron case, there is no change of variable allowing to eliminate the dependance in $\varepsilon$ of the scattering cross-section $\CrossSection_{ei}^{\textsc{i} \textsc{i}'}$, and we retain the set of variables $(\boldsymbol{C}_{e}, \boldsymbol{C}_{i}, \boldsymbol{\omega}_{ei}')$.

\paragraph{Expansion of the species velocities}

Conservation equations \eqref{KTScaledMomentumEnergy} for collisions between electrons and molecules of the $i^{\text{th}}$ heavy species yield the following asymptotic expansion:
\begin{align}
g_{ei}' & = g_{ei}'^{(0)} + \varepsilon ~ g_{ei}'^{(1)} + \varepsilon^{2} ~ g_{ei}'^{(2)} + O(\varepsilon^{3}),
\label{KTgeiprime} \\
\boldsymbol{C}_{e}' & = \boldsymbol{C}_{e}'^{(0)} + \varepsilon ~ \boldsymbol{C}_{e}'^{(1)} + \varepsilon^{2} ~ \boldsymbol{C}_{e}'^{(2)} + O(\varepsilon^{3}),
\label{KTCeprime} \\
\boldsymbol{C}_{i}' & = \boldsymbol{C}_{i}'^{(0)} + \varepsilon ~ \boldsymbol{C}_{i}'^{(1)} + \varepsilon^{2} ~ \boldsymbol{C}_{i}'^{(2)} + O(\varepsilon^{3}),
\label{KTCiprime}
\end{align}
where the coefficients for the amplitude of the relative velocity after collision $g_{ei}'$ read in terms of the variables $(\boldsymbol{C}_{e}, \boldsymbol{C}_{i}, \boldsymbol{\omega}_{ei}')$
\begin{equation}
\left\lbrace \begin{array}{l}
\displaystyle g_{ei}'^{(0)} = |\boldsymbol{C}_{e}| \Big( 1 - \frac{\Delta E_{\textsc{i} \textsc{i}'}}{\frac{1}{2}\Mass_{e}|\boldsymbol{C}_{e}|^{2}} \Big)^{\frac{1}{2}}, \\
\displaystyle g_{ei}'^{(1)} = - \frac{\boldsymbol{C}_{e} \cdot \boldsymbol{C}_{i}}{g_{ei}'^{(0)}}, \vphantom{\Big( 1 - \frac{\Delta E_{\textsc{i} \textsc{i}'}}{\frac{1}{2}\Mass_{e}|\boldsymbol{C}_{e}|^{2}} \Big)^{\frac{1}{2}}} \\
\displaystyle g_{ei}'^{(2)} = \frac{1}{2} \frac{|\boldsymbol{C}_{i}|^{2}}{g_{ei}'^{(0)}} \Big( 1 - \frac{\Delta E_{\textsc{i} \textsc{i}'}}{\frac{1}{2}\Mass_{i}|\boldsymbol{C}_{i}|^{2}} \Big) - \frac{1}{2} \frac{\left(\boldsymbol{C}_{e} \cdot \boldsymbol{C}_{i} \right)^{2}}{ \big(g_{ei}'^{(0)} \big)^{3}}, \vphantom{\Big( 1 - \frac{\Delta E_{\textsc{i} \textsc{i}'}}{\frac{1}{2}\Mass_{e}|\boldsymbol{C}_{e}|^{2}} \Big)^{\frac{1}{2}}}
\end{array}
\right.
\end{equation}
and the coefficients for the velocities after collisions $\boldsymbol{C}_{e}'$, $\boldsymbol{C}_{i}'$ are given by
\begin{equation}
\left\lbrace \begin{array}{lll}
\displaystyle \boldsymbol{C}_{e}'^{(0)} = g_{ei}'^{(0)} \boldsymbol{\omega}_{ei}', \qquad & \displaystyle \boldsymbol{C}_{e}'^{(1)} = \boldsymbol{C}_{i} + g_{ei}'^{(1)} \boldsymbol{\omega}_{ei}', & \displaystyle \boldsymbol{C}_{e}'^{(2)} = \boldsymbol{C}_{i}'^{(1)} + g_{ei}'^{(2)} \boldsymbol{\omega}_{ei}', \vphantom{\frac{\Mass_{e}}{\Mass_{i}}} \\
\displaystyle \boldsymbol{C}_{i}'^{(0)} = \boldsymbol{C}_{i}, & \displaystyle \boldsymbol{C}_{i}'^{(1)} = \frac{\Mass_{e}}{\Mass_{i}} \big( \boldsymbol{C}_{e} - \boldsymbol{C}_{e}'^{(0)} \big), \quad & \displaystyle \boldsymbol{C}_{i}'^{(2)} = - \frac{\Mass_{e}}{\Mass_{i}} \boldsymbol{C}_{e}'^{(1)}. \\
\end{array}
\right.
\end{equation}

\paragraph{Expansion of $\Scattering_{ei}$}

Relations \eqref{KTgeiprime}-\eqref{KTCiprime} allow one to determine, after a few calculations, the asymptotic expansion of the electron $i^\text{th}$-heavy-species scattering operator
\[
\Scattering_{ei} = \Scattering_{ei}^{0} + \varepsilon ~ \Scattering_{ei}^{1} + ~ \varepsilon^{2} ~ \Scattering_{ei}^{2} + ~ O(\varepsilon^{3}).
\]
The zeroth-order collision operator reads
\begin{align}
{} & \Scattering_{ei}^{0} \left( f_{e} , f_{i} \right) \left( \boldsymbol{C}_{e} \right) = 
\label{KTSei0} \\
{} & \qquad \sum_{\textsc{i} \in \QuantumSpace_{i}} \left( \int f_{i}(\boldsymbol{C}_{i},\textsc{i}) \, \mathrm{d} \boldsymbol{C}_{i} \right) \int \CrossSection_{ei}^{\textsc{i} \textsc{i}} ~ |\boldsymbol{C}_{e}| \big( f_{e}(|\boldsymbol{C}_{e}| \boldsymbol{\omega}_{ei}') - f_{e}(\boldsymbol{C}_{e}) \big) \, \mathrm{d} \boldsymbol{\omega}_{ei}', \nonumber
\end{align}
where
\begin{eqnarray*}
\CrossSection_{ei}^{\textsc{i} \textsc{i}} = \CrossSection_{ei}^{\textsc{i} \textsc{i}} \left( \Mass_{e} |\boldsymbol{C}_{e}|^{2}, \frac{\boldsymbol{C}_{e}}{|\boldsymbol{C}_{e}|} \cdot \boldsymbol{\omega}_{ei}' \right).
\end{eqnarray*}
The first-order term $\Scattering_{ei}^{1} \left( f_{e} , f_{i} \right) \left( \boldsymbol{C}_{e} \right)$ reads
\begin{align}
{} & \Scattering_{ei}^{1} \left( f_{e} , f_{i} \right) \left( \boldsymbol{C}_{e} \right) =
\label{KTSei1} \\
{} \nonumber \\
{} & + \sum\limits_{\substack{\textsc{i} \in \QuantumSpace_{i}}} \int f_{i} ( \boldsymbol{C}_{i}, \textsc{i} ) \boldsymbol{C}_{i} \, \mathrm{d} \boldsymbol{C}_{i} \cdot \int \CrossSection_{ei}^{\textsc{i} \textsc{i}} ~ |\boldsymbol{C}_{e}| \big( \boldsymbol{\partial}_{\boldsymbol{C}_{e}} f_{e} ( |\boldsymbol{C}_{e}| \boldsymbol{\omega}_{ei}' ) - \boldsymbol{\partial}_{\boldsymbol{C}_{e}} f_{e} ( \boldsymbol{C}_{e} ) \big) \, \mathrm{d} \boldsymbol{\omega}_{ei}' \nonumber \\
{} \nonumber \\
{} & - \sum\limits_{\substack{\textsc{i} \in \QuantumSpace_{i}}} \int f_{i} ( \boldsymbol{C}_{i}, \textsc{i} ) \boldsymbol{C}_{i} \, \mathrm{d} \boldsymbol{C}_{i} \cdot \boldsymbol{\partial}_{\boldsymbol{C}_{e}} \left[ \int \CrossSection_{ei}^{\textsc{i} \textsc{i}} ~ |\boldsymbol{C}_{e}| \big( f_{e} ( |\boldsymbol{C}_{e}| \boldsymbol{\omega}_{ei}' ) - f_{e} ( \boldsymbol{C}_{e} ) \big) \, \mathrm{d} \boldsymbol{\omega}_{ei}' \right]. \nonumber \\
{} \nonumber
\end{align}
Finally, the second-order term can be decomposed into an elastic and an inelastic contributions
\begin{equation}
\Scattering_{ei}^{2} = \Scattering_{ei}^{2,\text{el}} + \Scattering_{ei}^{2,\text{in}}.
\label{KTSei2}
\end{equation}
The elastic term reads
\begin{eqnarray}
\Scattering_{ei}^{2,\text{el}} \left( f_{e} , f_{i} \right) \left( \boldsymbol{C}_{e} \right) & = & \frac{\Mass_{e}}{\Mass_{i}} \sum_{\textsc{i} \in \QuantumSpace_{i}} \left( \int f_{i}(\boldsymbol{C}_{i},\textsc{i}) \, \mathrm{d} \boldsymbol{C}_{i} \right) \, K_{ei}^{\textsc{i},1}(\boldsymbol{C}_{e}) \\
{} & + & \frac{1}{2}  \sum_{\textsc{i} \in \QuantumSpace_{i}} \int f_{i}(\boldsymbol{C}_{i},\textsc{i}) \, \boldsymbol{C}_{i} \otimes \boldsymbol{C}_{i} \, \mathrm{d} \boldsymbol{C}_{i} \, : \, K_{ei}^{\textsc{i},2}(\boldsymbol{C}_{e}), \nonumber
\end{eqnarray}
where
\begin{eqnarray}
K_{ei}^{\textsc{i},1}(\boldsymbol{C}_{e}) & = & \boldsymbol{\partial}_{\boldsymbol{C}_{e}} \cdot \left[ \int \CrossSection_{ei}^{\textsc{i} \textsc{i}} ~ |\boldsymbol{C}_{e}| \left( \boldsymbol{C}_{e} - |\boldsymbol{C}_{e}| \boldsymbol{\omega}_{ei}' \right) f_{e} ( |\boldsymbol{C}_{e}| \boldsymbol{\omega}_{ei}' ) \, \mathrm{d} \boldsymbol{\omega}_{ei}' \right] \\
{} & - & \frac{1}{2} |\boldsymbol{C}_{e}| \boldsymbol{C}_{e} \cdot \int \boldsymbol{\partial}_{\boldsymbol{C}_{e}} \CrossSection_{ei}^{\textsc{i} \textsc{i}} ~ \big( f_{e} ( |\boldsymbol{C}_{e}| \boldsymbol{\omega}_{ei}' ) - f_{e} ( \boldsymbol{C}_{e} ) \big) \, \mathrm{d} \boldsymbol{\omega}_{ei}', \nonumber
\end{eqnarray}
and
\begin{eqnarray}
K_{ei}^{\textsc{i},2}(\boldsymbol{C}_{e}) & = & \boldsymbol{\partial}_{\boldsymbol{C}_{e} \boldsymbol{C}_{e}} \left[ \int \CrossSection_{ei}^{\textsc{i} \textsc{i}} ~ |\boldsymbol{C}_{e}| \, \big( f_{e} ( |\boldsymbol{C}_{e}| \boldsymbol{\omega}_{ei}' ) - f_{e} ( \boldsymbol{C}_{e} ) \big) \, \mathrm{d} \boldsymbol{\omega}_{ei}' \right] \\
{} & + & 2 \int \boldsymbol{\partial}_{\boldsymbol{C}_{e}} \left[ |\boldsymbol{C}_{e}| \CrossSection_{ei}^{\textsc{i} \textsc{i}} \right] \otimes \big( \boldsymbol{\partial}_{\boldsymbol{C}_{e}} f_{e} ( \boldsymbol{C}_{e} ) - \boldsymbol{\partial}_{\boldsymbol{C}_{e}} f_{e} ( |\boldsymbol{C}_{e}| \boldsymbol{\omega}_{ei}' ) \big) \, \mathrm{d} \boldsymbol{\omega}_{ei}' \nonumber \vphantom{\left[ \int \right]} \\
& + & |\boldsymbol{C}_{e}| \int \CrossSection_{ei}^{\textsc{i} \textsc{i}} \, \big( \boldsymbol{\partial}_{\boldsymbol{C}_{e} \boldsymbol{C}_{e}} f_{e} (  \boldsymbol{C}_{e} ) + \boldsymbol{\partial}_{\boldsymbol{C}_{e} \boldsymbol{C}_{e}} f_{e} ( |\boldsymbol{C}_{e}| \boldsymbol{\omega}_{ei}' ) \big) \, \mathrm{d} \boldsymbol{\omega}_{ei}' \nonumber \vphantom{\left[ \int \right]} \\
{} & - & 2 |\boldsymbol{C}_{e}| \int \CrossSection_{ei}^{\textsc{i} \textsc{i}} \, \boldsymbol{\partial}_{\boldsymbol{C}_{e} \boldsymbol{C}_{e}} f_{e} ( |\boldsymbol{C}_{e}| \boldsymbol{\omega}_{ei}' ) \cdot \boldsymbol{\omega}_{ei}' \otimes \frac{\boldsymbol{C}_{e}}{|\boldsymbol{C}_{e}|} \, \mathrm{d} \boldsymbol{\omega}_{ei}'. \nonumber \vphantom{\left[ \int \right]} \\
\nonumber
\end{eqnarray}
The inelastic term reads
\begin{align}
{} & \Scattering_{ei}^{2,\text{in}} \left( f_{e} , f_{i} \right) \left( \boldsymbol{C}_{e} \right) = \\
{} & \qquad \sum\limits_{\substack{\textsc{i}, \textsc{i}' \in \QuantumSpace_{i} \\ \textsc{i}' \neq \textsc{i}}} \int \CrossSection_{ei}^{\textsc{i} \textsc{i}'} |\boldsymbol{C}_{e}| \Big( f_{e}(\boldsymbol{C}_{e}'^{(0)}) f_{i}(\boldsymbol{C}_{i},\textsc{i}') \frac{a_{i \textsc{i}}}{a_{i \textsc{i}'}} - f_{e}(\boldsymbol{C}_{e})f_{i}(\boldsymbol{C}_{i},\textsc{i}) \Big) \, \mathrm{d} \boldsymbol{\omega}_{ei}' \mathrm{d} \boldsymbol{C}_{i}, \nonumber
\end{align}
where
\begin{align*}
\CrossSection_{ei}^{\textsc{i} \textsc{i}'} & = \CrossSection_{ei}^{\textsc{i} \textsc{i}'} \left( \Mass_{e} |\boldsymbol{C}_{e}|^{2}, \frac{\boldsymbol{C}_{e}}{|\boldsymbol{C}_{e}|} \cdot \boldsymbol{\omega}_{ei}' \right), \\
\boldsymbol{C}_{e}'^{(0)} & = |\boldsymbol{C}_{e}| \bigg( 1 - \frac{\Delta E_{\textsc{i} \textsc{i}'}}{\frac{1}{2}\Mass_{e}|\boldsymbol{C}_{e}|^{2}} \bigg)^{\frac{1}{2}} \boldsymbol{\omega}_{ei}'.
\end{align*}

\section{Chapman-Enskog Expansion of the Species Distribution Functions}
\label{SecKTChapEnsk}

In this section, we extend the classical procedure proposed by Enskog \cite{ChapmanCowling} \cite{FerzigerKaper} \cite{Giovangigli} to non-thermal polyatomic plasmas \cite{DegondLucquin} \cite{GrailleMaginMassot2009}. The species distribution functions are expanded in powers of the Knudsen number $\varepsilon$ and injected in the Boltzmann equations \eqref{KTScaled Boltzmann electron equation}-\eqref{KTScaled Boltzmann heavy-species equations}. Projection of the electron Boltzmann equation at order $\varepsilon^{-2}$ yields electron thermalization and isotropization in the heavy-species reference frame. As well, projection of the heavy-species Boltzmann equations at order $\varepsilon^{-1}$ yields thermalization of the heavy species. Euler type equations arise from the expansion of macroscopic equations at order $\varepsilon^{0}$, while expansion at order $\varepsilon^{1}$ yields Navier-Stokes type equations. As is classical, the closure of the equations is ensured by assuming that the perturbation of the distribution function from its Maxwellian equilibrium is orthogonal to the collisional invariants of the scattering operator \cite{ChapmanCowling} \cite{FerzigerKaper} \cite{Giovangigli}.

\subsection{Chapman-Enskog method}

We derive an approximate solution to the Boltzmann equations by expanding the species distribution functions as
\begin{align}
f_{e} & = f_{e}^{0} \left( 1 + \varepsilon \phi_{e} + \varepsilon^{2} \phi_{e}^{2} \right) + O(\varepsilon^{3}), \\
f_{i} & = f_{i}^{0} \left( 1 + \varepsilon \phi_{i} \right) + O(\varepsilon^{2}), ~ i \in \Heavy.
\end{align}
Traditionally, in the Chapman-Enskog's method \cite{ChapmanCowling} \cite{FerzigerKaper} \cite{Giovangigli}, the zeroth-order distribution function $f_{e}^{0}$ is assumed to yield the same local macroscopic properties as $f_{e}$ in the limit $\varepsilon \rightarrow 0$ \cite{GrailleMaginMassot2009}, namely
\begin{equation}
\left\lbrace \begin{array}{l}
\displaystyle \llangle f_{e}^{0}, \psi_{e}^{e} \rrangle_{e} = \density_{e}  \\
\displaystyle \llangle f_{e}^{0}, \psi_{e}^{n^{s}+4} \rrangle_{e} = \Energy_{e} \\
\end{array}
\right.
\label{KTZerothMacroscopicPropertiesElectron}
\end{equation}
for electrons, and as well
\begin{equation}
\left\lbrace \begin{array}{ll}
\displaystyle \llangle f_{\heavy}^{0}, \psi_{\heavy}^{j} \rrangle_{\heavy} = \density_{j}, & \displaystyle j \in \Heavy \vphantom{\psi_{\heavy}^{n^{s}+\nu}} \\
\displaystyle \llangle f_{\heavy}^{0}, \psi_{\heavy}^{n^{s}+\nu} \rrangle_{\heavy} = 0, & \displaystyle \nu \in \left\lbrace 1,2,3 \right\rbrace \\
\displaystyle \llangle f_{\heavy}^{0}, \psi_{\heavy}^{n^{s}+4} \rrangle_{\heavy} = \Energy_{\heavy} \\
\end{array}
\right.
\label{KTZerothMacroscopicPropertiesHeavy}
\end{equation}
for heavy species, where $\Energy_{e}$ and $\Energy_{\heavy}$ denote the respective internal energies per unit volume.

The Boltzmann equations \eqref{KTScaled Boltzmann electron equation},\eqref{KTScaled Boltzmann heavy-species equations} can be written in the compact form
\begin{align}
\Streaming_{e}(f_{e}) & = \frac{1}{\varepsilon^{2}} ~ \Scattering_{ee}\left(f_{e},f_{e}\right) + \frac{1}{\varepsilon^{2}} ~ \sum_{j \in \Heavy} \Scattering_{ej} \left(f_{e},f_{j} \right) + \varepsilon ~ \Chemistry_{e}\left(f\right),
\label{KTCompact Scaled Boltzmann electron equation} \\
\Streaming_{i}(f_{i}) & =  \frac{1}{\varepsilon^{2}} ~ \Scattering_{ie}\left(f_{i},f_{e}\right) + \frac{1}{\varepsilon} \sum_{j \in \Heavy} \Scattering_{ij} \left(f_{i},f_{j} \right) + \varepsilon ~ \Chemistry_{i}\left(f\right), \qquad i \in \Heavy.
\label{KTCompact Scaled Boltzmann heavy-species equations}
\end{align}
We have derived in the previous paragraph asymptotic expansions for the scattering operators $\Scattering_{ee}, \Scattering_{ej}, \Scattering_{je}, \Scattering_{ij}$, $i,j \in \Heavy$. We also expand the streaming operators $\Streaming$ as
\begin{align}
\Streaming_{e} & = \frac{1}{\varepsilon^{2}} ~ \Streaming_{e}^{-2} + \frac{1}{\varepsilon} ~ \Streaming_{e}^{-1} + \Streaming_{e}^{0} + \varepsilon ~ \Streaming_{e}^{1} + O(\varepsilon^{2}), \\
\Streaming_{i} & = \Streaming_{i}^{0} + \varepsilon ~ \Streaming_{i}^{1} + O(\varepsilon^{2}), \qquad i \in \Heavy, \vphantom{\frac{1}{\varepsilon^{2}}}
\end{align}
where
\begin{align}
 \Streaming_{e}^{-2}(f_{e}) = {} & \delta_{b1} \frac{q_{e}}{\Mass_{e}} \left[ \boldsymbol{C}_{e} \wedge \boldsymbol{B} \right] \cdot \boldsymbol{\partial}_{\boldsymbol{C}_{e}} f_{e}, \\
 \Streaming_{e}^{-1}(f_{e}) = {} & \boldsymbol{C}_{e} \cdot \boldsymbol{\partial_{x}} f_{e} + \delta_{b0} \frac{q_{e}}{\Mass_{e}} \left[ \boldsymbol{C}_{e} \wedge \boldsymbol{B} \right] \cdot \boldsymbol{\partial}_{\boldsymbol{C}_{e}} f_{e} + \delta_{b1} \frac{q_{e}}{\Mass_{e}} \left[ \boldsymbol{\Meanv}_{\heavy} \wedge \boldsymbol{B} \right] \cdot \boldsymbol{\partial}_{\boldsymbol{C}_{e}} f_{e} \\
{} & + \frac{q_{e}}{\Mass_{e}} \boldsymbol{E} \cdot \boldsymbol{\partial}_{\boldsymbol{C}_{e}} f_{e}, \nonumber \\
 \Streaming_{e}^{0}(f_{e}) = {} & \partial_{t}f_{e} + \boldsymbol{\Meanv}_{\heavy} \cdot \boldsymbol{\partial_{x}} f_{e} + \delta_{b(-1)} \frac{q_{e}}{\Mass_{e}} \left[ \boldsymbol{C}_{e} \wedge \boldsymbol{B} \right] \cdot \boldsymbol{\partial}_{\boldsymbol{C}_{e}} f_{e} + \delta_{b0} \frac{q_{e}}{\Mass_{e}} \left[ \boldsymbol{\Meanv}_{\heavy} \wedge \boldsymbol{B} \right] \cdot \boldsymbol{\partial}_{\boldsymbol{C}_{e}} f_{e} \\
{} & - (\boldsymbol{\partial}_{\boldsymbol{C}_{e}}f_{e} \otimes \boldsymbol{C}_{e}) : \boldsymbol{\partial_{x}} \boldsymbol{\Meanv}_{\heavy}, \nonumber \\
 \Streaming_{e}^{1}(f_{e}) = {} & \delta_{b(-2)} \frac{q_{e}}{\Mass_{e}} \left[ \boldsymbol{C}_{e} \wedge \boldsymbol{B} \right] \cdot \boldsymbol{\partial}_{\boldsymbol{C}_{e}} f_{e} + \delta_{b(-1)} \frac{q_{e}}{\Mass_{e}} \left[ \boldsymbol{\Meanv}_{\heavy} \wedge \boldsymbol{B} \right] \cdot \boldsymbol{\partial}_{\boldsymbol{C}_{e}} f_{e} - \frac{D \boldsymbol{\Meanv}_{\heavy}}{Dt} \cdot \boldsymbol{\partial}_{\boldsymbol{C}_{e}} f_{e},
\end{align}
and
\begin{align}
\Streaming_{i}^{0}(f_{i}) = {} & \partial_{t}f_{i} + (\boldsymbol{C}_{i} + \boldsymbol{\Meanv}_{\heavy}) \cdot \boldsymbol{\partial_{x}} f_{i} + \delta_{b1} \frac{q_{i}}{\Mass_{i}} \left[ (\boldsymbol{C}_{i} + \boldsymbol{\Meanv}_{\heavy}) \wedge \boldsymbol{B} \right] \cdot \boldsymbol{\partial}_{\boldsymbol{C}_{i}} f_{i} \\
{} & + \Big( \frac{q_{i}}{\Mass_{i}} \boldsymbol{E} - \frac{D \boldsymbol{\Meanv}_{\heavy}}{Dt} \Big) \cdot \boldsymbol{\partial}_{\boldsymbol{C}_{i}} f_{i} - (\boldsymbol{\partial}_{\boldsymbol{C}_{i}}f_{i} \otimes \boldsymbol{C}_{i}) : \boldsymbol{\partial_{x}} \boldsymbol{\Meanv}_{\heavy}, \nonumber \\
\Streaming_{i}^{1}(f_{i}) = {} & \delta_{b0} \frac{q_{i}}{\Mass_{i}} \left[ (\boldsymbol{C}_{i} + \boldsymbol{\Meanv}_{\heavy}) \wedge \boldsymbol{B} \right] \cdot \boldsymbol{\partial}_{\boldsymbol{C}_{i}} f_{i}.
\end{align}

Equations \eqref{KTCompact Scaled Boltzmann electron equation}, \eqref{KTCompact Scaled Boltzmann heavy-species equations} are then projected against collisional invariants in $\Invariants_{e}^{\varepsilon}$, $\Invariants_{\heavy}$. Making use of orthogonality properties \eqref{KTPairwiseInvariantsConservationElectron}, \eqref{KTPairwiseInvariantsConservationHeavy}, one obtains
\begin{align}
\llangle \psi_{e}^{e}, \Streaming_{e}(f_{e}) \rrangle_{e} = {} & \vphantom{\frac{1}{\varepsilon^{2}} \sum_{j \in \Heavy} \llangle \psi_{e}^{e}, \Scattering_{ej} \rrangle_{e} +} \varepsilon \llangle \psi_{e}^{e}, \Chemistry_{e}(f) \rrangle_{e},
\label{KTElectronMatterConservation} \\
\llangle \varepsilon \Mass_{e} C_{e \nu}, \Streaming_{e}(f_{e}) \rrangle_{e} = {} & \frac{1}{\varepsilon^{2}} \sum_{j \in \Heavy} \llangle \varepsilon \Mass_{e} C_{e \nu}, \Scattering_{ej} \rrangle_{e} + \varepsilon \llangle \varepsilon \Mass_{e} C_{e \nu}, \Chemistry_{e}(f) \rrangle_{e}, \quad \nu \in \left\lbrace 1,2,3 \right\rbrace,
\label{KTElectronMomentumConservation} \\
\llangle \psi_{e}^{n^{s}+4}, \Streaming_{e}(f_{e}) \rrangle_{e} = {} & \frac{1}{\varepsilon^{2}} \sum_{j \in \Heavy} \llangle \psi_{e}^{n^{s}+4}, \Scattering_{ej} \rrangle_{e} + \varepsilon \llangle \psi_{e}^{n^{s}+4}, \Chemistry_{e}(f) \rrangle_{e},
\label{KTElectronEnergyConservation}
\end{align}
for electrons, and
\begin{align}
\llangle \psi_{\heavy}^{i}, \Streaming_{\heavy}(f_{\heavy}) \rrangle_{\heavy} = {} & \vphantom{\frac{1}{\varepsilon^{2}} \llangle \psi_{\heavy}^{i}, \Scattering_{\heavy e} \rrangle_{\heavy} +} \varepsilon \llangle \psi_{\heavy}^{i}, \Chemistry_{\heavy}(f) \rrangle_{\heavy}, \qquad i \in \Heavy,
\label{KTHeavyMatterConservation} \\
\llangle \psi_{\heavy}^{n^{s}+\nu}, \Streaming_{\heavy}(f_{\heavy}) \rrangle_{\heavy} = {} & \frac{1}{\varepsilon^{2}} \llangle \psi_{\heavy}^{n^{s}+\nu}, \Scattering_{\heavy e} \rrangle_{\heavy} + \varepsilon \llangle \psi_{\heavy}^{n^{s}+\nu}, \Chemistry_{\heavy}(f) \rrangle_{\heavy}, \qquad \nu \in \left\lbrace 1,2,3 \right\rbrace,
\label{KTHeavyMomentumConservation} \\
\llangle \psi_{\heavy}^{n^{s}+4}, \Streaming_{\heavy}(f_{\heavy}) \rrangle_{\heavy} = {} & \frac{1}{\varepsilon^{2}} \llangle \psi_{\heavy}^{n^{s}+4}, \Scattering_{\heavy e} \rrangle_{\heavy} + \varepsilon \llangle \psi_{\heavy}^{n^{s}+4}, \Chemistry_{\heavy}(f) \rrangle_{\heavy},
\label{KTHeavyEnergyConservation}
\end{align}
for heavy species. Equations \eqref{KTElectronMatterConservation} and \eqref{KTHeavyMatterConservation} account for conservation of matter, equations \eqref{KTElectronMomentumConservation} and \eqref{KTHeavyMomentumConservation} for conservation of momentum, and equations \eqref{KTElectronEnergyConservation} and \eqref{KTHeavyEnergyConservation} for conservation of energy. These equations are completed with the cross-collision orthogonality relations \eqref{KTCrossInvariantsConservationNumber}, \eqref{KTCrossInvariantsConservationMomentum}, \eqref{KTCrossInvariantsConservationEnergy}, together with the macroscopic constraints \eqref{KTZerothMacroscopicPropertiesElectron}, \eqref{KTZerothMacroscopicPropertiesHeavy}.

\subsection{Electron thermalization}
\label{SubSecKTElectronThermalization}
We solve the electron Boltzmann equation \eqref{KTCompact Scaled Boltzmann electron equation} at order $\varepsilon^{-2}$, corresponding to the kinetic timescale $t_{e}^{0}$. We obtain the following equation for $f_{e}^{0}$:
\begin{equation}
\delta_{b1} ~ \frac{q_{e}}{\Mass_{e}} (\boldsymbol{C}_{e} \wedge \boldsymbol{B}) \cdot \boldsymbol{\partial}_{\boldsymbol{C}_{e}} f_{e}^{0} = \Scattering_{ee}\left(f_{e}^{0},f_{e}^{0}\right) + \sum_{j \in \Heavy} \Scattering_{ej}^{0} \left(f_{e}^{0},f_{j}^{0} \right).
\label{KTElectronBoltzmannEpsilon-2}
\end{equation}
Multiplying this equation by $\ln f_{e}^{0}$ and integrating over $\boldsymbol{C}_{e}$, we get
\[
\Gamma_{ee}^{0} \, + \, \sum_{j \in \Heavy} \Gamma_{ej}^{0} ~ + ~ \delta_{b1} ~ k_{\textsc{b}} ~ \frac{q_{e}}{\Mass_{e}} \int (\boldsymbol{C}_{e} \wedge \boldsymbol{B}) \cdot \boldsymbol{\partial}_{\boldsymbol{C}_{e}} f_{e}^{0} ~ \ln f_{e}^{0} \, \mathrm{d} \boldsymbol{C}_{e} = 0,
\]
where $\Gamma_{ee}^{0}$ and $\Gamma_{ej}^{0}$ are the zeroth-order entropy production rates for electron electron and electron $j^{\text{th}}$-heavy-species collisions, respectively
\begin{align}
\Gamma_{ee}^{0} & = - k_{\textsc{b}} \int \Scattering_{ee}\left(f_{e}^{0},f_{e}^{0}\right) (\boldsymbol{C}_{e}) ~ \ln \left( f_{e}^{0}(\boldsymbol{C}_{e}) \right) \, \mathrm{d} \boldsymbol{C}_{e}, \\
{} \nonumber \\
\Gamma_{ej}^{0} & = - k_{\textsc{b}} \int \Scattering_{ej}^{0} \left(f_{e}^{0},f_{j}^{0}\right) (\boldsymbol{C}_{e}) ~ \ln \left( f_{e}^{0}(\boldsymbol{C}_{e}) \right) \, \mathrm{d} \boldsymbol{C}_{e}, \qquad j \in \Heavy.
\end{align}
Noting that
\[
\boldsymbol{\partial}_{\boldsymbol{C}_{e}} f_{e}^{0} ~ \ln f_{e}^{0} = \boldsymbol{\partial}_{\boldsymbol{C}_{e}} \left[ f_{e}^{0} ~ \ln f_{e}^{0} - f_{e}^{0} \right],
\]
and integrating by parts we obtain
\begin{align*}
\int (\boldsymbol{C}_{e} \wedge \boldsymbol{B}) \cdot \boldsymbol{\partial}_{\boldsymbol{C}_{e}} f_{e}^{0} ~ \ln f_{e}^{0} \, \mathrm{d} \boldsymbol{C}_{e} = {} & \int (\boldsymbol{C}_{e} \wedge \boldsymbol{B}) \cdot \boldsymbol{\partial}_{\boldsymbol{C}_{e}} \left[ f_{e}^{0} ~ \ln f_{e}^{0} - f_{e}^{0} \right] \, \mathrm{d} \boldsymbol{C}_{e} \\
%= {} & \int \boldsymbol{\partial}_{\boldsymbol{C}_{e}} \cdot \left[ ( f_{e}^{0} ~ \ln f_{e}^{0} - f_{e}^{0} ) (\boldsymbol{C}_{e} \wedge \boldsymbol{B}) \right] \, \mathrm{d} \boldsymbol{C}_{e} \\
% & - \int ( f_{e}^{0} ~ \ln f_{e}^{0} - f_{e}^{0} ) ~ \boldsymbol{\partial}_{\boldsymbol{C}_{e}} \cdot \left[  \boldsymbol{C}_{e} \wedge \boldsymbol{B} \right] \, \mathrm{d} \boldsymbol{C}_{e} \\
= {} & \int \boldsymbol{\partial}_{\boldsymbol{C}_{e}} \cdot \left[ ( f_{e}^{0} ~ \ln f_{e}^{0} - f_{e}^{0} ) (\boldsymbol{C}_{e} \wedge \boldsymbol{B}) \right] \, \mathrm{d} \boldsymbol{C}_{e},
\end{align*}
since $\boldsymbol{\partial}_{\boldsymbol{C}_{e}} \cdot \left[  \boldsymbol{C}_{e} \wedge \boldsymbol{B} \right] = 0$. Now, for a given $R > 0$, if $\mathscr{B}(\boldsymbol{0},R)$ is the ball of center $\boldsymbol{0}$ and radius $R$ in the electron velocity space ($\boldsymbol{C}_{e} \in \mathbb{R}^{3}$), and $\mathscr{S}(0,R)$ is the corresponding sphere, then
\begin{align*}
 \int_{\mathscr{B}(\boldsymbol{0},R)} \boldsymbol{\partial}_{\boldsymbol{C}_{e}} \cdot \left[ ( f_{e}^{0} ~ \ln f_{e}^{0} - f_{e}^{0} ) (\boldsymbol{C}_{e} \wedge \boldsymbol{B}) \right] \, \mathrm{d} \boldsymbol{C}_{e} & = \int_{\mathscr{S}(\boldsymbol{0},R)} ( f_{e}^{0} ~ \ln f_{e}^{0} - f_{e}^{0} ) (\boldsymbol{C}_{e} \wedge \boldsymbol{B}) \cdot \frac{\boldsymbol{C}_{e}}{|\boldsymbol{C}_{e}|} \, \mathrm{d} \boldsymbol{C}_{e} \\
 \hphantom{\int_{\mathscr{B}(\boldsymbol{0},R)} \boldsymbol{\partial}_{\boldsymbol{C}_{e}} \cdot \left[ ( f_{e}^{0} ~ \ln f_{e}^{0} - f_{e}^{0} ) (\boldsymbol{C}_{e} \wedge \boldsymbol{B}) \right] \, \mathrm{d} \boldsymbol{C}_{e}} & = 0,
\end{align*}
and this holds for all $R >0$, so that
\[
\int \boldsymbol{\partial}_{\boldsymbol{C}_{e}} \cdot \left[ ( f_{e}^{0} ~ \ln f_{e}^{0} - f_{e}^{0} ) (\boldsymbol{C}_{e} \wedge \boldsymbol{B}) \right] \, \mathrm{d} \boldsymbol{C}_{e} = 0.
\]
More generally, the latter statement can also be proven for any kind of force instead of the Lorentz force $q_{e} \boldsymbol{C}_{e} \wedge \boldsymbol{B}$, upon assuming that the electron distribution function $f_{e}$ decreases sufficiently rapidly when $|\boldsymbol{C}_{e}| \to + \infty$ \cite{GraillePhD}. Thus, one gets finally
\begin{equation}
\Gamma_{ee}^{0} + \sum_{j \in \Heavy} \Gamma_{ej}^{0} = 0.
\label{KTElectronZerothOrderEntropy}
\end{equation}

Using the reciprocity relation \eqref{KTScattering Reciprocity Relation} and symmetrizing, a classical derivation \cite{GrailleMaginMassot2009} yields
\begin{equation}
 \Gamma_{ee}^{0} = \frac{k_{\textsc{b}}}{4} \int \CrossSection_{e \tilde{e}} g_{e \tilde{e}} \Omega(f_{e} \tilde{f}_{e}, f_{e}' \tilde{f}_{e}') \, \mathrm{d} \boldsymbol{\omega}_{e \tilde{e}}' \mathrm{d} \boldsymbol{C}_{e} \mathrm{d} \widetilde{\boldsymbol{C}}_{e},
\label{KTZerothKineticEntropyElectronElectron}
\end{equation}
where $\tilde{f}_{e} = f_{e}(t,\boldsymbol{x},\widetilde{\boldsymbol{C}}_{e})$ and $\tilde{f}_{e}' = f_{e}(t,\boldsymbol{x},\widetilde{\boldsymbol{C}}_{e}')$, and where
\begin{equation}
\Omega(x,y) = \ln \Big( \frac{x}{y} \Big) (x-y)
\label{KTOmega}
\end{equation}
is a nonnegative function. Similarly, the electron $j^{\text{th}}$-heavy-species entropy production term is expressed by means of \eqref{KTSei0} as
\begin{align*}
\Gamma_{ej}^{0} & = - k_{\textsc{b}} \int \Scattering_{ej}^{0} \left(f_{e}^{0},f_{j}^{0}\right) (\boldsymbol{C}_{e}) ~ \ln \left( f_{e}^{0}(\boldsymbol{C}_{e}) \right) \, \mathrm{d} \boldsymbol{C}_{e} \\
 & = - k_{\textsc{b}} \sum_{\textsc{j} \in \QuantumSpace_{j}} \left( \int f_{j}^{0}(\boldsymbol{C}_{j},\textsc{j}) \, \mathrm{d} \boldsymbol{C}_{j} \right) \\
 & \qquad \qquad \times \int \CrossSection_{ej}^{\textsc{j} \textsc{j}} ~ |\boldsymbol{C}_{e}| \left( f_{e}^{0}(|\boldsymbol{C}_{e}| \boldsymbol{\omega}_{ej}') - f_{e}^{0}(\boldsymbol{C}_{e}) \right) ~ \ln \left( f_{e}^{0}(\boldsymbol{C}_{e}) \right) \, \mathrm{d} \boldsymbol{\omega}_{ej}' \mathrm{d} \boldsymbol{C}_{e},
\end{align*}
which again by reciprocity relations and symmetrization is equal to
\begin{equation}
 \Gamma_{ej}^{0} = \frac{k_{\textsc{b}}}{2} \sum\limits_{\substack{\textsc{j} \in \QuantumSpace_{j}}} \left( \int f_{j}^{0} ( \boldsymbol{C}_{j}, \textsc{j} ) \, \mathrm{d} \boldsymbol{C}_{j} \right) \int \CrossSection_{ej}^{\textsc{j} \textsc{j}} |\boldsymbol{C}_{e}| ~ \Omega \left( f_{e}^{0} ( |\boldsymbol{C}_{e}| \boldsymbol{\omega}_{ej}' ), f_{e}^{0} ( \boldsymbol{C}_{e} ) \right) \, \mathrm{d} \boldsymbol{\omega}_{ej}' \mathrm{d} \boldsymbol{C}_{e} \geq 0. \label{KTZerothKineticEntropyElectronHeavy}
\end{equation}
A sum of positive terms is equal to zero if and only if each term of the sum is zero itself, thus the entropy production rates vanish: $\Gamma_{ee}^{0} = 0$, and $\Gamma_{ej}^{0} = 0$ for all $j \in \Heavy$.

Since $\Gamma_{ee}^{0} = 0$, one can see from expression \eqref{KTZerothKineticEntropyElectronElectron} and the definition of $\Omega$ that $\ln(f_{e}^{0})$ must be a collisional invariant associated with the electron electron scattering operator, i.e., must lie in the space $\Invariants_{e}^{\varepsilon}$, spanned by $\psi_{e}^{e} = 1$, $\psi_{e}^{n^{s}+\nu}~=~\varepsilon \Mass_{e} C_{e \nu}$, $\nu \in \left\lbrace 1,2,3 \right\rbrace$, and $\psi_{e}^{n^{s}+4} = \frac{1}{2} \Mass_{e} \boldsymbol{C}_{e} \cdot \boldsymbol{C}_{e}$. Similarly, one can see from expression \eqref{KTZerothKineticEntropyElectronHeavy} that $f_{e}^{0}$ has to be an isotropic function of $\boldsymbol{C}_{e}$, so that $\ln(f_{e}^{0})$ is an isotropic electron collisional invariant:
\[
\ln(f_{e}^{0})\in \Invariants_{e}^{0} =  \ran \left( 1, \frac{1}{2} \Mass_{e} \boldsymbol{C}_{e} \cdot \boldsymbol{C}_{e} \right).
\]

Using the macroscopic constraints \eqref{KTZerothMacroscopicPropertiesElectron} for conservation of matter and energy, we obtain finally the expression for the zeroth-order electron distribution function
\begin{equation}
f_{e}^{0} (\boldsymbol{C}_{e}) = \density_{e}  \left( \frac{\Mass_{e}}{2 \pi k_{\textsc{b}} \Temperature_{e}} \right)^{\frac{3}{2}} ~ \exp \left( - \frac{\Mass_{e}}{2 k_{\textsc{b}} \Temperature_{e}} ~  \boldsymbol{C}_{e} \cdot \boldsymbol{C}_{e} \right).
\label{KTMaxwellElectron}
\end{equation}
The electron population thus thermalizes to a quasi-equilibrium state described by a Maxwell-Boltzmann distribution at some temperature $\Temperature_{e}$, defined as
\begin{equation}
\Energy_{e} = \frac{3}{2} \density_{e} k_{\textsc{b}} \Temperature_{e}.
\end{equation}
The Maxwellian distribution \eqref{KTMaxwellElectron} is such that $\Scattering_{ee}(f_{e}^{0},f_{e}^{0}) = 0$ and $\Scattering_{ej}^{0}(f_{e}^{0},f_{j}^{0}) = 0$, for all $j \in \Heavy$.

Finally, if one defines the electron partial pressure $\pressure_{e}$ as
\begin{equation}
\pressure_{e} \, \mathbb{I} = \int \Mass_{e} \, \boldsymbol{C}_{e} \otimes \boldsymbol{C}_{e} \, f_{e}^{0} \, \mathrm{d} \boldsymbol{C}_{e} = \left( \frac{1}{3} \int \Mass_{e} \, \boldsymbol{C}_{e} \cdot \boldsymbol{C}_{e} \, f_{e}^{0} \, \mathrm{d} \boldsymbol{C}_{e} \right) \mathbb{I},
\end{equation}
where the latter equality comes from the isotropy of $f_{e}^{0}$, one retrieves the perfect gas law for the electrons
\begin{equation}
\rho_{e} = \frac{\pressure_{e} \MoleMass_{e}}{\RU \Temperature_{e}},
\label{KTElectronPerfectGas}
\end{equation}
where $\MoleMass_{e} = \Avogadro \Mass_{e}$ is the molar mass of the electron, and $\RU$ is the universal gas constant.

\subsection{Heavy-species thermalization}
\label{SubSecKTHeavyThermalization}

Now we solve the $i^{\text{th}}$-heavy-species equation \eqref{KTCompact Scaled Boltzmann heavy-species equations} at order $\varepsilon^{-1}$. Since $\Scattering_{ie}^{0} = 0	$ from \eqref{KTSie0}, this yields, for all $i \in \Heavy$
\begin{equation*}
\Scattering_{ie}^{1}(f_{i}^{0}, f_{e}^{0}) +  \sum_{j \in \Heavy} \Scattering_{ij}\left( f_{i}^{0},f_{j}^{0} \right) = 0.
\end{equation*}
Since $f_{e}^{0}$ is isotropic, the term $\Scattering_{ie}^{1}(f_{i}^{0}, f_{e}^{0})$ given in \eqref{KTSie1} vanishes, and hence
\begin{equation}
\sum_{j \in \Heavy} \Scattering_{ij}\left( f_{i}^{0},f_{j}^{0} \right) = 0.
\label{KTHeavyBoltzmannEpsilon-1}
\end{equation}
Multiplying this equation by $\ln(\beta_{i \textsc{i}} f_{i}^{0})$, integrating over $\boldsymbol{C}_{i}$, summing over $\textsc{i} \in \QuantumSpace_{i}$ and then over $i \in \Heavy$, we get
\begin{equation}
\Gamma_{\heavy \heavy}^{0} = 0,
\label{KTHeavyZerothOrderEntropy}
\end{equation}
where $\Gamma_{\heavy \heavy}^{0}$ is the zeroth-order entropy production rate associated with heavy-species collisions
\begin{equation}
\Gamma_{\heavy \heavy}^{0} = - k_{\textsc{b}} \sum_{i, j \in \Heavy} \sum_{\textsc{i} \in \QuantumSpace_{i}} \int \Scattering_{ij} \left(f_{i}^{0},f_{j}^{0}\right) (\boldsymbol{C}_{i}) ~ \ln \left( \beta_{i \textsc{i}} f_{i}^{0}(\boldsymbol{C}_{i}) \right) \, \mathrm{d} \boldsymbol{C}_{i}.
\end{equation}
From expression \eqref{KTScatteringHeavyHeavy} for $\Scattering_{ij}$, and using the reciprocity relation \eqref{KTScattering Reciprocity Relation} and symmetrization, the latter term can be written in the form
\begin{equation}
\Gamma_{\heavy \heavy}^{0} = \frac{k_{\textsc{b}}}{4} \sum_{i, j \in \Heavy} \sum_{\textsc{i}, \textsc{i}' \in \QuantumSpace_{i}} \sum_{\textsc{j}, \textsc{j}' \in \QuantumSpace_{j}} \int \CrossSection_{ij}^{\textsc{i}\textsc{j}\textsc{i}'\textsc{j}'} g_{ij} \, \Omega \Big( \frac{a_{i \textsc{i}} a_{j \textsc{j}}}{a_{i \textsc{i}'} a_{j \textsc{j}'}} f_{i}^{0'} f_{j}^{0'} , f_{i}^{0} f_{j}^{0} \Big) \, \mathrm{d} \boldsymbol{\omega}_{ij}' \mathrm{d} \boldsymbol{C}_{i} \mathrm{d} \boldsymbol{C}_{j}.
\end{equation}
Since $\Omega$ is nonnegative, each term in the sum has to be zero, i.e.,
\[
a_{i \textsc{i}} a_{j \textsc{j}} f_{i}^{0'} f_{j}^{0'} = a_{i \textsc{i}'} a_{j \textsc{j}'} f_{i}^{0} f_{j}^{0}, \qquad i,j \in \Heavy, \textsc{i}, \textsc{i}', \in \QuantumSpace_{i}, \textsc{j}, \textsc{j}' \in \QuantumSpace_{j}.
\]
In other words, $\big(\ln(\beta_{i \textsc{i}} f_{i}^{0}) \big)_{i \in \Heavy}$ is a collisional invariant of the heavy-species scattering operator:
\[
\big( \ln(\beta_{i \textsc{i}} f_{i}^{0}) \big)_{i \in \Heavy} \in \Invariants_{\heavy},
\]
i.e., there exists constants $\alpha_{i}$, $i \in \Heavy$, $\boldsymbol{w}$, and $\gamma$ such that
\[
\ln(\beta_{i \textsc{i}} f_{i}^{0}) = \alpha_{i} - \boldsymbol{w} \cdot \Mass_{i} \boldsymbol{C}_{i} - \gamma \big(\frac{1}{2}\Mass_{i} \boldsymbol{C}_{i} \cdot \boldsymbol{C}_{i} + E_{i \textsc{i}} \big), \qquad i \in \Heavy, \, \textsc{i} \in \QuantumSpace_{i}.
\]
The constants $\alpha_{i}$, $i \in \Heavy$, $\boldsymbol{w}$, and $\gamma$ are obtained from the macroscopic constraints \eqref{KTZerothMacroscopicPropertiesHeavy} for conservation of matter, momentum and energy, yielding the following expression for the zeroth-order heavy-species distribution function
\begin{equation}
f_{i}^{0} (\boldsymbol{C}_{i}, \textsc{i}) = \density_{i} \left( \frac{\Mass_{i}}{2 \pi k_{\textsc{b}} \Temperature_{\heavy}} \right)^{\frac{3}{2}} \frac{a_{i \textsc{i}}}{Q_{i}^{\text{int}}} ~ \exp \left( - \frac{\Mass_{i}}{2 k_{\textsc{b}} \Temperature_{\heavy}} ~ \boldsymbol{C}_{i} \cdot \boldsymbol{C}_{i} -\frac{E_{i \textsc{i}}}{k_{\textsc{b}} \Temperature_{\heavy}} \right), \, i \in \Heavy, \, \textsc{i} \in \QuantumSpace_{i},
\label{KTMaxwellHeavy}
\end{equation}
where $\Temperature_{\heavy}$ is the heavy-species temperature, given by
\begin{equation}
\frac{3}{2} \density_{\heavy} k_{\textsc{b}} \Temperature_{\heavy} = \sum_{i \in \Heavy} \sum_{\textsc{i} \in \QuantumSpace_{i}} \int \frac{1}{2} \Mass_{i} \boldsymbol{C}_{i} \cdot \boldsymbol{C}_{i} \, f_{i}^{0} \, \mathrm{d} \boldsymbol{C}_{i},
\end{equation}
and where we have introduced the partition function for internal energy of the $i^{\text{th}}$ species
\begin{equation}
Q_{i}^{\text{int}} = \sum_{\textsc{i} \in \QuantumSpace_{i}} a_{i \textsc{i}} \exp \left( - \frac{E_{i \textsc{i}}}{k_{\textsc{b}}\Temperature_{\heavy}} \right).
\end{equation}
Alternatively, one can write
\begin{equation}
f_{i}^{0} (\boldsymbol{C}_{i}, \textsc{i}) = \density_{i} \frac{1}{\beta_{i \textsc{i}} Q_{i}} ~ \exp \left( - \frac{\Mass_{i}}{2 k_{\textsc{b}} \Temperature_{\heavy}} ~ \boldsymbol{C}_{i} \cdot \boldsymbol{C}_{i} -\frac{E_{i \textsc{i}}}{k_{\textsc{b}} \Temperature_{\heavy}} \right), \qquad i \in \Heavy, \, \textsc{i} \in \QuantumSpace_{i},
\end{equation}
where the statistical weights $\beta_{i \textsc{i}}$ are given by
\begin{equation}
\beta_{i \textsc{i}} = \frac{h_{\textsc{p}}^{3}}{a_{i \textsc{i}} \Mass_{i}^{3}}, \qquad i \in \Heavy, \, \textsc{i} \in \QuantumSpace_{i},
\end{equation}
and where the translational and full partition functions per unit volume read
\begin{align}
Q_{i}^{\text{tr}} = \left( \frac{2 \pi \Mass_{i} k_{\textsc{b}} \Temperature_{\heavy}}{h_{\textsc{p}}^{2}} \right)^{\frac{3}{2}}, & \qquad Q_{i} = Q_{i}^{\text{tr}} Q_{i}^{\text{int}}, \qquad i \in \Heavy.
\end{align}
The Maxwellian distribution \eqref{KTMaxwellHeavy} is such that $\Scattering_{ij}(f_{i}^{0},f_{j}^{0}) = 0$, for $i,j \in \Heavy$.

As a consequence of expression \eqref{KTMaxwellHeavy}, the heavy-species internal energy per unit volume reads
\begin{equation}
\mathcal{E}_{\heavy} = \sum_{i \in \Heavy} \density_{i} (\frac{3}{2} k_{\textsc{b}} \Temperature_{\heavy} + \overline{E}_{i}),
\label{KTHeavyInternalEnergy}
\end{equation}
where
\begin{equation}
\overline{E}_{i} = \sum_{\textsc{i} \in \QuantumSpace_{i}} \frac{a_{i \textsc{i}} E_{i \textsc{i}}}{Q_{i}^{\text{int}}} \exp{ \left( - \frac{E_{i \textsc{i}}}{k_{\textsc{b}} \Temperature_{\heavy}}  \right)}
\end{equation}
is the mean excitation energy of the $i^{\text{th}}$ species.

Finally, defining the heavy-species partial pressure as
\begin{equation}
\pressure_{\heavy} \, \mathbb{I} = \sum_{i \in \Heavy} \sum_{\textsc{i} \in \QuantumSpace} \int \Mass_{i} \, \boldsymbol{C}_{i} \otimes \boldsymbol{C}_{i} \, f_{i}^{0} \, \mathrm{d} \boldsymbol{C}_{i} = \left( \frac{1}{3} \sum_{i \in \Heavy} \sum_{\textsc{i} \in \QuantumSpace} \int \Mass_{i} \, \boldsymbol{C}_{i} \cdot \boldsymbol{C}_{i} \, f_{i}^{0} \, \mathrm{d} \boldsymbol{C}_{i} \right) \mathbb{I},
\end{equation}
we retrieve the perfect gas law for heavy species
\begin{equation}
\rho_{\heavy} = \frac{\pressure_{\heavy} \overline{\MoleMass}_{\heavy}}{\RU \Temperature_{\heavy}},
\label{KTHeavyPerfectGas}
\end{equation}
where $\overline{\MoleMass}_{\heavy}$ is the mean heavy-species molar mass, given by
\begin{equation}
\frac{\rho_{\heavy}}{\overline{\MoleMass}_{\heavy}} = \sum_{i \in \Heavy} \frac{\rho_{i}}{\MoleMass_{i}},
\label{KTHeavyMeanMoleMass}
\end{equation}
$\MoleMass_{i} = \Avogadro \Mass_{i}$ being the molar mass of the $i^{\text{th}}$ heavy species.

\subsection{First-order perturbation function for electrons}
We introduce the electron linearized collision operator $\mathcal{F}_{e}$, defined as
\begin{equation}
\mathcal{F}_{e}(\phi_{e}) = - \frac{1}{f_{e}^{0}} \left[ \Scattering_{ee}(\phi_{e}f_{e}^{0}, f_{e}^{0}) + \Scattering_{ee}(f_{e}^{0}, \phi_{e}f_{e}^{0}) + \sum_{j \in \Heavy} \Scattering_{ej}^{0}(\phi_{e}f_{e}^{0}, f_{j}^{0}) \right].
\end{equation}
The sum of the first two terms on the right hand side can be computed from \eqref{KTScatteringElectronElectron} as follows:
\begin{multline*}
- \frac{1}{f_{e}^{0}} \left[ \Scattering_{ee}(\phi_{e}f_{e}^{0}, f_{e}^{0}) + \Scattering_{ee}(f_{e}^{0}, \phi_{e}f_{e}^{0}) \right] = \\
 - \frac{1}{f_{e}^{0}} \int g_{e \tilde{e}} \CrossSection_{e \tilde{e}} \left( \phi_{e}'f_{e}^{0'} \tilde{f}_{e}^{0'}- \phi_{e}f_{e}^{0} \tilde{f}_{e}^{0} + f_{e}^{0'} \tilde{\phi}_{e}' \tilde{f}_{e}^{0'} - f_{e}^{0} \tilde{\phi}_{e} \tilde{f}_{e}^{0} \right) \, \mathrm{d} \boldsymbol{\omega}_{e \tilde{e}}' \mathrm{d} \widetilde{\boldsymbol{C}}_{e},
\end{multline*}
where $\tilde{\psi}_{e} = \psi_{e}(t,\boldsymbol{x},\widetilde{\boldsymbol{C}}_{e})$ and $\tilde{\psi}_{e}' = \psi_{e}(t,\boldsymbol{x},\widetilde{\boldsymbol{C}}_{e}')$ for any function $\psi_{e}$ of $t$, $\boldsymbol{x}$ and $\boldsymbol{C}_{e}$. The conservation of energy for electron-electron collisions reads
\[
\frac{1}{2} \Mass_{e} |\boldsymbol{C}_{e}|^{2} + \frac{1}{2} \Mass_{e} |\widetilde{\boldsymbol{C}}_{e}|^{2} = \frac{1}{2} \Mass_{e} |\boldsymbol{C}_{e}'|^{2} + \frac{1}{2} \Mass_{e} |\widetilde{\boldsymbol{C}}_{e}'|^{2},
\]
so that from \eqref{KTMaxwellElectron}
\begin{equation}
f_{e}^{0'} \tilde{f}_{e}^{0'} = f_{e}^{0} \tilde{f}_{e}^{0},
\end{equation}
and thus
\[
- \frac{1}{f_{e}^{0}} \left[ \Scattering_{ee}(\phi_{e}f_{e}^{0}, f_{e}^{0}) + \Scattering_{ee}(f_{e}^{0}, \phi_{e}f_{e}^{0}) \right] = \\
 - \frac{1}{f_{e}^{0}} \int g_{e \tilde{e}} \CrossSection_{e \tilde{e}} f_{e}^{0} \tilde{f}_{e}^{0} \left( \phi_{e}' - \phi_{e} + \tilde{\phi}_{e}' - \tilde{\phi}_{e} \right) \, \mathrm{d} \boldsymbol{\omega}_{e \tilde{e}}' \mathrm{d} \widetilde{\boldsymbol{C}}_{e}.
\]
Similarly, from expression \eqref{KTSei0} of $\Scattering_{ej}^{0}$, the last term on the right hand side reads
\begin{align*}
 - \frac{1}{f_{e}^{0}} & \sum_{j \in \Heavy} \Scattering_{ej}^{0}(\phi_{e}f_{e}^{0}, f_{j}^{0}) \\
 & = - \frac{1}{f_{e}^{0}} \sum_{j \in \Heavy} \sum_{\textsc{j} \in \QuantumSpace_{j}} \left( \int f_{j}^{0}(\boldsymbol{C}_{j},\textsc{j}) \, \mathrm{d} \boldsymbol{C}_{j} \right) \int \CrossSection_{ej}^{\textsc{j} \textsc{j}} ~ |\boldsymbol{C}_{e}| \left( \phi_{e}f_{e}^{0}(|\boldsymbol{C}_{e}| \boldsymbol{\omega}_{ej}') - \phi_{e}f_{e}^{0}(\boldsymbol{C}_{e}) \right) \, \mathrm{d} \boldsymbol{\omega}_{ej}' \\
 & = - \frac{1}{f_{e}^{0}} \sum_{j \in \Heavy} \sum_{\textsc{j} \in \QuantumSpace_{j}} \density_{j} \frac{a_{j \textsc{j}} e^{-\epsilon_{j \textsc{j}}}}{Q_{j}^{\text{int}}} \int \CrossSection_{ej}^{\textsc{j} \textsc{j}} ~ |\boldsymbol{C}_{e}| \left( \phi_{e}f_{e}^{0}(|\boldsymbol{C}_{e}| \boldsymbol{\omega}_{ej}') - \phi_{e}f_{e}^{0}(\boldsymbol{C}_{e}) \right) \, \mathrm{d} \boldsymbol{\omega}_{ej}', \\
\end{align*}
where expression \eqref{KTMaxwellHeavy} for the zeroth-order $j^{\text{th}}$-heavy-species distribution function $f_{j}^{0}$ has been integrated over $\boldsymbol{C}_{j}$, and where we have introduced the reduced internal energy
\begin{equation}
\epsilon_{j \textsc{j}} = \frac{E_{j \textsc{j}}}{k_{\textsc{b}} \Temperature_{\heavy}}, \qquad j \in \Heavy, \textsc{j} \in \QuantumSpace_{j}.
\end{equation}
Since $f_{e}^{0}$ is isotropic, $f_{e}^{0}(|\boldsymbol{C}_{e}| \boldsymbol{\omega}_{ej}') = f_{e}^{0}(\boldsymbol{C}_{e})$, and the electron linearized collision operator finally reads
\begin{align}
\mathcal{F}_{e}(\phi_{e}) = & - \int g_{e \tilde{e}} \CrossSection_{e \tilde{e}} \tilde{f}_{e}^{0} \left( \phi_{e}' + \tilde{\phi}_{e}' - \phi_{e} - \tilde{\phi}_{e} \right) \, \mathrm{d} \boldsymbol{\omega}_{e \tilde{e}}' \mathrm{d} \widetilde{\boldsymbol{C}}_{e}
\label{KTLinearizedElectronCollisionOperator} \\
{} & - \sum\limits_{j \in \Heavy \vphantom{\textsc{j} \in \QuantumSpace_{j}}} \sum\limits_{\vphantom{j \in \Heavy} \textsc{j} \in \QuantumSpace_{j}} \density_{j} \frac{a_{j \textsc{j}} e^{-\epsilon_{j \textsc{j}}}}{Q_{j}^{\text{int}}} \int \CrossSection_{ej}^{\textsc{j} \textsc{j}} ~ | \boldsymbol{C}_{e} | \left( \phi_{e}(|\boldsymbol{C}_{e}| \boldsymbol{\omega}_{ej}') - \phi_{e}(\boldsymbol{C}_{e}) \right) \, \mathrm{d} \boldsymbol{\omega}_{ej}'. \nonumber
\end{align}
The kernel of $\mathcal{F}_{e}$ coincides with the set of electron collisional invariants $\Invariants_{e}^{0}$. Indeed, if $\mathcal{F}_{e}(\phi_{e}) = 0$, then multiplying expression \eqref{KTLinearizedElectronCollisionOperator} by $\phi_{e} f_{e}^{0}$, integrating over $\boldsymbol{C}_{e}$, and using the reciprocity relation \eqref{KTScattering Reciprocity Relation} and symmetrization one obtains
\begin{eqnarray*}
& \phi_{e}(|\boldsymbol{C}_{e}| \boldsymbol{\omega}_{ej}') = \phi_{e}(\boldsymbol{C}_{e}) \\
& \phi_{e}(\boldsymbol{C}_{e}') + \phi_{e}(\widetilde{\boldsymbol{C}}_{e}') = \phi_{e}(\boldsymbol{C}_{e}) + \phi_{e}(\widetilde{\boldsymbol{C}}_{e}),
\end{eqnarray*}
for all $\boldsymbol{C}_{e}$, $\widetilde{\boldsymbol{C}}_{e}$, $\boldsymbol{\omega}_{e\tilde{e}}'$, $\boldsymbol{\omega}_{ej}'$, i.e., $\phi_{e} \in \Invariants_{e}^{0}$.

Furthermore, $\mathcal{F}_{e}$ is isotropic, i.e., it transforms a tensor constructed from $\boldsymbol{C}_{e}$ into another tensor of the same type \cite{Waldmann} \cite{GraillePhD} \cite{GiovangigliGraille2009}. This will be of great importance for the calculation of transport coefficients. We also introduce the associated integral bracket operator:
\begin{equation}
\llbracket \xi_{e}, \zeta_{e} \rrbracket_{e} = \llangle f_{e}^{0} \xi_{e}, \mathcal{F}_{e}(\zeta_{e}) \rrangle_{e}, \\
\end{equation}
which can be expressed in the form
\begin{align}
 \llbracket \xi_{e}, \zeta_{e} \rrbracket_{e} & = \frac{1}{4} \int g_{e \tilde{e}} \CrossSection_{e \tilde{e}} f_{e}^{0} \tilde{f}_{e}^{0} \big( \xi_{e}' + \tilde{\xi}_{e}' - \xi_{e} - \tilde{\xi}_{e} \big) \big( \zeta_{e}' + \tilde{\zeta}_{e}' - \zeta_{e} - \tilde{\zeta}_{e} \big) \, \mathrm{d} \boldsymbol{\omega}_{e \tilde{e}}' \mathrm{d} \boldsymbol{C}_{e} \mathrm{d} \widetilde{\boldsymbol{C}}_{e}
 \label{KTElectronBracket} \\
{} & + \frac{1}{2} \sum\limits_{j \in \Heavy \vphantom{\textsc{j} \in \QuantumSpace_{j}}} \sum\limits_{\vphantom{j \in \Heavy} \textsc{j} \in \QuantumSpace_{j}} \density_{j} \frac{a_{j \textsc{j}} e^{-\epsilon_{j \textsc{j}}}}{Q_{j}^{\text{int}}}
\nonumber \\
{} & \quad \times \int \CrossSection_{ej}^{\textsc{j} \textsc{j}} ~ | \boldsymbol{C}_{e} | f_{e}^{0}(\boldsymbol{C}_{e}) \left( \xi_{e}(|\boldsymbol{C}_{e}| \boldsymbol{\omega}_{ej}') - \xi_{e}(\boldsymbol{C}_{e}) \right) \left( \zeta_{e}(|\boldsymbol{C}_{e}| \boldsymbol{\omega}_{ej}') - \zeta_{e}(\boldsymbol{C}_{e}) \right) \, \mathrm{d} \boldsymbol{\omega}_{ej}' \mathrm{d} \boldsymbol{C}_{e}. \nonumber
\end{align}
From expression \eqref{KTElectronBracket}, the bracket operator $\llbracket \cdot \rrbracket_{e}$ is readily seen to be symmetric, i.e.,
\begin{equation}
\llbracket \xi_{e}, \zeta_{e} \rrbracket_{e} = \llbracket \zeta_{e}, \xi_{e} \rrbracket_{e},
\end{equation}
positive semi-definite:
\begin{equation}
\llbracket \xi_{e}, \xi_{e} \rrbracket_{e} \geq 0,
\end{equation}
and its kernel is seen to coincide with the kernel of $\mathcal{F}_{e}$:
\begin{equation}
\llbracket \xi_{e}, \xi_{e} \rrbracket_{e} = 0 \Leftrightarrow \mathcal{F}_{e}(\xi_{e}) = 0 \Leftrightarrow \xi_{e} \in \Invariants_{e}^{0}.
\end{equation}

The electron Boltzmann equation \eqref{KTCompact Scaled Boltzmann electron equation} is now projected at order $\varepsilon^{-1}$. As long as $f_{e}^{0}$ is an isotropic function of $\boldsymbol{C}_{e}$, respectively $f_{j}^{0}$ is an isotropic function of $\boldsymbol{C}_{j}$, from \eqref{KTSei0} the term $\Scattering_{ej}^{0}(f_{e}^{0},\phi_{j}f_{j}^{0})$ is shown to vanish:
\begin{equation}
\Scattering_{ej}^{0}(f_{e}^{0},\phi_{j}f_{j}^{0}) = 0,
\end{equation}
respectively from \eqref{KTSei1}
\begin{equation}
\Scattering_{ej}^{1}(f_{e}^{0},f_{j}^{0}) = 0.
\end{equation}
Thus, the first-order electron perturbation function $\phi_{e}$ is solution to the following linear equation:
\begin{equation}
\mathcal{F}_{e}(\phi_{e}) + \delta_{b1} \frac{q_{e}}{\Mass_{e}} (\boldsymbol{C}_{e} \wedge \boldsymbol{B}) \cdot \boldsymbol{\partial}_{\boldsymbol{C}_{e}} \phi_{e} = \Psi_{e},
\label{KTFirstOrderElectronPerturbationEquation}
\end{equation}
where
\begin{equation}
\Psi_{e} = - \Streaming_{e}^{-1}(\ln{f_{e}^{0}}).
\end{equation}
Equation \eqref{KTFirstOrderElectronPerturbationEquation} must be completed with constraints \eqref{KTZerothMacroscopicPropertiesElectron} in order to be well posed
\begin{equation}
\llangle \phi_{e} f_{e}^{0}, \psi_{e}^{l} \rrangle_{e} = 0, \quad l \in \left\lbrace e, n^{s}+4 \right\rbrace.
\end{equation}
Indeed, the streaming operator $\Psi_{e} = - \Streaming_{e}^{-1}(\ln{f_{e}^{0}})$ on the right hand side of \eqref{KTFirstOrderElectronPerturbationEquation} is orthogonal to the electron collisional invariants \cite{GrailleMaginMassot2009}, and the second term on the left hand side of \eqref{KTFirstOrderElectronPerturbationEquation} is orthogonal to the electron isotropic collisional invariants \cite{GiovangigliGraille2003} \cite{GraillePhD} \cite{GrailleMaginMassot2009}, since if $f_{e}^{0}$ decreases sufficiently rapidly as $|\boldsymbol{C}_{e}| \to + \infty$, then
\begin{align*}
\int f_{e}^{0} \, \psi_{e}^{l} \, (\boldsymbol{C}_{e} \wedge \boldsymbol{B}) \cdot \boldsymbol{\partial}_{\boldsymbol{C}_{e}} \phi_{e} \, \mathrm{d} \boldsymbol{C}_{e} & = - \int f_{e}^{0} \phi_{e} ~ \boldsymbol{\partial}_{\boldsymbol{C}_{e}} \cdot (\psi_{e}^{l} \boldsymbol{C}_{e} \wedge \boldsymbol{B}) \, \mathrm{d} \boldsymbol{C}_{e}, \\
& = - \int f_{e}^{0} \phi_{e} \, (\boldsymbol{C}_{e} \wedge \boldsymbol{B}) \cdot \boldsymbol{\partial}_{\boldsymbol{C}_{e}} \psi_{e}^{l} \, \mathrm{d} \boldsymbol{C}_{e}, \\
& = 0 \vphantom{\int},
\end{align*}
for $l \in \left\lbrace e, n^{s}+4 \right\rbrace$. Besides, the relation
\begin{align*}
\llangle f_{e}^{0} \phi_{e}, (\boldsymbol{C}_{e} \wedge \boldsymbol{B}) \cdot \boldsymbol{\partial}_{\boldsymbol{C}_{e}} \phi_{e} \rrangle_{e} & = \int f_{e}^{0} \phi_{e} \, (\boldsymbol{C}_{e} \wedge \boldsymbol{B}) \cdot \boldsymbol{\partial}_{\boldsymbol{C}_{e}} \phi_{e} \, \mathrm{d} \boldsymbol{C}_{e}, \\
& = \int f_{e}^{0} \, (\boldsymbol{C}_{e} \wedge \boldsymbol{B}) \cdot \boldsymbol{\partial}_{\boldsymbol{C}_{e}} \Big( \frac{1}{2} \phi_{e}^{2} \Big) \, \mathrm{d} \boldsymbol{C}_{e}, \\
& = 0, \vphantom{\int}
\end{align*}
ensures that the set of solutions of the homogeneous linear equation associated with the linear equation \eqref{KTFirstOrderElectronPerturbationEquation} coincides with the kernel of $\mathcal{F}_{e}$, $\Invariants_{e}^{0}$.

\subsection{Inelastic collision cross-sections}
\label{SubSecKTScalingInelastic}

We can now discuss in more details the choice of scaling adopted for $\CrossSection_{j e}^{\textsc{j} \textsc{j}'}$, $j \in \Heavy$, $\textsc{j}, \textsc{j}' \in \QuantumSpace_{j}$, $\textsc{j} \neq \textsc{j}'$. We consider the two possible alternative scaling $\CrossSection_{\heavy e}^{\text{in},0} = \CrossSection^{0}$ and $\CrossSection_{\heavy e}^{\text{in},0} = \varepsilon \CrossSection^{0}$, instead of $\CrossSection_{\heavy e}^{\text{in},0} = \varepsilon^{2} \CrossSection^{0}$. Other things being equal, in the former case the electron thermalization requires that the electron temperature be equal to the heavy-species temperature $\Temperature_{e} = \Temperature_{\heavy}$, while in the latter case the heavy-species thermalization induces $\Temperature_{\heavy} = \Temperature_{e}$.

\subsubsection{Electron thermalization}
First, if the inelastic reference collision scaled as
\begin{equation}
\CrossSection_{\heavy e}^{\text{in},0} = \CrossSection^{0},
\end{equation}
that is to say if there was only one relevant characteristic cross-section, then the right hand side of the electron Boltzmann equation at order $\varepsilon^{-2}$ \eqref{KTElectronBoltzmannEpsilon-2} would contain the additional term $\sum_{j \in \Heavy} \Scattering_{ej}^{2,\text{in}} \left(f_{e}^{0},f_{j}^{0} \right)$. Thus, equation \eqref{KTElectronZerothOrderEntropy} for electron thermalization would be replaced by
\begin{equation}
\Gamma_{ee}^{0} + \sum_{j \in \Heavy} {\Gamma_{ej}^{0}} + \sum_{j \in \Heavy} {\Gamma_{ej}^{2,\text{in}}} = 0,
\label{KTElectronZerothOrderEntropyInelastic}
\end{equation}
where the entropy production rate due to inelastic scattering of electron by the $j^{\text{th}}$ heavy species reads
\begin{equation}
{\Gamma_{ej}^{2,\text{in}}} = - k_{\textsc{b}} \int \Scattering_{ej}^{2,\text{in}} \left(f_{e}^{0}(\boldsymbol{C}_{e}),f_{j}^{0}(\boldsymbol{C}_{j}) \right) \ln{\left( f_{e}^{0}(\boldsymbol{C}_{e}) \right)} \, \mathrm{d} \boldsymbol{C}_{e},
\end{equation}
that is
\begin{align}
{\Gamma_{ej}^{2,\text{in}}} & = - k_{\textsc{b}} \sum\limits_{\substack{\textsc{j}, \textsc{j}' \in \QuantumSpace_{j} \\ \textsc{j}' \neq \textsc{j}}} \int \CrossSection_{ej}^{\textsc{j} \textsc{j}'} |\boldsymbol{C}_{e}| \Big( f_{e}^{0}(\boldsymbol{C}_{e}'^{(0)}) f_{j}^{0}(\boldsymbol{C}_{j},\textsc{j}') \frac{\beta_{j \textsc{j}'}}{\beta_{j \textsc{j}}} - f_{e}^{0}(\boldsymbol{C}_{e})f_{j}^{0}(\boldsymbol{C}_{j},\textsc{j}) \Big) \\
 & \hphantom{= - k_{\textsc{b}} \sum\limits_{\substack{\textsc{j}, \textsc{j}' \in \QuantumSpace_{j} \\ \textsc{j}' \neq \textsc{j}}} \int \CrossSection_{ej}^{\textsc{j} \textsc{j}'} |\boldsymbol{C}_{e}|} \times \ln{\left( f_{e}^{0}(\boldsymbol{C}_{e}) \right)} \, \mathrm{d} \boldsymbol{\omega}_{ej}' \mathrm{d} \boldsymbol{C}_{e} \mathrm{d} \boldsymbol{C}_{j},
 \nonumber
\end{align}
where $\boldsymbol{C}_{e}'^{(0)} = |\boldsymbol{C}_{e}| \big( 1 - \Delta E_{\textsc{j} \textsc{j}'} / \frac{1}{2}\Mass_{e}|\boldsymbol{C}_{e}|^{2} \big)^{1/2} \boldsymbol{\omega}_{ej}'$ is the zeroth-order electron velocity after collision. Using the reciprocity relation \eqref{KTScattering Reciprocity Relation} in the form
\begin{equation}
\beta_{j \textsc{j}'} |\boldsymbol{C}_{e}| \CrossSection_{ej}^{\textsc{j} \textsc{j}'} \, \mathrm{d} \boldsymbol{\omega}_{ej}' \mathrm{d} \boldsymbol{C}_{e} \mathrm{d} \boldsymbol{C}_{j} = \beta_{j \textsc{j}} |\boldsymbol{C}_{e}'^{(0)}| \CrossSection_{ej}^{\textsc{j}' \textsc{j}} \, \mathrm{d} \boldsymbol{\omega}_{ej} \mathrm{d} \boldsymbol{C}_{e}'^{(0)} \mathrm{d} \boldsymbol{C}_{j},
\label{KTReciprocityInelastic}
\end{equation}
and symmetrizing, ${\Gamma_{ej}^{2,\text{in}}}$ may be expressed as
\begin{equation}
{\Gamma_{ej}^{2,\text{in}}} = k_{\textsc{b}} \sum\limits_{\substack{\textsc{j}, \textsc{j}' \in \QuantumSpace_{j} \\ \textsc{j}' \neq \textsc{j}}} \int \CrossSection_{ej}^{\textsc{j} \textsc{j}'} |\boldsymbol{C}_{e}| \big[ \ln{\left( f_{e}^{0}(\boldsymbol{C}_{e}) \right)} - \ln{\big( f_{e}^{0}(\boldsymbol{C}_{e}'^{(0)}) \big)} \big] f_{e}^{0}(\boldsymbol{C}_{e})f_{j}^{0}(\boldsymbol{C}_{j},\textsc{j}) \, \mathrm{d} \boldsymbol{\omega}_{ej}' \mathrm{d} \boldsymbol{C}_{e} \mathrm{d} \boldsymbol{C}_{j}.
\end{equation}
%and then as
%\begin{align}
% {\Gamma_{ej}^{2,\text{in}}} & = \frac{k_{\textsc{b}}}{2} \sum\limits_{\substack{\textsc{j}, \textsc{j}' \in \QuantumSpace_{j} \\ \textsc{j}' \neq \textsc{j}}} a_{j \textsc{j}} \int \CrossSection_{ej}^{\textsc{j} \textsc{j}'} |\boldsymbol{C}_{e}| \big[ \ln{\big( f_{e}^{0}(\boldsymbol{C}_{e}'^{(0)}) \big)} - \ln{\left( f_{e}^{0}(\boldsymbol{C}_{e}) \right)} \big] \\
% & \hphantom{= \frac{k_{\textsc{b}}}{2} \sum\limits_{\substack{\textsc{j}, \textsc{j}' \in \QuantumSpace_{j} \\ \textsc{j}' \neq \textsc{j}}}} \times \Big( f_{e}^{0}(\boldsymbol{C}_{e}'^{(0)}) \frac{f_{j}^{0}(\boldsymbol{C}_{j},\textsc{j}')}{a_{j \textsc{j}'}} - f_{e}^{0}(\boldsymbol{C}_{e}) \frac{f_{j}^{0}(\boldsymbol{C}_{j},\textsc{j})}{a_{j \textsc{j}}} \Big) \, \mathrm{d} \boldsymbol{\omega}_{ej}' \mathrm{d} \boldsymbol{C}_{e} \mathrm{d} \boldsymbol{C}_{j},
% \nonumber
%\end{align}
Now, for the alternative scaling $\CrossSection_{\heavy e}^{\text{in},0} = \CrossSection^{0}$, the projection of the heavy species Boltzmann equation \eqref{KTScaled Boltzmann heavy-species equations} at order $\varepsilon^{-2}$ would be non trivial, and would yield
\begin{align}
0 = \Scattering_{je}^{2,\text{in}}(f_{j}^{0},f_{e}^{0}) & = \sum\limits_{\substack{\textsc{j}' \in \QuantumSpace_{j} \\ \textsc{j}' \neq \textsc{j}}} \int \CrossSection_{je}^{\textsc{j} \textsc{j}'} |\boldsymbol{C}_{e}| \Big( f_{e}^{0}(\boldsymbol{C}_{e}'^{(0)}) f_{j}^{0}(\boldsymbol{C}_{j},\textsc{j}') \frac{\beta_{j \textsc{j}'}}{\beta_{j \textsc{j}}} - f_{e}^{0}(\boldsymbol{C}_{e})f_{j}^{0}(\boldsymbol{C}_{j},\textsc{j}) \Big) \\
 & \hphantom{= \sum\limits_{\substack{\textsc{j}' \in \QuantumSpace_{j} \\ \textsc{j}' \neq \textsc{j}}} \int \CrossSection_{je}^{\textsc{j} \textsc{j}'} |\boldsymbol{C}_{e}|} \times \, \mathrm{d} \boldsymbol{\omega}_{je}' \mathrm{d} \boldsymbol{C}_{e} \mathrm{d} \boldsymbol{C}_{j}.
 \nonumber
\end{align}
Multiplying the latter equation by $\ln{( \beta_{j \textsc{j}} f_{j}^{0}(\boldsymbol{C}_{j}, \textsc{j}) )}$, integrating over $\boldsymbol{C}_{j}$ summing over $\textsc{j} \in \QuantumSpace_{j}$, and noting that $\boldsymbol{\omega}_{je} = - \boldsymbol{\omega}_{ej}$, $\boldsymbol{\omega}_{je}' = - \boldsymbol{\omega}_{ej}'$, and that $\CrossSection_{je}^{\textsc{j} \textsc{j}'} = \CrossSection_{ej}^{\textsc{j} \textsc{j}'}$, one would obtain
\begin{align}
0 & = \sum\limits_{\substack{\textsc{j}, \textsc{j}' \in \QuantumSpace_{j} \\ \textsc{j}' \neq \textsc{j}}} \frac{1}{\beta_{j \textsc{j}}} \int \CrossSection_{ej}^{\textsc{j} \textsc{j}'} |\boldsymbol{C}_{e}| \ln{\big( \beta_{j \textsc{j}} f_{j}^{0}(\boldsymbol{C}_{j}, \textsc{j}) \big)} \\
 & \hphantom{= \frac{k_{\textsc{b}}}{2} \sum\limits_{\substack{\textsc{j}, \textsc{j}' \in \QuantumSpace_{j} \\ \textsc{j}' \neq \textsc{j}}}} \times \big( \beta_{j \textsc{j}'} f_{e}^{0}(\boldsymbol{C}_{e}'^{(0)}) f_{j}^{0}(\boldsymbol{C}_{j},\textsc{j}') - \beta_{j \textsc{j}} f_{e}^{0}(\boldsymbol{C}_{e}) f_{j}^{0}(\boldsymbol{C}_{j},\textsc{j}) \big) \, \mathrm{d} \boldsymbol{\omega}_{ej}' \mathrm{d} \boldsymbol{C}_{e} \mathrm{d} \boldsymbol{C}_{j},
 \nonumber
\end{align}
which by reciprocity and symmetry would yield
\begin{align}
0 & = \sum\limits_{\substack{\textsc{j}, \textsc{j}' \in \QuantumSpace_{j} \\ \textsc{j}' \neq \textsc{j}}} \frac{1}{\beta_{j \textsc{j}}} \int \CrossSection_{ej}^{\textsc{j} \textsc{j}'} |\boldsymbol{C}_{e}| \big[ \ln{\big( \beta_{j \textsc{j}'} f_{j}^{0}(\boldsymbol{C}_{j}, \textsc{j}') \big)} - \ln{\big( \beta_{j \textsc{j}} f_{j}^{0}(\boldsymbol{C}_{j}, \textsc{j}) \big)} \big]  \\
 & \hphantom{= \frac{k_{\textsc{b}}}{2} \sum\limits_{\substack{\textsc{j}, \textsc{j}' \in \QuantumSpace_{j} \\ \textsc{j}' \neq \textsc{j}}}} \times \big( \beta_{j \textsc{j}'} f_{e}^{0}(\boldsymbol{C}_{e}'^{(0)}) f_{j}^{0}(\boldsymbol{C}_{j},\textsc{j}') - \beta_{j \textsc{j}} f_{e}^{0}(\boldsymbol{C}_{e}) f_{j}^{0}(\boldsymbol{C}_{j},\textsc{j}) \big) \, \mathrm{d} \boldsymbol{\omega}_{ej}' \mathrm{d} \boldsymbol{C}_{e} \mathrm{d} \boldsymbol{C}_{j},
 \nonumber
\end{align}
so that the entropy production rate due to inelastic scattering of electron by the $j^{\text{th}}$ heavy species would read
\begin{align}
 {\Gamma_{ej}^{2,\text{in}}} & = \frac{k_{\textsc{b}}}{2} \sum\limits_{\substack{\textsc{j}, \textsc{j}' \in \QuantumSpace_{j} \\ \textsc{j}' \neq \textsc{j}}} \frac{1}{\beta_{j \textsc{j}}} \int \CrossSection_{ej}^{\textsc{j} \textsc{j}'} |\boldsymbol{C}_{e}| \Omega \big( \beta_{j \textsc{j}'} f_{e}^{0}(\boldsymbol{C}_{e}'^{(0)}) f_{j}^{0}(\boldsymbol{C}_{j},\textsc{j}'), \beta_{j \textsc{j}} f_{e}^{0}(\boldsymbol{C}_{e}) f_{j}^{0}(\boldsymbol{C}_{j},\textsc{j}) \big) \\
 & \hphantom{= \frac{k_{\textsc{b}}}{2} \sum\limits_{\substack{\textsc{j}, \textsc{j}' \in \QuantumSpace_{j} \\ \textsc{j}' \neq \textsc{j}}} \frac{1}{\beta_{j \textsc{j}}} \int \CrossSection_{ej}^{\textsc{j} \textsc{j}'} |\boldsymbol{C}_{e}|} \, \mathrm{d} \boldsymbol{\omega}_{ej}' \mathrm{d} \boldsymbol{C}_{e} \mathrm{d} \boldsymbol{C}_{j},
 \nonumber
\end{align}
where $\Omega$ is the function defined in \eqref{KTOmega}. Since $\Omega$ is nonnegative, the $\varepsilon^{-2}$ electron Boltzmann equation \eqref{KTElectronZerothOrderEntropyInelastic} would yield the thermalization of electrons \eqref{KTMaxwellElectron} as in subsection \ref{SubSecKTElectronThermalization}. Besides, the electron $j^{\text{th}}$-heavy-species inelastic entropy production rate would vanish
\begin{equation}
\Gamma_{ej}^{2,\text{in}} = 0, \quad j \in \Heavy,
\end{equation}
which would require
\begin{equation}
a_{j \textsc{j}} f_{e}^{0}(\boldsymbol{C}_{e}'^{(0)}) f_{j}^{0}(\boldsymbol{C}_{j},\textsc{j}') = a_{j \textsc{j}'} f_{e}^{0}(\boldsymbol{C}_{e}) f_{j}^{0}(\boldsymbol{C}_{j},\textsc{j}), \quad j \in \Heavy, \textsc{j},\textsc{j}' \in \QuantumSpace_{j}, \textsc{j}' \neq \textsc{j},
\end{equation}
for $\Omega$ is nonnegative. From expression \eqref{KTMaxwellElectron} for the electron Maxwellian distribution function, the latter equation is rewritten
\begin{equation}
a_{j \textsc{j}} f_{j}^{0}(\boldsymbol{C}_{j},\textsc{j}') = \exp{\Big( - \frac{\Delta E_{\textsc{j} \textsc{j}'}}{k_{\textsc{b}} \Temperature_{e}} \Big)} a_{j \textsc{j}'} f_{j}^{0}(\boldsymbol{C}_{j},\textsc{j}), \quad j \in \Heavy, \textsc{j},\textsc{j}' \in \QuantumSpace_{j}, \textsc{j}' \neq \textsc{j}.
\label{KTEquilibriumForInelastic}
\end{equation}
Assuming that the zeroth-order heavy-species distribution functions $f_{j}^{0}$, $j \in \Heavy$, would still be Maxwellian of the form \eqref{KTMaxwellHeavy}, equation \eqref{KTEquilibriumForInelastic} would be equivalent to
\begin{align*}
 \frac{\Delta E_{\textsc{j} \textsc{j}'}}{k_{\textsc{b}} \Temperature_{\heavy}} = \frac{\Delta E_{\textsc{j} \textsc{j}'}}{k_{\textsc{b}} \Temperature_{e}}, \quad j \in \Heavy, \textsc{j}, \textsc{j}' \in \QuantumSpace_{j}, \textsc{j}' \neq \textsc{j},
\end{align*}
which would imply $\Temperature_{e} = \Temperature_{\heavy}$.

\subsubsection{Heavy-species thermalization}
In the preceding paragraph it was shown that, other things being equal, the inelastic collisions between electrons and heavy species must be at least one order of magnitude slower than the corresponding elastic collisions. We now consider the alternative scaling
\begin{equation}
\CrossSection_{\heavy e}^{\text{in},0} = \varepsilon \CrossSection^{0}.
\end{equation}
For such a scaling electron thermalization would proceed as in \ref{SubSecKTElectronThermalization}, since equation \eqref{KTElectronZerothOrderEntropy} would remain unchanged, and hence the zeroth-order electron distribution function $f_{e}^{0}$ would be a Maxwellian of the form \eqref{KTMaxwellElectron}.

The heavy-species Boltzmann equation at order $\varepsilon^{-1}$ \eqref{KTHeavyBoltzmannEpsilon-1} would contain the additional term $\Scattering_{ie}^{2,\text{in}} \left(f_{i}^{0},f_{e}^{0} \right)$, so that equation \eqref{KTHeavyZerothOrderEntropy} for heavy-species thermalization would be replaced by
\begin{equation}
\Gamma_{\heavy \heavy}^{0} + \Gamma_{\heavy e}^{2,\text{in}} = 0,
\label{KTHeavyFirstOrderEntropyInelasticHeavy}
\end{equation}
where we have introduced the entropy production rate due to inelastic scattering of heavy species by electrons
\begin{equation}
\Gamma_{\heavy e}^{2,\text{in}} = - k_{\textsc{b}} \sum_{i \in \Heavy} \sum_{\textsc{i} \in \QuantumSpace_{i}} \int \ln{\big( \beta_{i \textsc{i}} f_{i}^{0}(\boldsymbol{C}_{i}, \textsc{i}) \big)} \Scattering_{ie}^{2,\text{in}} \left(f_{i}^{0},f_{e}^{0} \right) \, \mathrm{d} \boldsymbol{C}_{i}.
\end{equation}
From expression \eqref{KTSie2in} for $\Scattering_{ie}^{2,\text{in}}$, this inelastic entropy production rate is expanded as
\begin{align}
\Gamma_{\heavy e}^{2,\text{in}} & = - k_{\textsc{b}} \sum_{i \in \Heavy} \sum_{\textsc{i} \in \QuantumSpace_{i}} \int \CrossSection_{ie}^{\textsc{i} \textsc{i}'} |\boldsymbol{C}_{e}| \ln{\big( \beta_{i \textsc{i}} f_{i}^{0}(\boldsymbol{C}_{i}, \textsc{i}) \big)}
\nonumber \\
 & \hphantom{= - k_{\textsc{b}} \sum_{i \in \Heavy} \sum_{\textsc{i} \in \QuantumSpace_{i}} \int} \times \Big( f_{i}^{0} (\boldsymbol{C}_{i}, \textsc{i}') f_{e}^{0} (\boldsymbol{C}_{e}'^{(0)}) \frac{\beta_{i \textsc{i}'}}{\beta_{i \textsc{i}}} - f_{i}^{0}(\boldsymbol{C}_{i}, \textsc{i}) f_{e}^{0}(\boldsymbol{C}_{e}) \Big) \, \mathrm{d} \boldsymbol{\omega}_{ie}' \mathrm{d} \boldsymbol{C}_{i} \mathrm{d} \boldsymbol{C}_{e},
 \nonumber \\
 & = \frac{k_{\textsc{b}}}{2} \sum_{i \in \Heavy} \sum_{\textsc{i} \in \QuantumSpace_{i}} \int \CrossSection_{ie}^{\textsc{i} \textsc{i}'} |\boldsymbol{C}_{e}| \big[ \ln{\big( \beta_{i \textsc{i}'} f_{i}^{0}(\boldsymbol{C}_{i}, \textsc{i}') \big)} - \ln{\big( \beta_{i \textsc{i}} f_{i}^{0}(\boldsymbol{C}_{i}, \textsc{i}) \big)} \big]
\label{KTGammahe2in} \\
 & \hphantom{= - k_{\textsc{b}} \sum_{i \in \Heavy} \sum_{\textsc{i} \in \QuantumSpace_{i}} \int} \times \Big( f_{i}^{0} (\boldsymbol{C}_{i}, \textsc{i}') f_{e}^{0} (\boldsymbol{C}_{e}'^{(0)}) \frac{\beta_{i \textsc{i}'}}{\beta_{i \textsc{i}}} - f_{i}^{0}(\boldsymbol{C}_{i}, \textsc{i}) f_{e}^{0}(\boldsymbol{C}_{e}) \Big) \, \mathrm{d} \boldsymbol{\omega}_{ie}' \mathrm{d} \boldsymbol{C}_{i} \mathrm{d} \boldsymbol{C}_{e},
 \nonumber
\end{align}
where we have used the reciprocity relation \eqref{KTReciprocityInelastic} and symmetrization.

Now, for the scaling $\CrossSection_{\heavy e}^{\text{in},0} = \varepsilon \CrossSection^{0}$ proposed, equation \eqref{KTFirstOrderElectronPerturbationEquation} obtained from the projection of the Boltzmann equation at order $\varepsilon^{-1}$ would remain valid, but the source term $\Psi_{e}$ would read
\begin{equation}
f_{e}^{0} \Psi_{e} = - \Streaming_{e}^{-1}(f_{e}^{0}) + \sum_{j \in \Heavy} \Scattering_{ej}^{1}(f_{e}^{0},f_{j}^{0}) + \sum_{j \in \Heavy} \Scattering_{ej}^{2,\text{in}}(f_{e}^{0},f_{j}^{0}),
\label{KTFirstOrderElectronPerturbationEquationInelastic}
\end{equation}
where $f_{j}^{0}$ would remain unknown at this point. Since the streaming operator, the electron linearized collision operator $\mathcal{F}_{e}$, and the term $(\boldsymbol{C}_{e} \wedge \boldsymbol{B}) \cdot \boldsymbol{\partial}_{\boldsymbol{C}_{e}} \phi_{e}$ are orthogonal to the electron isotropic collisional invariants $\Invariants_{e}^{0}$
\begin{align}
\llangle \Streaming_{e}^{-1}(f_{e}), \psi_{e}^{l} \rrangle_{e} = 0, \quad l \in \left\lbrace e, n^{s}+4 \right\rbrace, \\
\llangle f_{e}^{0} \mathcal{F}_{e}(\phi_{e}), \psi_{e}^{l} \rrangle_{e} = 0, \quad l \in \left\lbrace e, n^{s}+4 \right\rbrace, \\
\llangle f_{e}^{0} (\boldsymbol{C}_{e} \wedge \boldsymbol{B}) \cdot \boldsymbol{\partial}_{\boldsymbol{C}_{e}} \phi_{e}, \psi_{e}^{l} \rrangle_{e} = 0, \quad l \in \left\lbrace e, n^{s}+4 \right\rbrace,
\end{align}
and since $\ln{f_{e}^{0}}$ belongs to $\Invariants_{e}^{0}$, multiplying \eqref{KTFirstOrderElectronPerturbationEquationInelastic} by $\ln{f_{e}^{0}}$ and integrating over $\boldsymbol{C}_{e}$ one would obtain
\begin{equation}
\sum_{j \in \Heavy} \llangle \Scattering_{ej}^{1}(f_{e}^{0},f_{j}^{0}), \ln{f_{e}^{0}} \rrangle_{e} + \sum_{j \in \Heavy} \llangle \Scattering_{ej}^{2,\text{in}}(f_{e}^{0},f_{j}^{0}), \ln{f_{e}^{0}} \rrangle_{e} = 0.
\end{equation}
The first term on the left hand side would be shown to vanish since from expression \eqref{KTSei1} $\Scattering_{ej}^{1}(f_{e}^{0},f_{j}^{0}) = 0$ as long as $f_{e}^{0}$ is given by a Maxwellian of the form \eqref{KTMaxwellElectron}. Thus, the second term on the left hand side would be zero:
\begin{align}
0 & = \sum_{j \in \Heavy} \llangle \Scattering_{ej}^{2,\text{in}}(f_{e}^{0},f_{j}^{0}), \ln{f_{e}^{0}} \rrangle_{e}
\nonumber \\
 & =  \sum_{j \in \Heavy} \sum_{\textsc{j} \in \QuantumSpace_{j}} \int \CrossSection_{je}^{\textsc{j} \textsc{j}'} |\boldsymbol{C}_{e}| \ln{\big( f_{e}^{0}(\boldsymbol{C}_{e}) \big)}
\nonumber \\
 & \hphantom{= - k_{\textsc{b}} \sum_{j \in \Heavy} \sum_{\textsc{j} \in \QuantumSpace_{j}} \int} \times \Big( f_{j}^{0} (\boldsymbol{C}_{j}, \textsc{j}') f_{e}^{0} (\boldsymbol{C}_{e}'^{(0)}) \frac{\beta_{j \textsc{j}'}}{\beta_{j \textsc{j}}} - f_{j}^{0}(\boldsymbol{C}_{j}, \textsc{j}) f_{e}^{0}(\boldsymbol{C}_{e}) \Big) \, \mathrm{d} \boldsymbol{\omega}_{ej}' \mathrm{d} \boldsymbol{C}_{j} \mathrm{d} \boldsymbol{C}_{e},
  \nonumber \\
 & = - \frac{1}{2}  \sum_{j \in \Heavy} \sum_{\textsc{j} \in \QuantumSpace_{j}} \int \CrossSection_{je}^{\textsc{j} \textsc{j}'} |\boldsymbol{C}_{e}| \big[ \ln{\big( f_{e}^{0}(\boldsymbol{C}_{e}'^{(0)}) \big)} - \ln{\big( f_{e}^{0}(\boldsymbol{C}_{e}) \big)} \big]
\nonumber \\
 & \hphantom{= - k_{\textsc{b}} \sum_{j \in \Heavy} \sum_{\textsc{j} \in \QuantumSpace_{j}} \int} \times \Big( f_{j}^{0} (\boldsymbol{C}_{j}, \textsc{j}') f_{e}^{0} (\boldsymbol{C}_{e}'^{(0)}) \frac{\beta_{j \textsc{j}'}}{\beta_{j \textsc{j}}} - f_{j}^{0}(\boldsymbol{C}_{j}, \textsc{j}) f_{e}^{0}(\boldsymbol{C}_{e}) \Big) \, \mathrm{d} \boldsymbol{\omega}_{je}' \mathrm{d} \boldsymbol{C}_{j} \mathrm{d} \boldsymbol{C}_{e}.
  \nonumber
\end{align}
Combining the latter equation with equation \eqref{KTGammahe2in}, the heavy-species electron inelastic entropy production rate would finally read
\begin{align}
\Gamma_{\heavy e}^{2,\text{in}} & = \frac{k_{\textsc{b}}}{2} \sum_{i \in \Heavy} \sum_{\textsc{i} \in \QuantumSpace_{i}} \int \CrossSection_{ie}^{\textsc{i} \textsc{i}'} \, |\boldsymbol{C}_{e}| \, \Omega \Big( \frac{\beta_{i \textsc{i}'}}{\beta_{i \textsc{i}}} f_{i}^{0}(\boldsymbol{C}_{i}, \textsc{i}') f_{e}^{0} (\boldsymbol{C}_{e}'^{(0)}), f_{i}^{0}(\boldsymbol{C}_{i}, \textsc{i}) f_{e}^{0}(\boldsymbol{C}_{e}) \Big) \, \mathrm{d} \boldsymbol{\omega}_{ie}' \mathrm{d} \boldsymbol{C}_{i} \mathrm{d} \boldsymbol{C}_{e}.
 \label{KTGammahe2inSymm}
\end{align}
Since $\Omega$ is nonnegative, equation \eqref{KTHeavyFirstOrderEntropyInelasticHeavy} would induce the thermalization of the heavy species as in subsection \ref{SubSecKTHeavyThermalization}. Additionnally, one would retrieve \eqref{KTEquilibriumForInelastic}, and thus $\Temperature_{\heavy}$ would equal $\Temperature_{e}$ as in the preceding paragraph.

A possible extension of the present theory where some of the heavy species internal modes thermalize at $\Temperature_{e}$ is discussed in the conclusion.

\subsection{Zeroth-order macroscopic equations for electrons}
Equations \eqref{KTElectronMatterConservation}, \eqref{KTElectronMomentumConservation}, \eqref{KTElectronEnergyConservation} are now expanded at order $\varepsilon^{0}$. Conservation of matter and energy yield a system of Navier-Stokes type drift-diffusion equations for electrons, from which the equation for conservation of momentum uncouples.

Equation \eqref{KTElectronMatterConservation} for conservation of matter yields at order $\varepsilon^{0}$ the following mass conservation equation for electrons:
\begin{equation}
\partial_{t} \rho_{e} + \boldsymbol{\partial_{x}} \cdot \left( \rho_{e} \boldsymbol{\Meanv}_{\heavy} + \rho_{e} \boldsymbol{\MeanV}_{e}^{0} \right) = 0,
\end{equation}
where we have introduced the electron zeroth-order diffusion velocity
\begin{equation}
\density_{e} \boldsymbol{\MeanV}_{e}^{0} = \int \boldsymbol{C}_{e} \phi_{e} f_{e}^{0} \, \mathrm{d} \boldsymbol{C}_{e}.
\end{equation}
On the other hand, the energy conservation equation \eqref{KTElectronEnergyConservation} yields at order $\varepsilon^{0}$
\begin{equation}
\partial_{t} \Energy_{e} + \boldsymbol{\partial_{x}} \cdot (\Energy_{e} \boldsymbol{\Meanv}_{\heavy}) = - \pressure_{e} \boldsymbol{\partial_{x}} \cdot \boldsymbol{\Meanv}_{\heavy} - \boldsymbol{\partial_{x}} \cdot \boldsymbol{\HeatFlux}_{e}^{0} + \boldsymbol{\Current}_{e}^{0} \cdot \boldsymbol{E}' + \Delta E_{e \heavy}^{0},	
\end{equation}
where $\pressure_{e} = \density_{e} k_{\textsc{b}} \Temperature_{e}$ is the electron partial pressure, $\boldsymbol{\Current}_{e}^{0} = \density_{e} q_{e} \boldsymbol{\MeanV}_{e}^{0}$ is the zeroth-order electron conduction current density in the heavy-species reference frame, and $\boldsymbol{E}'$ is the electric field expressed in the heavy-species reference frame
\begin{equation}
\boldsymbol{E}' = \boldsymbol{E} + \delta_{b1} \boldsymbol{\Meanv}_{\heavy} \wedge \boldsymbol{B}.
\end{equation}
We have denoted by $\boldsymbol{\HeatFlux}_{e}^{0}$ the zeroth-order electron heat flux
\begin{equation}
\boldsymbol{\HeatFlux}_{e}^{0} = \int \phi_{e} f_{e}^{0} ~ \big(\frac{1}{2}\Mass_{e} \boldsymbol{C}_{e} \cdot \boldsymbol{C}_{e} \big) ~ \boldsymbol{C}_{e} \, \mathrm{d} \boldsymbol{C}_{e},
\end{equation}
while $\Delta E_{e \heavy}^{0}$ is an energy exchange term due to scattering collisions between electrons and heavy species
\begin{equation}
\Delta E_{e \heavy}^{0} = \sum_{j \in \Heavy} \llangle \psi_{e}^{n^{s}+4}, \Scattering_{e j}^{2} (f_{e}^{0}, f_{j}^{0}) \rrangle_{e},
\end{equation}
where $\Scattering_{e j}^{2}$ is the second-order electron $j^{\text{th}}$-heavy-species scattering source term \eqref{KTSei2}. The zeroth-order electron mean velocity in the inertial reference frame is defined as
\begin{equation}
\boldsymbol{\Meanv}_{e}^{0} = \boldsymbol{\Meanv}_{\heavy} + \boldsymbol{\MeanV}_{e}^{0}.
\end{equation}

Finally, momentum conservation equation \eqref{KTElectronMomentumConservation} yields the momentum relation
\begin{equation}
\boldsymbol{\partial_{x}} \pressure_{e} = \density_{e} q_{e} \boldsymbol{E} + \delta_{b1} \boldsymbol{\current}_{e}^{0} \wedge \boldsymbol{B} + \boldsymbol{\Force}_{e \heavy}^{0},
\label{KTZerothElectronMomentumRelation}
\end{equation}
where $\boldsymbol{\current}_{e}^{0} = \density_{e} q_{e} \boldsymbol{\Meanv}_{e}^{0}$ is the zeroth-order electron current density in the inertial reference frame, and
\begin{equation}
\boldsymbol{\Force}_{e \heavy}^{0} = \sum_{j \in \Heavy} \llangle \Scattering_{ej}^{0}(\phi_{e} f_{e}^{0},f_{j}^{0}) , \Mass_{e} \boldsymbol{C}_{e} \rrangle_{e}
\label{KTFeh0}
\end{equation}
is the average force exerted by the heavy species on electrons due to scattering collisions, which can be expressed as
\begin{equation}
\boldsymbol{\Force}_{e \heavy}^{0} = \sum_{j \in \Heavy} \density_{j} \boldsymbol{\Force}_{e j}^{0},
\end{equation}
where $\boldsymbol{\Force}_{e j}^{0}$ is the average force exerted by the $j^{\text{th}}$ heavy species on electrons. We can further decompose
\begin{equation}
\boldsymbol{\Force}_{e j}^{0} = \sum_{\textsc{j} \in \QuantumSpace_{j}} \frac{a_{j \textsc{j}} e^{- \epsilon_{j \textsc{j}}}}{Q_{j}^{\text{int}}} \boldsymbol{\Force}_{e j}^{\textsc{j},0},
\end{equation}
where $\boldsymbol{\Force}_{e j}^{\textsc{j},0}$ is the average force exerted on electrons by molecules of the $j^{\text{th}}$ heavy species in the $\textsc{j}^{\text{th}}$ internal state:
\begin{equation}
\boldsymbol{\Force}_{e j}^{\textsc{j},0} = - \Mass_{e} \int \Sigma_{\textsc{j}\textsc{j}}^{(1)} (|\boldsymbol{C}_{e}|^{2}) |\boldsymbol{C}_{e}| \phi_{e}(\boldsymbol{C}_{e}) f_{e}^{0}(\boldsymbol{C}_{e}) \boldsymbol{C}_{e} \, \mathrm{d} \boldsymbol{C}_{e}.
\end{equation}
Equation \eqref{KTZerothElectronMomentumRelation} thus provides an expression for the average force $\boldsymbol{\Force}_{e \heavy}^{0}$ in terms of the macroscopic variable gradients and external fields, which will be useful in the derivation of the heavy-species momentum equation.

\subsection{Zeroth-order macroscopic equations for the heavy species}
We now expand equations \eqref{KTHeavyMatterConservation}, \eqref{KTHeavyMomentumConservation}, \eqref{KTHeavyEnergyConservation} at order $\varepsilon^{0}$. First, equation \eqref{KTHeavyMatterConservation} yields the mass conservation equation
\begin{equation}
\partial_{t} \rho_{i} + \boldsymbol{\partial_{x}} \cdot (\rho_{i} \boldsymbol{\Meanv}_{\heavy}) = 0, \qquad i \in \Heavy. \label{KTEulerMassHeavy}
\end{equation}
Then, expansion of equation \eqref{KTHeavyMomentumConservation} at order $\varepsilon^{0}$ reads
\begin{equation}
\partial_{t} \left( \rho_{\heavy} \boldsymbol{\Meanv}_{\heavy} \right) + \boldsymbol{\partial_{x}} \cdot \left(  \rho_{\heavy} \boldsymbol{\Meanv}_{\heavy} \otimes \boldsymbol{\Meanv}_{\heavy} + \pressure_{\heavy} \, \mathbb{I} \right) = \density_{\heavy} q_{\heavy} \boldsymbol{E} + \delta_{b1} ~ \boldsymbol{\current}_{\heavy}^{0} \wedge \boldsymbol{B} + \boldsymbol{\Force}_{\heavy e}^{0}, \label{KTZerothMomentumHeavy}
\end{equation}
where $\pressure_{\heavy} = \density_{\heavy}  k_{\textsc{b}} \Temperature_{\heavy}$ is the heavy-species partial pressure, $\density_{\heavy} = \sum_{j \in \Heavy} \density_{j}$ the heavy-species density, $\density_{\heavy} q_{\heavy} = \sum_{j \in \Heavy} \density_{j} q_{j}$ the heavy-species charge density, $\boldsymbol{\current}_{\heavy}^{0} = \density_{\heavy} q_{\heavy} \boldsymbol{\Meanv}_{\heavy}$ the zeroth-order heavy-species current density in the inertial reference frame, and
\begin{equation}
\boldsymbol{\Force}_{\heavy e}^{0} = \llangle \Scattering_{\heavy e}^{1}(f_{\heavy}^{0}, \phi_{e} f_{e}^{0}) + \Scattering_{\heavy e}^{2}(f_{\heavy}^{0}, f_{e}^{0}) , \Mass_{\heavy} \boldsymbol{C}_{\heavy} \rrangle_{\heavy}
\end{equation}
is the average force exerted by electrons on the heavy species due to scattering collisions. Indeed, we recall that from \eqref{KTSie0} $\Scattering_{ie}^{0} = 0$, and from \eqref{KTSie1} it is readily seen that
\begin{equation}
\Scattering_{ie}^{1}(\phi_{i} f_{i}^{0},f_{e}^{0}) = 0,
\end{equation}
as long as $f_{e}^{0}$ is isotropic. Due to orthogonality relation \eqref{KTCrossInvariantsConservationMomentum}, the following reciprocity relation holds:
\begin{equation}
\boldsymbol{\Force}_{\heavy e}^{0} = - \boldsymbol{\Force}_{e \heavy}^{0} \label{KTMomentumExchange}.
\end{equation}
As a consequence, making use of the electron momentum relation \eqref{KTZerothElectronMomentumRelation}, the momentum conservation equation \eqref{KTZerothMomentumHeavy} for the heavy species can be rewritten
\begin{equation}
\partial_{t} \left( \rho_{\heavy} \boldsymbol{\Meanv}_{\heavy} \right) + \boldsymbol{\partial_{x}} \cdot \left(  \rho_{\heavy} \boldsymbol{\Meanv}_{\heavy} \otimes \boldsymbol{\Meanv}_{\heavy} + \pressure \, \mathbb{I} \right) = \density q \boldsymbol{E} + \delta_{b1} ~ \boldsymbol{\current}^{0} \wedge \boldsymbol{B}, \label{KTEulerMomentumHeavy}
\end{equation}
where $\pressure = \pressure_{\heavy} + \pressure_{e}$ is the total pressure, $\density = \density_{\heavy} + \density_{e}$ the total density, $\density q = \density_{\heavy} q_{\heavy} + \density_{e} q_{e}$ the total charge density, and $\boldsymbol{\current}^{0} = \boldsymbol{\current}_{\heavy}^{0} + \boldsymbol{\current}_{e}^{0}$ is the zeroth-order current density in the inertial reference frame.

Finally, equation \eqref{KTHeavyEnergyConservation} yields the following energy conservation equation at order zero:
\begin{equation}
\partial_{t} \Energy_{\heavy} + \boldsymbol{\partial_{x}} \cdot \left( \Energy_{\heavy} \boldsymbol{\Meanv}_{\heavy} \right) = - \pressure_{\heavy} ~ \boldsymbol{\partial_{x}} \cdot \boldsymbol{\Meanv}_{\heavy} + \Delta E_{\heavy e}^{0}, \label{KTEulerEnergyHeavy}
\end{equation}
where $\Delta E_{\heavy e}^{0} = \llangle \Scattering_{\heavy e}^{2}(f_{\heavy}^{0},f_{e}^{0}), \psi_{\heavy}^{n^{s}+4} \rrangle_{\heavy}$ is an energy exchange term due to scattering collisions between heavy species and electrons. Given orthogonality property \eqref{KTCrossInvariantsConservationEnergy}, energy exchange terms are shown to be symmetric as momentum exchange terms in \eqref{KTMomentumExchange}:
\begin{equation}
\Delta E_{\heavy e}^{0} = - \Delta E_{e \heavy}^{0}.
\end{equation}
The energy exchange term splits into an elastic and an inelastic contributions:
\begin{equation}
\Delta E_{\heavy e}^{0} = \Delta E_{\heavy e}^{0,\text{el}} + \Delta E_{\heavy e}^{0,\text{in}}.
\label{KTDeltahe0}
\end{equation}
The elastic term $\Delta E_{\heavy e}^{0,\text{el}} = \llangle \Scattering_{\heavy e}^{2,\text{el}}(f_{\heavy}^{0},f_{e}^{0}), \psi_{\heavy}^{n^{s}+4} \rrangle_{\heavy}$ is computed from expression \eqref{KTSie2el} for the second-order $i^{\text{th}}$-heavy-species electron elastic scattering source term $\Scattering_{ie}^{2,\text{el}}$, and reads
\begin{equation}
\Delta E_{\heavy e}^{0,\text{el}} = \frac{3}{2} \density_{\heavy} k_{\textsc{b}} (\Temperature_{e}-\Temperature_{\heavy})  \frac{1}{\tau^{\text{el}}},
\label{KTDeltaEhe0el}
\end{equation}
where $\tau^{\text{el}}$ is the average collision time at which elastic energy transfer occurs:
\begin{align}
\frac{1}{\tau^{\text{el}}} = {} & \frac{2}{3} \sum_{i \in \Heavy} \frac{\density_{i}}{\density_{\heavy}} \frac{\Mass_{e}}{\Mass_{i}} \nu_{ie}^{\text{el}}, \\
\nu_{ie}^{\text{el}} = {} & \sum_{\textsc{i} \in \QuantumSpace_{i}} \frac{a_{i \textsc{i}} e^{- E_{i \textsc{i}}/k_{\textsc{b}}\Temperature_{\heavy}}}{Q_{i}^{\text{int}}} \nu_{ie}^{\textsc{i} \textsc{i}}, \\
\nu_{ie}^{\textsc{i} \textsc{i}} = {} & \frac{\Mass_{e}}{k_{\textsc{b}} \Temperature_{e}} \int \Sigma_{\textsc{i} \textsc{i}}^{(1)}(|\boldsymbol{C}_{e}|^{2}) \, |\boldsymbol{C}_{e}|^{3} \, f_{e}^{0}(\boldsymbol{C}_{e}) \, \mathrm{d} \boldsymbol{C}_{e}.
\end{align}
The inelastic contribution can be expressed using \eqref{KTSie2in} as
\begin{equation}
\Delta E_{\heavy e}^{0,\text{in}} = \frac{1}{2} \sum_{i \in \Heavy} \sum\limits_{\substack{\textsc{i}, \textsc{i}' \in \QuantumSpace_{i} \\ \textsc{i}' \neq \textsc{i}}} \Delta E_{\textsc{i} \textsc{i}'} \frac{\density_{i} a_{i \textsc{i}}}{Q_{i}^{\text{int}}} \nu_{ie}^{\textsc{i} \textsc{i}'} \bigg( e^{- E_{i \textsc{i}} / k_{\textsc{b}} \Temperature_{\heavy}} - \exp{ \Big( \frac{\Delta E_{\textsc{i} \textsc{i}'}}{k_{\textsc{b}} \Temperature_{e}} \Big)} e^{- E_{i \textsc{i}'} / k_{\textsc{b}} \Temperature_{\heavy}} \bigg),
\label{KTDeltaEhe0in}
\end{equation}
where $\nu_{ie}^{\textsc{i} \textsc{i}'}$ is the collision frequency between a molecule of the $i^{\text{th}}$ heavy species with initial quantum state $\textsc{i}$ and final quantum state $\textsc{i}'$ and an electron:
\begin{equation}
\nu_{ie}^{\textsc{i} \textsc{i}'} = \int \Sigma_{\textsc{i} \textsc{i}'}^{(0)}(|\boldsymbol{C}_{e}|^{2}) \, |\boldsymbol{C}_{e}| \, f_{e}^{0}(\boldsymbol{C}_{e}) \, \mathrm{d} \boldsymbol{C}_{e}, \qquad \textsc{i}' \neq \textsc{i}.
\end{equation}
Expression \eqref{KTDeltaEhe0in} can be rewritten in terms of the temperature difference $\Temperature_{e} - \Temperature_{\heavy}$
\begin{equation}
\Delta E_{\heavy e}^{0,\text{in}} = \frac{3}{2} \density_{\heavy} k_{\textsc{b}} (\Temperature_{e}-\Temperature_{\heavy})  \frac{1}{\tau^{\text{in}}},
\end{equation}
where $\tau^{\text{in}}$ is the average collision time at which inelastic energy transfer occurs:
\begin{align}
\frac{1}{\tau^{\text{in}}} = {} &  \frac{2}{3} \frac{\Temperature_{\heavy}}{\Temperature_{e}} \sum_{i \in \Heavy} \frac{\density_{i}}{\density_{\heavy}} \nu_{ie}^{\text{in}}, \\
\nu_{ie}^{\text{in}} = {} & \sum\limits_{\substack{\textsc{i}, \textsc{i}' \in \QuantumSpace_{i} \\ \textsc{i}' \neq \textsc{i}}} \frac{\left(\Delta \epsilon_{\textsc{i} \textsc{i}'} \right)^{2}}{2} \frac{a_{i \textsc{i}} e^{- \epsilon_{i \textsc{i}}}}{Q_{i}^{\text{int}}} g \Big(\Delta \epsilon_{\textsc{i} \textsc{i}'} \big( 1 - \frac{\Temperature_{\heavy}}{\Temperature_{e}} \big) \Big) \nu_{ie}^{\textsc{i} \textsc{i}'},
\end{align}
where $\Delta \epsilon_{\textsc{i} \textsc{i}'} = \epsilon_{i \textsc{i}'} - \epsilon_{i \textsc{i} \vphantom{i \textsc{i}'}} = \frac{\Delta E_{\textsc{i} \textsc{i}'}}{k_{\textsc{b}} \Temperature_{\heavy}}$, and where we have introduced the function $g$ defined by
\begin{equation}
\left\lbrace
\begin{array}{lr}
\displaystyle g(u) = \frac{1-e^{-u}}{u}, & u \neq 0, \\
\displaystyle g(0) = 1. \vphantom{\frac{1-e^{-u}}{u}} & {}
\end{array}
\right.
\end{equation}
We can also formulate $\Delta E_{\heavy e}^{0}$ as
\begin{equation}
\Delta E_{\heavy e}^{0} =  \frac{3}{2} \density_{\heavy} k_{\textsc{b}} (\Temperature_{e}-\Temperature_{\heavy})  \frac{1}{\tau},
\end{equation}
where
\begin{equation}
\frac{1}{\tau} = \frac{1}{\tau^{\text{el}}} + \frac{1}{\tau^{\text{in}}}.
\end{equation}
\subsection{First-order perturbation for the heavy species}

As for electrons, we introduce the heavy-species linearized collision operator $\mathcal{F}_{\heavy} = \left( \mathcal{F}_{i} \right)_{i \in \Heavy}$, defined as
\begin{equation}
\mathcal{F}_{i}(\phi_{\heavy}) = - \frac{1}{f_{i}^{0}} \sum_{j \in \Heavy} \left[ \Scattering_{ij}(\phi_{i}f_{i}^{0}, f_{j}^{0}) + \Scattering_{ij}(f_{i}^{0}, \phi_{j}f_{j}^{0}) \right], \qquad i \in \Heavy.
\end{equation}
Again, using the reciprocity relation \eqref{KTScattering Reciprocity Relation} and symmetrization \cite{ChapmanCowling} \cite{FerzigerKaper} \cite{GrailleMaginMassot2009}, the linearized collision operator is expressed as
\begin{equation}
f_{i}^{0} \mathcal{F}_{i}(\phi_{\heavy}) = \sum\limits_{j \in \Heavy \vphantom{\textsc{j}' \in \QuantumSpace_{j}}} \sum\limits_{\vphantom{j \in \Heavy} \textsc{i}' \in \QuantumSpace_{i}} \sum\limits_{\vphantom{j \in \Heavy} \textsc{j},\textsc{j}' \in \QuantumSpace_{j}} \int g_{ij} \CrossSection_{ij}^{\textsc{i} \textsc{j} \textsc{i}' \textsc{j}'} f_{i}^{0} f_{j}^{0} \left( \phi_{i} + \phi_{j}  - \phi_{i}' - \phi_{j}' \right) \, \mathrm{d} \boldsymbol{\omega}_{ij}' \mathrm{d} \boldsymbol{C}_{j}.
\end{equation}
The kernel of $\mathcal{F}_{\heavy}$ coincides with the set of heavy-species collisional invariants $\Invariants_{\heavy}$. We also introduce the associated integral bracket operator
\begin{equation}
\llbracket \xi_{\heavy}, \zeta_{\heavy} \rrbracket_{\heavy} = \llangle f_{\heavy}^{0} \xi_{\heavy}, \mathcal{F}_{\heavy}(\zeta_{\heavy}) \rrangle_{\heavy} = \sum_{i \in \Heavy} \sum_{\textsc{i} \in \QuantumSpace_{i}} \int f_{i}^{0} \xi_{i} \mathcal{F}_{i}(\zeta_{\heavy}) \, \mathrm{d} \boldsymbol{C}_{i},
\end{equation}
which can be expressed in the form
\begin{align}
\llbracket \xi_{\heavy}, \zeta_{\heavy} \rrbracket_{\heavy} & = \frac{1}{4} \sum_{i, j \in \Heavy} \sum_{\textsc{i}, \textsc{i}' \in \QuantumSpace_{i}} \sum_{\textsc{j},\textsc{j}' \in \QuantumSpace_{j}}
 \label{KTHeavyBracket} \\
 & \quad \times \int g_{ij} \CrossSection_{ij}^{\textsc{i} \textsc{j} \textsc{i}' \textsc{j}'} f_{i}^{0} f_{j}^{0} \left( \xi_{i}' + \xi_{j}'  - \xi_{i} - \xi_{j} \right) \left( \zeta_{i}' + \zeta_{j}'  - \zeta_{i} - \zeta_{j} \right) \, \mathrm{d} \boldsymbol{\omega}_{ij}' \mathrm{d} \boldsymbol{C}_{j} \mathrm{d} \boldsymbol{C}_{i}.
\nonumber
\end{align}
From expression \eqref{KTHeavyBracket}, the bracket operator $\llbracket \cdot \rrbracket_{\heavy}$ is shown to be symmetric, i.e.,
\begin{equation}
\llbracket \xi_{\heavy}, \zeta_{\heavy} \rrbracket_{\heavy} = \llbracket \zeta_{\heavy}, \xi_{\heavy} \rrbracket_{\heavy}, \label{KTHeavyBracketSymmetry}
\end{equation}
positive semi-definite:
\begin{equation}
\llbracket \xi_{\heavy}, \xi_{\heavy} \rrbracket_{\heavy} \geq 0, \label{KTHeavyBracketPositivity}
\end{equation}
and its kernel is seen to coincide with the kernel of $\mathcal{F}_{\heavy}$:
\begin{equation}
\llbracket \xi_{\heavy}, \xi_{\heavy} \rrbracket_{\heavy} = 0 \Leftrightarrow \mathcal{F}_{\heavy}(\xi_{\heavy}) = 0 \Leftrightarrow \xi_{\heavy} \in \Invariants_{\heavy}. \label{KTHeavyBracketKernel}
\end{equation}

Projecting the heavy-species Boltzmann equations \eqref{KTCompact Scaled Boltzmann heavy-species equations} at order $\varepsilon^{0}$, the first-order heavy-species perturbation function $\phi_{\heavy}$ is shown to be solution to the linear equation
\begin{equation}
\mathcal{F}_{i}(\phi_{\heavy}) = \Psi_{i}, \qquad i \in \Heavy,
\label{KTFirstOrderHeavyPerturbationEquation}
\end{equation}
where
\begin{equation}
\Psi_{i} = - \Streaming_{i}^{0}(\ln{f_{i}^{0}}) + \frac{1}{f_{i}^{0}} \Scattering_{ie}^{1}(f_{i}^{0},\phi_{e} f_{e}^{0}) + \frac{1}{f_{i}^{0}} \Scattering_{ie}^{2}(f_{i}^{0},f_{e}^{0}).
\label{KTPsiHeavy}
\end{equation}
Indeed, $\Scattering_{ie}^{0} = 0$, and $\Scattering_{ie}^{1}(\phi_{i} f_{i}^{0}, f_{e}^{0}) = 0$ since $f_{e}^{0}$ is isotropic. Equation \eqref{KTFirstOrderHeavyPerturbationEquation} must be completed with the constraints \eqref{KTZerothMacroscopicPropertiesHeavy} in order to be well posed \cite{ChapmanCowling} \cite{FerzigerKaper} \cite{Giovangigli}
\begin{equation}
\llangle \phi_{\heavy} f_{\heavy}^{0}, \psi_{\heavy}^{l} \rrangle_{\heavy} = 0, \quad l \in \Heavy \cup \left\lbrace n^{s}+1, \ldots, n^{s}+4 \right\rbrace.
\label{KTFirstOrderHeavyPerturbationConstraints}
\end{equation}

\subsection{Second-order perturbation for electrons}
We project the electron Boltzmann equation \eqref{KTCompact Scaled Boltzmann electron equation} at order $\varepsilon^{0}$ and after a few calculations, the second-order electron perturbation function $\phi_{e}^{2}$ is found to be solution to the following linear integral equation \cite{GrailleMaginMassot2009}:
\begin{equation}
f_{e}^{0} \mathcal{F}_{e}(\phi_{e}^{2}) + \delta_{b1} \frac{q_{e}}{\Mass_{e}} (\boldsymbol{C}_{e} \wedge \boldsymbol{B}) \cdot \boldsymbol{\partial}_{\boldsymbol{C}_{e}} (\phi_{e}^{2}f_{e}^{0}) = \Psi_{e}^{2},
\label{KTSecondOrderElectronPerturbationEquation}
\end{equation}
where
\begin{align}
\Psi_{e}^{2} = & - \Streaming_{e}^{0}(f_{e}^{0}) - \Streaming_{e}^{-1}(\phi_{e}f_{e}^{0}) + \Scattering_{ee}(\phi_{e} f_{e}^{0},\phi_{e}f_{e}^{0}) \vphantom{\sum_{j \in \Heavy}} \\
{} & + \sum_{j \in \Heavy} \left[ \Scattering_{ej}^{0}(\phi_{e}f_{e}^{0},\phi_{j}f_{j}^{0}) + \Scattering_{ej}^{1}(f_{e}^{0},\phi_{j}f_{j}^{0}) + \Scattering_{ej}^{2}(f_{e}^{0},f_{j}^{0}) \right]. \nonumber
\end{align}
Indeed, by \eqref{KTSei0} $\Scattering_{ej}^{0}(f_{e}^{0},\phi_{j}^{2}f_{j}^{0}) = 0$ since $f_{e}^{0}$ is isotropic, and by \eqref{KTSei1} $\Scattering_{ej}^{1}(\phi_{e}f_{e}^{0},f_{j}^{0}) = 0$ since $f_{j}^{0}$ is isotropic. Equation \eqref{KTSecondOrderElectronPerturbationEquation} is completed with the constraints \eqref{KTZerothMacroscopicPropertiesElectron} in order to be well posed
\begin{equation}
\llangle \phi_{e}^{2} f_{e}^{0}, \psi_{e}^{l} \rrangle_{e} = 0, \quad l \in \left\lbrace e, n^{s}+4 \right\rbrace.
\end{equation}

\subsection{First-order macroscopic equations for electrons}
Equations \eqref{KTElectronMatterConservation}, \eqref{KTElectronMomentumConservation}, \eqref{KTElectronEnergyConservation} are now expanded at order $\varepsilon^{1}$. The following first-order drift-diffusion equations for electrons follow
\begin{gather}
\partial_{t} \rho_{e} + \boldsymbol{\partial_{x}} \cdot \left( \rho_{e} \boldsymbol{\Meanv}_{\heavy} + \rho_{e} \boldsymbol{\MeanV}_{e}^{0} + \varepsilon \, \rho_{e} \boldsymbol{\MeanV}_{e}^{1} \right) = \varepsilon \, \Mass_{e} \mathfrak{w}_{e}, \label{KTNavierStokesMassElectron} \\
\partial_{t} \Energy_{e} + \boldsymbol{\partial_{x}} \cdot (\Energy_{e} \boldsymbol{\Meanv}_{\heavy}) = - \pressure_{e} \boldsymbol{\partial_{x}} \cdot \boldsymbol{\Meanv}_{\heavy} - \varepsilon \, \boldsymbol{\partial_{x}} \boldsymbol{\Meanv}_{\heavy}: \boldsymbol{\Pi}_{e} - \boldsymbol{\partial_{x}} \cdot \boldsymbol{\HeatFlux}_{e}^{0} - \varepsilon \, \boldsymbol{\partial_{x}} \cdot \boldsymbol{\HeatFlux}_{e}^{1} \label{KTNavierStokesEnergyElectron} \\
 + \boldsymbol{\Current}_{e}^{0} \cdot \boldsymbol{E}' + \varepsilon \, \boldsymbol{\Current}_{e}^{1} \cdot \boldsymbol{E}' + \varepsilon \, \delta_{b0} \, \boldsymbol{\Current}_{e}^{0} \cdot (\boldsymbol{\Meanv}_{\heavy} \wedge \boldsymbol{B}) + \Delta E_{e \heavy}^{0} + \varepsilon \, \Delta E_{e \heavy}^{1}, \nonumber
\end{gather}
where we have introduced the electron first-order diffusion velocity, heat flux, viscous tensor, and the electron first-order conduction current density in the heavy-species reference frame, respectively
\begin{eqnarray}
\density_{e} \boldsymbol{\MeanV}_{e}^{1} & = & \int \phi_{e}^{2} f_{e}^{0} \, \boldsymbol{C}_{e} \, \mathrm{d} \boldsymbol{C}_{e}, \\
\boldsymbol{\HeatFlux}_{e}^{1} &  = & \int \phi_{e}^{2} f_{e}^{0} \, \big(\frac{1}{2}\Mass_{e} \boldsymbol{C}_{e} \cdot \boldsymbol{C}_{e} \big) \, \boldsymbol{C}_{e} \, \mathrm{d} \boldsymbol{C}_{e}, \\
\boldsymbol{\Pi}_{e} & = & \int \phi_{e} f_{e}^{0} \, \Mass_{e} \, \boldsymbol{C}_{e} \otimes \boldsymbol{C}_{e} \, \mathrm{d} \boldsymbol{C}_{e}, \\
\boldsymbol{\Current}_{e}^{1} & = & \density_{e} q_{e} \boldsymbol{\MeanV}_{e}^{1}. \vphantom{\int}
\end{eqnarray}

We have denoted by $\mathfrak{w}_{e}$ the zeroth-order molecular production rate of electrons due to chemically reactive collisions:
\begin{equation}
\mathfrak{w}_{e} = \llangle \psi_{e}^{e}, \Chemistry_{e}(f^{0}) \rrangle_{e} = \int \Chemistry_{e}(f^{0}) \, \mathrm{d} \boldsymbol{C}_{e},
\end{equation}
and $\Delta E_{e \heavy}^{1}$ is the first-order energy exchange term arising from chemically reactive collisions on the one hand, and scattering collisions on the other hand:
\begin{equation}
\Delta E_{e \heavy}^{1} = \Delta E_{e \heavy}^{1,\text{chem}} + \Delta E_{e \heavy}^{1,\text{scatt}}.
\label{KTDeltaEeh1}
\end{equation}
The reactive term reads
\begin{equation}
\Delta E_{e \heavy}^{1,\text{chem}} = \int \big(\frac{1}{2}\Mass_{e} \boldsymbol{C}_{e} \cdot \boldsymbol{C}_{e} \big) \, \Chemistry_{e}(f^{0}) \, \mathrm{d} \boldsymbol{C}_{e},
\end{equation}
and the scattering term is in turn decomposed into an elastic and an inelastic contribution
\begin{equation}
\Delta E_{e \heavy}^{1,\text{scatt}} = \Delta E_{e \heavy}^{1,\text{el}} + \Delta E_{e \heavy}^{1,\text{in}},
\label{KTDeltaEeh1scatt}
\end{equation}
where
\begin{align}
\Delta E_{e \heavy}^{1,\text{el}} & = - \llangle \psi_{\heavy}^{n^{s}+4}, \Scattering_{\heavy e}^{1} ( \phi_{\heavy} f_{\heavy}^{0}, \phi_{e}f_{e}^{0}) + \Scattering_{\heavy e}^{2,\text{el}} ( \phi_{\heavy} f_{\heavy}^{0}, f_{e}^{0}) + \Scattering_{\heavy e}^{2,\text{el}} ( f_{\heavy}^{0}, \phi_{e}f_{e}^{0}) \rrangle_{\heavy}, \\
\Delta E_{e \heavy}^{1,\text{in}} & = - \llangle \psi_{\heavy}^{n^{s}+4}, \Scattering_{\heavy e}^{2,\text{in}} ( \phi_{\heavy} f_{\heavy}^{0}, f_{e}^{0}) + \Scattering_{\heavy e}^{2,\text{in}} ( f_{\heavy}^{0}, \phi_{e}f_{e}^{0}) \rrangle_{\heavy}.
\end{align}

Finally, the momentum electron conservation equation \eqref{KTElectronMomentumConservation} yields the following first-order momentum relation for electrons:
\begin{equation}
\boldsymbol{\partial_{x}} \pressure_{e} + \varepsilon \, \boldsymbol{\partial_{x}} \cdot \boldsymbol{\Pi}_{e} = \density_{e} q_{e} \boldsymbol{E} + \varepsilon \, \delta_{b0} \, \boldsymbol{\current}_{e}^{0} \wedge \boldsymbol{B} + \delta_{b1} \, (\boldsymbol{\current}_{e}^{0}+ \varepsilon \, \boldsymbol{\Current}_{e}^{1}) \wedge \boldsymbol{B} + \boldsymbol{\Force}_{e \heavy}^{0} + \varepsilon \, \boldsymbol{\Force}_{e \heavy}^{1},
\label{KTFirstElectronMomentumRelation}
\end{equation}
where $\boldsymbol{\Force}_{e \heavy}^{1}$ is the first-order average force exerted by the heavy species on electrons:
\begin{equation}
\boldsymbol{\Force}_{e \heavy}^{1} = \sum_{j \in \Heavy} \llangle \Mass_{e} \boldsymbol{C}_{e}, \Scattering_{ej}^{0}(\phi_{e}^{2}f_{e}^{0},f_{j}^{0}) + \Scattering_{ej}^{0}(\phi_{e}f_{e}^{0}, \phi_{j}f_{j}^{0}) + \Scattering_{ej}^{1}(f_{e}^{0}, \phi_{j}f_{j}^{0}) + \Scattering_{ej}^{2}(f_{e}^{0},f_{j}^{0}) \rrangle_{e}.
\label{KTFeh1}
\end{equation}

\subsection{First-order macroscopic equations for the heavy species}
Proceeding as for the zeroth-order macroscopic equations, we obtain the following set of Navier-Stokes type equations for the heavy species:
\begin{gather}
\partial_{t} \rho_{i} + \boldsymbol{\partial_{x}} \cdot (\rho_{i} \boldsymbol{\Meanv}_{\heavy} + \varepsilon \, \rho_{i} \boldsymbol{\MeanV}_{i}) = \varepsilon \, \Mass_{i} \mathfrak{w}_{i}, \qquad i \in \Heavy, \label{KTNavierStokesMassHeavy} \\
\partial_{t} \left( \rho_{\heavy} \boldsymbol{\Meanv}_{\heavy} \right) + \boldsymbol{\partial_{x}} \cdot \left(  \rho_{\heavy} \boldsymbol{\Meanv}_{\heavy} \otimes \boldsymbol{\Meanv}_{\heavy} + \pressure \, \mathbb{I} \right) = - \varepsilon \, \boldsymbol{\partial_{x}} \cdot ( \boldsymbol{\Pi}_{e} + \boldsymbol{\Pi}_{\heavy}) + \density q \boldsymbol{E} \label{KTNavierStokesMomentumHeavy} \\
 + \varepsilon \, \delta_{b0} \, \boldsymbol{\current}^{0} \wedge \boldsymbol{B} + \delta_{b1} \, \boldsymbol{\current}^{1} \wedge \boldsymbol{B}, \nonumber \\
\partial_{t} \Energy_{\heavy} + \boldsymbol{\partial_{x}} \cdot \left( \Energy_{\heavy} \boldsymbol{\Meanv}_{\heavy} \right) = - \pressure_{\heavy} \, \boldsymbol{\partial_{x}} \cdot \boldsymbol{\Meanv}_{\heavy} - \varepsilon \, \boldsymbol{\partial_{x}} \boldsymbol{\Meanv}_{\heavy}: \boldsymbol{\Pi}_{\heavy} - \varepsilon \, \boldsymbol{\partial_{x}} \cdot \boldsymbol{\HeatFlux}_{\heavy} \label{KTNavierStokesEnergyHeavy} \\
+ \varepsilon \, \boldsymbol{\Current}_{\heavy} \cdot \boldsymbol{E}' + \Delta E_{\heavy e}^{0} + \varepsilon \, \Delta E_{\heavy e}^{1}, \nonumber
\end{gather}
where we have introduced the diffusion velocity of the $i^{\text{th}}$ heavy species $\boldsymbol{\MeanV}_{i}$, the heavy-species viscous tensor $\boldsymbol{\Pi}_{\heavy}$, and the heavy-species heat flux $\boldsymbol{\HeatFlux}_{\heavy}$:
\begin{eqnarray}
\density_{i} \boldsymbol{\MeanV}_{i} & = & \sum_{\textsc{i} \in \QuantumSpace_{i}} \int \phi_{i}f_{i}^{0} \boldsymbol{C}_{i} \, \mathrm{d} \boldsymbol{C}_{i}, \quad i \in \Heavy, \\
\boldsymbol{\Pi}_{\heavy} & = & \sum_{j \in \Heavy} \sum_{\textsc{j} \in \QuantumSpace_{j}} \int \phi_{j} f_{j}^{0} \, \Mass_{j} \, \boldsymbol{C}_{j} \otimes \boldsymbol{C}_{j} \, \mathrm{d} \boldsymbol{C}_{j}, \\
\boldsymbol{\HeatFlux}_{\heavy} & = & \sum_{j \in \Heavy} \sum_{\textsc{j} \in \QuantumSpace_{j}} \int \phi_{j}f_{j}^{0} \Big( \frac{1}{2}\Mass_{j} \boldsymbol{C}_{j} \cdot \boldsymbol{C}_{j} + E_{j \textsc{j}} \Big) \boldsymbol{C}_{j} \, \mathrm{d} \boldsymbol{C}_{j}.
\end{eqnarray}
We denote by $\boldsymbol{\Current}_{\heavy}$ the heavy-species conduction current density in the heavy-species reference frame, $\boldsymbol{\current}_{\heavy}^{1}$ the first-order heavy-species current density in the inertial reference frame, and $\boldsymbol{\current}_{e}^{1}$ the first-order electron current density in the inertial reference frame:
\begin{align}
 \boldsymbol{\Current}_{\heavy} & = \sum_{j \in \Heavy} \density_{j} q_{j} \boldsymbol{\MeanV}_{j}, \\
 \boldsymbol{\current}_{\heavy}^{1} & = \sum_{j \in \Heavy} \density_{j} q_{j} (\boldsymbol{\Meanv}_{\heavy} + \varepsilon \boldsymbol{\MeanV}_{j}) = \density_{\heavy} q_{\heavy} \boldsymbol{\Meanv}_{\heavy} + \varepsilon \boldsymbol{\Current}_{\heavy}, \\
 \boldsymbol{\current}_{e}^{1} & = \density_{e} q_{e} (\boldsymbol{\Meanv}_{\heavy} + \boldsymbol{\MeanV}_{e}^{0} + \varepsilon \boldsymbol{\MeanV}_{e}^{1}) = \density_{e} q_{e} \boldsymbol{\Meanv}_{e} + \varepsilon \boldsymbol{\Current}_{e}^{1},
\end{align}
where $\boldsymbol{\Current}_{e}^{1} = \density_{e} q_{e} \boldsymbol{\MeanV}_{e}^{1}$ is the first-order electron conduction current density in the heavy-species reference frame. The zeroth-order and first-order current density in the inertial reference frame, $\boldsymbol{\current}^{0}$ and $\boldsymbol{\current}^{1}$, respectively, are given by
\begin{align}
\boldsymbol{\current}^{0} & = \boldsymbol{\current}_{\heavy}^{0} + \boldsymbol{\current}_{e}^{0}, \\
\boldsymbol{\current}^{1} & = \boldsymbol{\current}_{\heavy}^{1} + \boldsymbol{\current}_{e}^{1}.
\end{align}

Finally, $\mathfrak{w}_{i}, i \in \Heavy$ is the zeroth-order molecular production rate of the $i^{\text{th}}$ species due to chemically reactive collisions:
\begin{equation}
\mathfrak{w}_{i} = \llangle \psi_{\heavy}^{i}, \Chemistry_{\heavy}(f^{0}) \rrangle_{\heavy} = \sum_{\textsc{i} \in \QuantumSpace_{i}} \int \Chemistry_{i}(f^{0}) \, \mathrm{d} \boldsymbol{C}_{i},
\end{equation}
and $\Delta E_{\heavy e}^{1}$ is the first-order energy exchange term, involving both scattering and reactive energy transfer. Again, due to relation \eqref{KTCrossInvariantsConservationEnergy}, energy exchange terms are symmetric:
\begin{equation}
\Delta E_{\heavy e}^{1} = - \Delta E_{e \heavy}^{1}.
\end{equation}

\subsection{Chemistry source terms}
For $k \in \Species$, the chemically reactive source term reads
\begin{equation}
\mathfrak{w}_{k} = \sum_{\textsc{k} \in \QuantumSpace_{k}} \int \Chemistry_{k}(f^{0}) \, \mathrm{d} \boldsymbol{C}_{k}.
\end{equation}
This term can be expressed as follows. We first recall the decomposition \eqref{KTChemistrySource}
\begin{equation}
\Chemistry_{k}(f) = \sum_{r \in \Reaction} \Chemistry_{k}^{r}(f),
\end{equation}
where $\Chemistry_{k}^{r}(f)$ was expressed in \eqref{KTChemistrySourcePerReaction}. This yields the following decomposition:
\begin{equation}
\mathfrak{w}_{k} = \sum_{r \in \Reaction} \left( \nu_{k}^{r\mathrm{b}} - \nu_{k}^{r\mathrm{f}} \right) \tau_{r},
\label{KTOmegak}
\end{equation}
where $\tau_{r}$ is the rate of progress of the $r^{\text{th}}$ reaction, which can be written
\begin{equation}
\tau_{r} = \ChemK_{r}^{\mathrm{f}} \prod_{l \in \Species} \density_{l}^{\nu_{l}^{r\mathrm{f}}} - \ChemK_{r}^{\mathrm{b}} \prod_{k \in \Species} \density_{k}^{\nu_{k}^{r\mathrm{b}}},
\label{KTRateof progress}
\end{equation}
where the forward and backward constant rates associated with the $r^{\text{th}}$ reaction read
\begin{align}
\ChemK_{r}^{\mathrm{f}} & = \frac{1}{\prod_{l \in \Species} Q_{l}^{\nu_{l}^{r\mathrm{f}}}} \sum\limits_{\textsc{f}^{r}, \textsc{b}^{r}} \int \frac{1}{\prod\limits_{l \in \Forward^{r}} \beta_{l \textsc{l}}} \exp{ \bigg( - \sum_{l \in \Forward^{r}} \frac{\frac{1}{2} \Mass_{l} \boldsymbol{C}_{l} \cdot \boldsymbol{C}_{l} + E_{l \textsc{l}}}{k_{\textsc{b}} \Temperature_{l}} \bigg)} \\
 & \qquad \qquad \qquad \qquad \times \TransitionProbabilities_{\Forward^{r} \Backward^{r}}^{\textsc{f}^{r} \textsc{b}^{r}} \, \prod_{l \in \Forward^{r}} \mathrm{d} \boldsymbol{C}_{l} \prod_{k \in \Backward^{r}} \mathrm{d} \boldsymbol{C}_{k}, \nonumber \\
\ChemK_{r}^{\mathrm{b}} & = \frac{1}{\prod_{k \in \Species} Q_{k}^{\nu_{k}^{r\mathrm{b}}}} \sum\limits_{ \textsc{f}^{r}, \textsc{b}^{r}} \int \frac{1}{\prod\limits_{l \in \Forward^{r}} \beta_{l \textsc{l}}} \exp{ \bigg( - \sum_{k \in \Backward^{r}} \frac{\frac{1}{2} \Mass_{k} \boldsymbol{C}_{k} \cdot \boldsymbol{C}_{k} + E_{k \textsc{k}}}{k_{\textsc{b}} \Temperature_{k}} \bigg)} \\
& \qquad \qquad \qquad \qquad \times \TransitionProbabilities_{\Forward^{r} \Backward^{r}}^{\textsc{f}^{r} \textsc{b}^{r}} \, \prod_{l \in \Forward^{r}} \mathrm{d} \boldsymbol{C}_{l} \prod_{k \in \Backward^{r}} \mathrm{d} \boldsymbol{C}_{k}. \nonumber
\end{align}
From equation \eqref{KTMassStoichio}, the production rates $\mathfrak{w}_{k}$, $k \in \Species$ satisfy the mass conservation constraint
\begin{equation}
\sum_{k \in \Species} \Mass_{k} \mathfrak{w}_{k} = 0.
\label{KTMassConservationChemistry}
\end{equation}

We now distinguish between two cases. First, if the reaction does not involve any electron, one retrieves the law of mass action. Indeed, in that case, all the species temperatures equal $\Temperature_{\heavy}$, and the following relation for conservation of energy holds:
\begin{equation}
\sum_{l \in \Forward^{r}} \Big( \frac{1}{2}\Mass_{l} \boldsymbol{C}_{l} \cdot \boldsymbol{C}_{l} + E_{l \textsc{l}} \Big) = \sum_{k \in \Backward^{r}} \Big( \frac{1}{2}\Mass_{k} \boldsymbol{C}_{k} \cdot \boldsymbol{C}_{k} + E_{k \textsc{k}} \Big),
\end{equation}
so that $\ChemK_{r}^{\mathrm{f}}$ and $\ChemK_{r}^{\mathrm{b}}$ read
\begin{align}
\ChemK_{r}^{\mathrm{f}} & = \frac{\ChemK_{r}}{\prod_{j \in \Heavy} Q_{j}^{\nu_{j}^{r\mathrm{f}}}} \\
\ChemK_{r}^{\mathrm{b}} & = \frac{\ChemK_{r}}{\prod_{j \in \Heavy} Q_{j}^{\nu_{j}^{r\mathrm{b}}}},
\end{align}
where
\begin{align}
\ChemK_{r} & = \frac{1}{\prod\limits_{l \in \Forward^{r}} \beta_{l \textsc{l}}} \sum\limits_{ \textsc{f}^{r} , \textsc{b}^{r}} \int \exp{ \bigg( - \sum_{l \in \Forward^{r}} \frac{\frac{1}{2} \Mass_{l} \boldsymbol{C}_{l} \cdot \boldsymbol{C}_{l} + E_{l \textsc{l}}}{k_{\textsc{b}} \Temperature_{\heavy}} \bigg)} \\
 & \qquad \qquad \qquad \qquad \times \TransitionProbabilities_{\Forward^{r} \Backward^{r}}^{\textsc{f}^{r} \textsc{b}^{r}} \, \prod_{l \in \Forward^{r}} \mathrm{d} \boldsymbol{C}_{l} \prod_{k \in \Backward^{r}} \mathrm{d} \boldsymbol{C}_{k}, \nonumber \\
 & = \frac{1}{\prod\limits_{l \in \Forward^{r}} \beta_{l \textsc{l}}} \sum\limits_{ \textsc{f}^{r}, \textsc{b}^{r}} \int \exp{ \bigg( - \sum_{k \in \Backward^{r}} \frac{\frac{1}{2} \Mass_{k} \boldsymbol{C}_{k} \cdot \boldsymbol{C}_{k} + E_{k \textsc{k}}}{k_{\textsc{b}} \Temperature_{\heavy}} \bigg)} \nonumber \\
& \qquad \qquad \qquad \qquad \times \TransitionProbabilities_{\Forward^{r} \Backward^{r}}^{\textsc{f}^{r} \textsc{b}^{r}} \, \prod_{l \in \Forward^{r}} \mathrm{d} \boldsymbol{C}_{l} \prod_{k \in \Backward^{r}} \mathrm{d} \boldsymbol{C}_{k}. \nonumber
\end{align}
If we introduce the equilibrium constant of the $r^{\text{th}}$ reaction
\begin{equation}
\ChemK_{r}^{\mathrm{e}} = \prod_{j \in \Heavy} Q_{j}^{\nu_{j}^{r\mathrm{b}}- \nu_{j}^{r\mathrm{f}}},
\label{KTEquilibriumConstant}
\end{equation}
we then have finally
\begin{equation}
\ChemK_{r}^{\mathrm{e}} = \frac{\ChemK_{r}^{\mathrm{f}}}{\ChemK_{r}^{\mathrm{b}}}.
\label{KTMassAction}
\end{equation}

On the other hand, when the reaction $r$ involves one or more electrons, the equilibrium constant $\ChemK_{r}^{\mathrm{e}}$ is undefined, since in general $\Temperature_{e} \neq \Temperature_{\heavy}$. However, in specific cases, depending on the form of the reaction considered, e.g., electron impact ionization or ion impact ionization, a generalized law of mass action can be derived where the equilibrium constant might depend on either of the temperatures $\Temperature_{e}$ or $\Temperature_{\heavy}$. For example, for an electron impact ionization in the form
\begin{equation}
\Molecule_{e} + \Molecule_{n} \rightleftharpoons \Molecule_{i} + \Molecule_{e} + \Molecule_{e}, \label{KTIonizationReaction}
\end{equation}
where the subscripts $e$, $n$, and $i$ refer to the electron, the neutral species and the corresponding positive ion, respectively, the law of mass action obtained for the monoatomic case in \cite{GrailleMaginMassot2008} is a generalization of the well-known \og Saha \fg{} equation, and reads
\begin{equation}
\ChemK_{r}^{\mathrm{e}} = \frac{\ChemK_{r}^{\mathrm{f}}}{\ChemK_{r}^{\mathrm{b}}} = \left( \frac{\Mass_{i}}{\Mass_{n}} \right)^{\frac{3}{2}} \left( \frac{2 \pi \Mass_{e} k_{\textsc{b}} \Temperature_{e}}{h_{\textsc{p}}^{2}} \right)^{\frac{3}{2}} \exp{\left( - \frac{\Delta \Energy_{\text{ioniz}}}{k_{\textsc{b}} \Temperature_{e}} \right)},
\label{KTSaha}
\end{equation}
where $\Delta \Energy_{\text{ioniz}}$ is the ionization energy associated with reaction \eqref{KTIonizationReaction}. The first two factors of equation \eqref{KTSaha} correspond to the translational partition functions of the respective species at their respective temperatures, and the term in the exponential factor corresponds to the ionization energy divided by the electron temperature. Indeed, the colliding electron is providing the energy required for the ionization to occur \cite{GrailleMaginMassot2008}. Equation \eqref{KTSaha} can also be derived from non-equilibrium thermodynamics \cite{vandeSanden} \cite{GiordanoCapitelli2002}.

Finally, the energy exchange term $\Delta E_{e \heavy}^{1, \text{chem}} = - \Delta E_{\heavy e}^{1, \text{chem}}$ associated with chemically reactive collisions can be decomposed in the form
\begin{equation}
\Delta E_{e \heavy}^{1, \text{chem}} = \sum_{r \in \Reaction} \Delta E_{r} = \sum_{r \in \Reaction_{e}} \Delta E_{r},
\end{equation}
where the sum has been reduced to the set of reactions involving one or more electrons $\Reaction_{e}$. Upon introducing the net average energy $\Delta \Energy_{e r}$ gained by electrons during reaction $r$, defined as
\begin{align}
\Delta \Energy_{e r} = {} & \frac{1}{\tau_{r}} \sum\limits_{ \textsc{f}^{r}, \textsc{b}^{r}} \int \big(\nu_{e}^{r\mathrm{b}}\frac{1}{2} \Mass_{e} \boldsymbol{C}_{e}^{\mathrm{b}} \cdot \boldsymbol{C}_{e}^{\mathrm{b}} - \nu_{e}^{r\mathrm{f}} \frac{1}{2} \Mass_{e} \boldsymbol{C}_{e}^{\mathrm{f}} \cdot \boldsymbol{C}_{e}^{\mathrm{f}}\big) \\
 & \times \big( \prod\limits_{j \in \Forward^{r}} f_{j} ~ - ~ \prod_{k \in \Backward^{r}} f_{k} ~ \frac{\prod\limits_{k \in \Backward^{r}} \beta_{k \textsc{k}}}{\prod\limits_{j \in \Forward^{r}} \beta_{j \textsc{j}}} \big) \TransitionProbabilities_{\Forward^{r} \Backward^{r}}^{\textsc{f}^{r} \textsc{b}^{r}} \, \prod_{j \in \Forward^{r}} \mathrm{d} \boldsymbol{C}_{j} \prod_{k \in \Backward^{r}} \mathrm{d} \boldsymbol{C}_{k},
\nonumber \end{align}
where $\tau_{r}$ is the rate of progress of the $r^{\text{th}}$ reaction, and where $\boldsymbol{C}_{e}^{\mathrm{f}}$, $\boldsymbol{C}_{e}^{\mathrm{b}}$ denote the electron velocities, taken as a reactant or a product respectively. The chemistry energy exchange term then reads
\begin{equation}
\Delta E_{e \heavy}^{1, \text{chem}} = \sum_{r \in \Reaction_{e}} \Delta \Energy_{e r} \tau_{r}.
\label{KTDeltaEeh1chem}
\end{equation}
As a first approximation, due to the strong mass disparity between electrons and heavy species, the net energy $\nu_{e}^{r\mathrm{f}}\frac{1}{2} \Mass_{e} \boldsymbol{C}_{e}^{\mathrm{f}} \cdot \boldsymbol{C}_{e}^{\mathrm{f}} - \nu_{e}^{r\mathrm{b}} \frac{1}{2} \Mass_{e} \boldsymbol{C}_{e}^{\mathrm{b}} \cdot \boldsymbol{C}_{e}^{\mathrm{b}}$ lost by electrons during the $r^{\text{th}}$ electron collision reaction can be taken constant, independent of the velocities $\boldsymbol{C}_{e}^{\mathrm{f}}$, $\boldsymbol{C}_{e}^{\mathrm{b}}$, and equal to the threshold energy of the collision process considered \cite{GrailleMaginMassot2008}, so that the net average energy lost $-\Delta \Energy_{e r}$ can be identified with the threshold energy of the $r^{\text{th}}$ reaction. This assumption is customary in practical applications \cite{NienhuisPhD} \cite{KalacheNovikova2004} \cite{BolsigPlusDocumentation}.

\section{Transport Fluxes}
\label{SecKTTransCoef}

In this section, we derive an expression for the transport fluxes, namely the diffusion velocities $\boldsymbol{\MeanV}_{i}$, $i \in \Heavy$, $\boldsymbol{\MeanV}_{e}^{0}$, $\boldsymbol{\MeanV}_{e}^{1}$, the viscous tensors $\boldsymbol{\Pi}_{\heavy}$, $\boldsymbol{\Pi}_{e}$, and the heat fluxes $\boldsymbol{\HeatFlux}_{\heavy}$,  $\boldsymbol{\HeatFlux}_{e}^{0}$, $\boldsymbol{\HeatFlux}_{e}^{1}$, in terms of macroscopic variable gradients. These expressions involve transport coefficients, which are also stated in terms of bracket products of the perturbed distribution functions. For the sake of simplicity, we assume that the plasma is weakly magnetized, i.e., $b = 0$.

\subsection{Electron transport coefficients}
In the case when $b = 0$, equation \eqref{KTFirstOrderElectronPerturbationEquation} for the first-order electron perturbation $\phi_{e}$ becomes
\begin{equation}
\mathcal{F}_{e}(\phi_{e}) = \Psi_{e},
\label{KTFirstOrderElectronPerturbationEquationWeakB}
\end{equation}
where
\begin{equation}
\Psi_{e} = - \Streaming_{e}^{-1}(\ln{f_{e}^{0}}),
\end{equation}
under the constraints
\begin{equation}
\llangle \phi_{e} f_{e}^{0}, \psi_{e}^{l} \rrangle_{e} = 0, \quad l \in \left\lbrace e, n^{s}+4 \right\rbrace.
\end{equation}
Given expression \eqref{KTMaxwellElectron} for $f_{e}^{0}$, $\Psi_{e}$ can be decomposed into
\begin{equation}
\Psi_{e} = - \mathbf{\Psi}_{e}^{D_{e}} \cdot \left( \boldsymbol{\partial_{x}} \pressure_{e} - \density_{e} q_{e} \boldsymbol{E} \right) - \mathbf{\Psi}_{e}^{\hat{\lambda}_{e}} \cdot \boldsymbol{\partial_{x}} \left( \frac{1}{k_{\textsc{b}}\Temperature_{e}} \right),
\end{equation}
where
\begin{align}
\mathbf{\Psi}_{e}^{D_{e}} & = \frac{1}{\pressure_{e}} \boldsymbol{C}_{e} \vphantom{\left( \frac{5}{2}k_{\textsc{b}} \Temperature_{e} - \frac{1}{2} \Mass_{e} \boldsymbol{C}_{e} \cdot \boldsymbol{C}_{e} \right)}, \\
\mathbf{\Psi}_{e}^{\hat{\lambda}_{e}} & = \left( \frac{5}{2}k_{\textsc{b}} \Temperature_{e} - \frac{1}{2} \Mass_{e} \boldsymbol{C}_{e} \cdot \boldsymbol{C}_{e} \right) \boldsymbol{C}_{e}.
\end{align}
Making use of the linearity of the electron linearized collision operator $\mathcal{F}_{e}$ \cite{GiovangigliGraille2009} \cite{GrailleMaginMassot2009}, we can expand the first-order perturbation function $\phi_{e}$ as follows:
\begin{equation}
\phi_{e} = - \mathbf{\Phi}_{e}^{D_{e}} \cdot \left( \boldsymbol{\partial_{x}} \pressure_{e} - \density_{e} q_{e} \boldsymbol{E} \right) - \mathbf{\Phi}_{e}^{\hat{\lambda}_{e}} \cdot \boldsymbol{\partial_{x}} \left( \frac{1}{k_{\textsc{b}}\Temperature_{e}} \right).
\end{equation}
For each $\mu = D_{e}, \hat{\lambda}_{e}$, the function $\mathbf{\Phi}_{e}^{\mu}$ is solution to the system of equations
\begin{align}
& \mathcal{F}_{e}(\mathbf{\Phi}_{e}^{\mu}) = \mathbf{\Psi}_{e}^{\mu}, \\
&  \llangle f_{e}^{0} \mathbf{\Phi}_{e}^{\mu}, \psi_{e}^{l} \rrangle_{e} = 0, \qquad l \in \left\lbrace e, n^{s}+4 \right\rbrace,
\end{align}
which is well posed since for both values of $\mu$
\begin{equation}
\llangle f_{e}^{0} \mathbf{\Psi}_{e}^{\mu}, \psi_{e}^{l} \rrangle_{e} = 0, \qquad l \in \left\lbrace e, n^{s}+4 \right\rbrace.
\end{equation}
Additionally, because of the linearity and the space isotropy of $\mathcal{F}_{e}$, $\mathbf{\Phi}_{e}^{D_{e}}$ and $\mathbf{\Phi}_{e}^{\hat{\lambda}_{e}}$ can be taken in the form
\begin{align}
\displaystyle \mathbf{\Phi}_{e}^{D_{e}} = \phi_{e}^{D_{e}} ~ \boldsymbol{C}_{e}, \\
\displaystyle \mathbf{\Phi}_{e}^{\hat{\lambda}_{e}} = \phi_{e}^{\hat{\lambda}_{e}} ~ \boldsymbol{C}_{e},
\end{align}
where $\phi_{e}^{D_{e}}$, $\phi_{e}^{\hat{\lambda}_{e}}$ are scalar isotropic functions of $\boldsymbol{C}_{e} \cdot \boldsymbol{C}_{e}$.

Thanks to this decomposition of $\phi_{e}$, the electron viscous tensor $\boldsymbol{\Pi}_{e}$ can be shown to vanish:
\begin{equation}
\boldsymbol{\Pi}_{e} = 0.
\end{equation}

Defining the zeroth-order electron self-diffusion coefficient $D_{ee}^{0}$ and the zeroth-order electron electron-temperature thermal diffusion coefficient $\theta_{ee}^{0}$ by
\begin{align}
D_{ee}^{0} & = \frac{\pressure k_{\textsc{b}}\Temperature_{e}}{3} \llbracket \mathbf{\Phi}_{e}^{D_{e}}, \mathbf{\Phi}_{e}^{D_{e}} \rrbracket_{e}, \\
\theta_{ee}^{0} & = - \frac{1}{3} \llbracket \mathbf{\Phi}_{e}^{\hat{\lambda}_{e}}, \mathbf{\Phi}_{e}^{D_{e}} \rrbracket_{e},
\end{align}
respectively, the zeroth-order electron diffusion velocity is expressed in the form
\begin{equation}
\boldsymbol{\MeanV}_{e}^{0} = - D_{ee}^{0} \boldsymbol{\hat{d}}_{e} - \theta_{ee}^{0} \boldsymbol{\partial_{x}} \ln{\Temperature_{e}},
\label{KTFirstOrderElectronDiffusionVelocity}
\end{equation}
where we have introduced the unconstrained electron diffusion driving force
\begin{equation}
\boldsymbol{\hat{d}}_{e} = \frac{1}{\pressure} \left( \boldsymbol{\partial_{x}} \pressure_{e} - \density_{e} q_{e} \boldsymbol{E} \right).
\end{equation}
Finally, upon defining the zeroth-order electron self-partial-thermal-conductivity
\begin{equation}
\hat{\lambda}_{ee}^{0} = \frac{1}{3 k_{\textsc{b}} \Temperature_{e}} \llbracket \mathbf{\Phi}_{e}^{\hat{\lambda}_{e}}, \mathbf{\Phi}_{e}^{\hat{\lambda}_{e}} \rrbracket_{e},
\end{equation}
the zeroth-order electron heat flux $\boldsymbol{\HeatFlux}_{e}^{0}$ is found in the form
\begin{equation}
\boldsymbol{\HeatFlux}_{e}^{0} = - \pressure \theta_{ee}^{0} \boldsymbol{\hat{d}}_{e} - \hat{\lambda}_{ee}^{0} \boldsymbol{\partial_{x}} \ln{\Temperature_{e}} + \Big( \frac{5}{2}k_{\textsc{b}}\Temperature_{e} \Big) \density_{e} \boldsymbol{\MeanV}_{e}^{0}.
\label{KTFirstOrderElectronHeatFlux}
\end{equation}

\subsection{Heavy-species transport coefficients}
We recall that from \eqref{KTFirstOrderHeavyPerturbationEquation} and \eqref{KTFirstOrderHeavyPerturbationConstraints} the first-order heavy-species perturbation function $\phi_{\heavy}$ is solution to the following constrained linear system of integral equations:
\begin{align}
& \mathcal{F}_{i}(\phi_{\heavy}) = \Psi_{i}, \qquad i \in \Heavy,
\label{KTHeavyCVDLinearizedBoltzmann} \\
&  \llangle f_{\heavy}^{0} \phi_{\heavy}, \psi_{\heavy}^{j} \rrangle_{\heavy} = 0, \qquad j \in \Heavy \cup \left\lbrace n^{s}+1, \ldots , n^{s}+4 \right\rbrace,
\label{KTHeavyCVDLinearizedBoltzmannConstraint}
\end{align}
where $\Psi_{i}$, $i \in \Heavy$, is given by \eqref{KTPsiHeavy}.
After a lengthy calculation, $\Psi_{i}$ can be decomposed into
\begin{align}
\Psi_{i} = & - \mathbf{\Psi}_{i}^{\eta_{\heavy}} : \boldsymbol{\partial_{x}} \boldsymbol{\Meanv}_{\heavy} - \frac{1}{3} \Psi_{i}^{\kappa_{\heavy}} ~ (\boldsymbol{\partial_{x}} \cdot \boldsymbol{\Meanv}_{\heavy}) - \sum_{j \in \Heavy} \mathbf{\Psi}_{i}^{D_{j}} \cdot \left( \boldsymbol{\partial_{x}} \pressure_{j} - \density_{j} q_{j} \boldsymbol{E}  \right) - \mathbf{\Psi}_{i}^{D_{e}} \cdot \left( \boldsymbol{\partial_{x}} \pressure_{e} - \density_{e} q_{e} \boldsymbol{E}  \right) 
\nonumber \\
 & - \mathbf{\Psi}_{i}^{\hat{\lambda}_{\heavy}} \cdot \boldsymbol{\partial_{x}} \left( \frac{1}{k_{\textsc{b}} \Temperature_{\heavy}} \right) - \mathbf{\Psi}_{i}^{\hat{\lambda}_{e}} \cdot \boldsymbol{\partial_{x}} \left( \frac{1}{k_{\textsc{b}} \Temperature_{e}} \right) - \Psi_{i}^{\Theta} \left( \Temperature_{e} - \Temperature_{\heavy} \right), \quad i \in \Heavy.
\end{align}
The expansion coefficients are given by
\begin{align}
\mathbf{\Psi}_{i}^{\eta_{\heavy}} & = \frac{\Mass_{i}}{k_{\textsc{b}} \Temperature_{\heavy}} \left[ \boldsymbol{C}_{i} \otimes \boldsymbol{C}_{i} - \frac{1}{3} \boldsymbol{C}_{i} \cdot \boldsymbol{C}_{i} \, \mathbb{I} \right] \vphantom{\sum_{j \in \Heavy} \left[ \frac{1}{2}] \right]}
\label{KTPsiiEtah} \\
\Psi_{i}^{\kappa_{\heavy}} & = \frac{2 c^{\text{int}}}{c_{v}k_{\textsc{b}} \Temperature_{\heavy}} \left[ \frac{1}{2} \Mass_{i} \boldsymbol{C}_{i} \cdot \boldsymbol{C}_{i} - \frac{3}{2} k_{\textsc{b}} \Temperature_{\heavy} \right] + \frac{2 c_{v}^{\text{tr}}}{c_{v}k_{\textsc{b}} \Temperature_{\heavy}} (\overline{E}_{i} - E_{i \textsc{i}}) \vphantom{\sum_{j \in \Heavy} \left[ \frac{1}{2}] \right]}
\label{KTPsiiKappah} \\
\mathbf{\Psi}_{i}^{D_{j}} & = \frac{1}{\pressure_{i}} \left[ \delta_{ij} - Y^{\heavy}_{i} \right] \boldsymbol{C}_{i} \vphantom{\sum_{j \in \Heavy} \left[ \frac{1}{2}] \right]}
\label{KTPsiiDj} \\
\mathbf{\Psi}_{i}^{D_{e}} & = \frac{\Mass_{e}}{3} \sum_{j \in \Heavy} \sum_{\textsc{j} \in \QuantumSpace_{j}} \density_{j} \frac{1}{\pressure_{i}} \left( \delta_{ij} \delta_{\textsc{i} \textsc{j}} - Y^{\heavy}_{i} \frac{a_{j \textsc{j}} e^{- \epsilon_{j \textsc{j}}}}{Q_{j}^{\text{int}}} \right) \boldsymbol{C}_{i} \vphantom{\sum_{j \in \Heavy} \left[ \frac{1}{2}] \right]}
\label{KTPsiiDe} \\
 & \quad \times \int \Sigma_{\textsc{j} \textsc{j}}^{(1)} (|\boldsymbol{C}_{e}|^{2}) |\boldsymbol{C}_{e}| f_{e}^{0}(\boldsymbol{C}_{e}) ~ \boldsymbol{C}_{e} \cdot \mathbf{\Phi}_{e}^{D_{e}} \, \mathrm{d} \boldsymbol{C}_{e}
\nonumber \vphantom{\sum_{j \in \Heavy} \left[ \frac{1}{2}] \right]} \\
\mathbf{\Psi}_{i}^{\hat{\lambda}_{e}} & = \frac{\Mass_{e}}{3} \sum_{j \in \Heavy} \sum_{\textsc{j} \in \QuantumSpace_{j}} \density_{j} \frac{1}{\pressure_{i}} \left( \delta_{ij} \delta_{\textsc{i} \textsc{j}} - Y^{\heavy}_{i} \frac{a_{j \textsc{j}} e^{- \epsilon_{j \textsc{j}}}}{Q_{j}^{\text{int}}} \right) \boldsymbol{C}_{i}
\label{KTPsiiLambdahate} \\ 
 & \quad \times \int \Sigma_{\textsc{j} \textsc{j}}^{(1)} (|\boldsymbol{C}_{e}|^{2}) |\boldsymbol{C}_{e}| f_{e}^{0}(\boldsymbol{C}_{e}) ~ \boldsymbol{C}_{e}\cdot \mathbf{\Phi}_{e}^{\hat{\lambda}_{e}} \, \mathrm{d} \boldsymbol{C}_{e} \vphantom{\sum_{j \in \Heavy} \left[ \frac{1}{2}] \right]} 
\nonumber \vphantom{\sum_{j \in \Heavy} \left[ \frac{1}{2}] \right]} \\
\mathbf{\Psi}_{i}^{\hat{\lambda}_{\heavy}} & = \left( \frac{5}{2}k_{\textsc{b}} \Temperature_{\heavy} - \frac{1}{2}\Mass_{i} \boldsymbol{C}_{i} \cdot \boldsymbol{C}_{i} + \overline{E}_{i} - E_{i \textsc{i}} \right) \boldsymbol{C}_{i} \vphantom{\sum_{j \in \Heavy} \left[ \frac{1}{2}] \right]}
\label{KTPsiiLambdahath} \\
\Psi_{i}^{\Theta} & = \frac{3}{2} \frac{1}{\tau} \frac{1}{c_{v} \Temperature_{\heavy}^{2}} \left( \frac{1}{2}\Mass_{i} \boldsymbol{C}_{i} \cdot \boldsymbol{C}_{i} + E_{i \textsc{i}} - \overline{E}_{i} - \frac{3}{2}k_{\textsc{b}} \Temperature_{\heavy} \right) \vphantom{\sum_{j \in \Heavy} \left[ \frac{1}{2}] \right]}
\label{KTPsiiTheta} \\
 & + \frac{2}{3} \frac{\Mass_{e}}{\Mass_{i}} \frac{1}{k_{\textsc{b}} \Temperature_{\heavy}^{2}} \left( \frac{3}{2}k_{\textsc{b}} \Temperature_{\heavy} - \frac{1}{2}\Mass_{i} \boldsymbol{C}_{i} \cdot \boldsymbol{C}_{i} \right) \nu_{ie}^{\textsc{i} \textsc{i}} + \frac{1}{k_{\textsc{b}} \Temperature_{e} \Temperature_{\heavy}} \sum\limits_{\substack{ \textsc{i}' \in \QuantumSpace_{i} \\ \textsc{i}' \neq \textsc{i} }} \Delta E_{\textsc{i} \textsc{i}'} ~ g_{\textsc{i} \textsc{i}'} \nu_{ie}^{\textsc{i} \textsc{i}'}, \nonumber
\end{align}
where $Y^{\heavy}_{i}$ is the mass fraction of the $i^{\text{th}}$ species with respect to the heavy species, which is proportional to the mass fraction $Y_{i} = \frac{\rho_{i}}{\rho}$ of the $i^{\text{th}}$ species with respect to the whole mixture:
\begin{equation}
Y^{\heavy}_{i} = \frac{\rho_{i}}{\rho_{\heavy}} = \frac{\rho}{\rho_{\heavy}} Y_{i}, \qquad i \in \Heavy,
\end{equation}
and where
\begin{equation}
g_{\textsc{i} \textsc{i}'} = g \Big(\Delta \epsilon_{\textsc{i} \textsc{i}'} \big( 1 - \frac{\Temperature_{\heavy}}{\Temperature_{e}} \big) \Big).
\end{equation}
We also denote by $X^{\heavy}_{i}$ the mole fraction of the $i^{\text{th}}$ species with respect to the heavy species. If $\MoleMass_{i}$ is the molar mass of the $i^{\text{th}}$ heavy species, and $\overline{\MoleMass}_{\heavy}$ is the mean molar mass of the heavy species, given by \eqref{KTHeavyMeanMoleMass}, then
\begin{equation}
X^{\heavy}_{i} = \frac{\density_{i}}{\density_{\heavy}} = \frac{\overline{\MoleMass}_{\heavy}}{\MoleMass_{i}} Y^{\heavy}_{i}, \qquad i \in \Heavy.
\end{equation}

In \eqref{KTPsiiEtah}-\eqref{KTPsiiTheta}, the symbol $c_{v}^{\text{tr}}$ denotes the translational constant-volume specific heat per molecule, $c^{\text{int}}$ the heavy-species internal specific heat per molecule, $c_{v}$ the heavy-species constant-volume specific heat per molecule
\begin{equation}
\left\lbrace
\begin{array}{l}
\displaystyle c_{v}^{\text{tr}} = \frac{3}{2} k_{\textsc{b}} \vphantom{\sum_{i \in \Heavy}} \\
\displaystyle c^{\text{int}} = \sum_{i \in \Heavy} X^{\heavy}_{i} c_{i}^{\text{int}} \\
\displaystyle c_{v} = c_{v}^{\text{tr}} + c^{\text{int}}, \vphantom{\sum_{i \in \Heavy}}
\end{array}
\right.
\label{KTSpecificHeatConstantVolume}
\end{equation}
where $c_{i}^{\text{int}}$ denotes the internal heat capacity of the $i^{\text{th}}$ species
\begin{equation}
c_{i}^{\text{int}} = \frac{d \overline{E}_{i}}{d \Temperature}, \qquad i \in \Heavy.
\end{equation}

As for the first-order electron perturbation, making use of the linearity of the linearized collision operator, the perturbation functions $\phi_{i}, i \in \Heavy$ can be decomposed in the form
\begin{align}
\phi_{i} = & - \mathbf{\Phi}_{i}^{\eta_{\heavy}} : \boldsymbol{\partial_{x}} \boldsymbol{\Meanv}_{\heavy} - \frac{1}{3} \phi_{i}^{\kappa_{\heavy}} ~ (\boldsymbol{\partial_{x}} \cdot \boldsymbol{\Meanv}_{\heavy}) - \sum_{j \in \Heavy} \mathbf{\Phi}_{i}^{D_{j}} \cdot \left( \boldsymbol{\partial_{x}} \pressure_{j} - \density_{j} q_{j} \boldsymbol{E}  \right) - \mathbf{\Phi}_{i}^{D_{e}} \cdot \left( \boldsymbol{\partial_{x}} \pressure_{e} - \density_{e} q_{e} \boldsymbol{E}  \right)
\nonumber \\
& - \mathbf{\Phi}_{i}^{\hat{\lambda}_{\heavy}} \cdot \boldsymbol{\partial_{x}} \left( \frac{1}{k_{\textsc{b}} \Temperature_{\heavy}} \right) - \mathbf{\Phi}_{i}^{\hat{\lambda}_{e}} \cdot \boldsymbol{\partial_{x}} \left( \frac{1}{k_{\textsc{b}} \Temperature_{e}} \right) - \phi_{i}^{\Theta} \left( \Temperature_{e} - \Temperature_{\heavy} \right), \qquad i \in \Heavy.
\end{align}
Note that one can also expand the coefficient $\Psi_{i}^{\Theta}$ as
\begin{equation}
\Psi_{i}^{\Theta} = - \frac{1}{2 \Temperature_{\heavy}} \frac{1}{\tau} \Psi_{i}^{\kappa_{\heavy}} + \Psi_{i}^{\Theta^{\text{el}}} + \Psi_{i}^{\Theta^{\text{in}}},
\end{equation}
where the elastic and inelastic contributions are given by
\begin{align}
\Psi_{i}^{\Theta^{\text{el}}} & = \frac{1}{k_{\textsc{b}} \Temperature_{\heavy}^{2}} \left( \frac{1}{\tau^{\text{el}}} - \frac{2}{3} \frac{\Mass_{e}}{\Mass_{i}} \nu_{ie}^{\textsc{i} \textsc{i}} \right) \left( \frac{1}{2} \Mass_{i} \boldsymbol{C}_{i} \cdot \boldsymbol{C}_{i} - \frac{3}{2} k_{\textsc{b}} \Temperature_{\heavy} \right) \\
\Psi_{i}^{\Theta^{\text{in}}} & = \frac{1}{k_{\textsc{b}} \Temperature_{\heavy}^{2}} \frac{1}{\tau^{\text{in}}} \left( \frac{1}{2} \Mass_{i} \boldsymbol{C}_{i} \cdot \boldsymbol{C}_{i} - \frac{3}{2} k_{\textsc{b}} \Temperature_{\heavy} \right) + \frac{1}{k_{\textsc{b}} \Temperature_{e} \Temperature_{\heavy}} \sum_{\textsc{i}' \in \QuantumSpace_{i}} \Delta E_{\textsc{i} \textsc{i}'} g_{\textsc{i} \textsc{i}'} \nu_{ie}^{\textsc{i} \textsc{i}'},
\end{align}
and the corresponding decomposition for $\phi_{i}^{\Theta}$ reads
\begin{equation}
\phi_{i}^{\Theta} = - \frac{1}{2 \Temperature_{\heavy}} \frac{1}{\tau} \phi_{i}^{\kappa_{\heavy}} + \phi_{i}^{\Theta^{\text{el}}} + \phi_{i}^{\Theta^{\text{in}}}, \qquad i \in \Heavy.
\end{equation}
For each value of $\mu = \eta_{\heavy}, \kappa_{\heavy}, D_{j}, j \in \Heavy, D_{e}, \hat{\lambda}_{\heavy}, \hat{\lambda}_{e}, \Theta, \Theta^{\text{el}}, \Theta^{\text{in}}$, the functional $\phi_{\heavy}^{\mu}$ is solution of the following constrained linear system of integral equations:
\begin{align}
& \mathcal{F}_{i}(\phi_{\heavy}^{\mu}) = \Psi_{i}^{\mu}, \qquad i \in \Heavy \\
&  \llangle f_{\heavy}^{0} \phi_{\heavy}^{\mu}, \psi_{\heavy}^{j} \rrangle_{\heavy} = 0, \qquad j \in \Heavy \cup \left\lbrace n^{s}+1, \ldots , n^{s}+4 \right\rbrace, \label{KTPhiOrtho}
\end{align}
which is well posed since
\begin{equation}
\llangle f_{\heavy}^{0} \Psi_{\heavy}^{\mu}, \psi_{\heavy}^{j} \rrangle_{\heavy} = 0, \qquad j \in \Heavy \cup \left\lbrace n^{s}+1, \ldots , n^{s}+4 \right\rbrace.
\end{equation}
Furthermore, because of the isotropy of the linearized collision operator $\mathcal{F}_{\heavy}$, each $\phi_{\heavy}^{\mu}$ is of the same tensorial type as $\Psi_{\heavy}^{\mu}$ \cite{GraillePhD}.

Upon defining the heavy-species diffusion coefficients, the heavy-species electron diffusion coefficients, the heavy-species heavy-temperature thermal diffusion coefficients, and the heavy-species electron-temperature thermal diffusion coefficients by
\begin{align}
D_{i j} & = \frac{\pressure k_{\textsc{b}} \Temperature_{\heavy}}{3} \llbracket \mathbf{\Phi}_{\heavy}^{D_{i}}, \mathbf{\Phi}_{\heavy}^{D_{j}} \rrbracket_{\heavy}, \qquad i \in \Heavy, j \in \Heavy,
\label{KTDij} \\
D_{i e} & = D_{e i} =  \frac{\pressure k_{\textsc{b}} \Temperature_{\heavy}}{3} \llbracket \mathbf{\Phi}_{\heavy}^{D_{i}}, \mathbf{\Phi}_{\heavy}^{D_{e}} \rrbracket_{\heavy}, \qquad i \in \Heavy,
\label{KTDie} \\
\theta_{i \heavy} & = \theta_{\heavy i} = - \frac{1}{3} \llbracket \mathbf{\Phi}_{\heavy}^{D_{i}}, \mathbf{\Phi}_{\heavy}^{\hat{\lambda}_{\heavy}} \rrbracket_{\heavy}, \qquad i \in \Heavy,
\label{KTThetaih} \\
\theta_{i e} & = \theta_{e i} = - \frac{1}{3} \frac{\Temperature_{\heavy}}{\Temperature_{e}} \llbracket \mathbf{\Phi}_{\heavy}^{D_{i}}, \mathbf{\Phi}_{\heavy}^{\hat{\lambda}_{e}} \rrbracket_{\heavy}, \qquad i \in \Heavy,
\label{KTThetaie}
\end{align}
the heavy-species diffusion velocities are expressed in the form
\begin{equation}
\boldsymbol{\MeanV}_{i} = - \sum_{j \in \Heavy} D_{i j} \boldsymbol{\hat{d}}_{j} - D_{ie} \boldsymbol{\hat{d}}_{e} - \theta_{i \heavy} \boldsymbol{\partial_{x}} \ln{\Temperature_{\heavy}} - \theta_{i e} \boldsymbol{\partial_{x}} \ln{\Temperature_{e}}, \quad i \in \Heavy,
\label{KTHeavyDiffusionVelocities}
\end{equation}
where we have introduced the heavy-species diffusion driving forces
\begin{equation}
\boldsymbol{\hat{d}}_{i} = \frac{1}{\pressure} \left( \boldsymbol{\partial_{x}} \pressure_{i} - \density_{i} q_{i} \boldsymbol{E} \right), \quad i \in \Heavy.
\end{equation}
We also define the shear viscosity, the volume viscosity, and the thermal non-equilibrium viscosity, respectively:
\begin{align}
\eta_{\heavy} & = \frac{k_{\textsc{b}} \Temperature_{\heavy}}{10} \llbracket \mathbf{\Phi}_{\heavy}^{\eta_{\heavy}}, \mathbf{\Phi}_{\heavy}^{\eta_{\heavy}} \rrbracket_{\heavy}
\label{KTEtah} \\
\kappa_{\heavy} & = \frac{k_{\textsc{b}} \Temperature_{\heavy}}{9} \llbracket \phi_{\heavy}^{\kappa_{\heavy}}, \phi_{\heavy}^{\kappa_{\heavy}} \rrbracket_{\heavy}
\label{KTKappah} \\
\zeta & = \frac{k_{\textsc{b}} \Temperature_{\heavy}}{3} \llbracket \phi_{\heavy}^{\Theta}, \phi_{\heavy}^{\kappa_{\heavy}} \rrbracket_{\heavy},
\label{KTZeta}
\end{align}
so that the viscous tensor reads
\begin{equation}
\boldsymbol{\Pi}_{\heavy} = - \eta_{\heavy} \, \Big( \boldsymbol{\partial_{x}} \boldsymbol{\Meanv}_{\heavy} + (\boldsymbol{\partial_{x}} \boldsymbol{\Meanv}_{\heavy})^{\textsc{t}} - \frac{2}{3} (\boldsymbol{\partial_{x}} \cdot \boldsymbol{\Meanv}_{\heavy}) \, \mathbb{I} \Big) - \kappa_{\heavy} (\boldsymbol{\partial_{x}} \cdot \boldsymbol{\Meanv}_{\heavy}) \, \mathbb{I} - \zeta (\Temperature_{e}-\Temperature_{\heavy}) \, \mathbb{I}.
\label{KTViscousTensor}
\end{equation}

Finally, if we introduce the heavy-species self-partial-thermal-conductivity, the heavy electron partial thermal conductivity, respectively
\begin{align}
\hat{\lambda}_{\heavy \heavy} & = \frac{1}{3 k_{\textsc{b}} \Temperature_{\heavy}} \llbracket \mathbf{\Phi}_{\heavy}^{\hat{\lambda}_{\heavy}}, \mathbf{\Phi}_{\heavy}^{\hat{\lambda}_{\heavy}} \rrbracket_{\heavy},
\label{KTLambdahathh} \\
\hat{\lambda}_{\heavy e} & = \hat{\lambda}_{e \heavy} = \frac{1}{3 k_{\textsc{b}} \Temperature_{e}} \llbracket \mathbf{\Phi}_{\heavy}^{\hat{\lambda}_{\heavy}}, \mathbf{\Phi}_{\heavy}^{\hat{\lambda}_{e}} \rrbracket_{\heavy},
\label{KTLambdahathe}
\end{align}
and the electron heavy-temperature thermal diffusion coefficient
\begin{equation}
\theta_{\heavy e} = \theta_{e \heavy} = - \frac{1}{3} \llbracket \mathbf{\Phi}_{\heavy}^{\hat{\lambda}_{\heavy}}, \mathbf{\Phi}_{\heavy}^{D_{e}} \rrbracket_{\heavy},
\label{KTThetahe}
\end{equation}
the heavy-species heat flux can be expressed as
\begin{equation}
\boldsymbol{\HeatFlux}_{\heavy} = - \pressure \sum_{j \in \Heavy} \theta_{\heavy j} \boldsymbol{\hat{d}}_{j} - \pressure \theta_{\heavy e} \boldsymbol{\hat{d}}_{e} - \hat{\lambda}_{\heavy \heavy} \boldsymbol{\partial_{x}} \ln{\Temperature_{\heavy}} - \hat{\lambda}_{\heavy e} \boldsymbol{\partial_{x}} \ln{\Temperature_{e}} + \sum_{j \in \Heavy} \big( \frac{5}{2}k_{\textsc{b}} \Temperature_{\heavy} + \overline{E}_{j} \big) \density_{j} \boldsymbol{\MeanV}_{j}.
\label{KTHeavyHeatFlux}
\end{equation}

The second and fourth terms of the respective expressions for the heavy-species diffusion velocities \eqref{KTHeavyDiffusionVelocities} and the heavy-species heat flux \eqref{KTHeavyHeatFlux} are the \og heavy-species Kolesnikov diffusion fluxes \fg{}. They arise from the coupling between heavy species and electrons as first described by Kolesnikov \cite{GrailleMaginMassot2009} \cite{Kolesnikov1974}.

We also introduce the new bracket
\begin{equation}
\llbracket \boldsymbol{\xi}_{\heavy}, \boldsymbol{\zeta}_{e} \rrbracket_{\heavy e} = \sum_{j \in \Heavy} \sum_{\textsc{j} \in \QuantumSpace_{j}} \frac{1}{3} \frac{\Mass_{e}}{k_{\textsc{b}} \Temperature_{e}} \left( \int \boldsymbol{\xi}_{j} \cdot \boldsymbol{C}_{j} ~ f_{j}^{0} \, \mathrm{d} \boldsymbol{C}_{j} \right) \int f_{e}^{0} | \boldsymbol{C}_{e}| \Sigma_{\textsc{j} \textsc{j}}^{(1)} (|\boldsymbol{C}_{e}|^{2}) ~ \boldsymbol{\zeta}_{e} \cdot \boldsymbol{C}_{e} \, \mathrm{d} \boldsymbol{C}_{e},
\end{equation}
which is non trivial when $\boldsymbol{\xi}_{\heavy}$, $\boldsymbol{\zeta}_{e}$ are vectors but would be trivial for scalars or traceless symmetric tensors of rank $2$. The following relations hold:
\begin{align}
D_{i e} & = \frac{\pressure k_{\textsc{b}} \Temperature_{\heavy}}{3} \llbracket \mathbf{\Phi}_{\heavy}^{D_{i}}, \mathbf{\Phi}_{\heavy}^{D_{e}} \rrbracket_{\heavy} = \frac{\pressure k_{\textsc{b}} \Temperature_{e}}{3} \llbracket \mathbf{\Phi}_{\heavy}^{D_{i}}, \mathbf{\Phi}_{e}^{D_{e}} \rrbracket_{\heavy e}, \qquad i \in \Heavy, \\
\theta_{i e} & = - \frac{1}{3} \frac{\Temperature_{\heavy}}{\Temperature_{e}} \llbracket \mathbf{\Phi}_{\heavy}^{D_{i}}, \mathbf{\Phi}_{\heavy}^{\hat{\lambda}_{e}} \rrbracket_{\heavy} = - \frac{1}{3} \llbracket \mathbf{\Phi}_{\heavy}^{D_{i}}, \mathbf{\Phi}_{e}^{\hat{\lambda}_{e}} \rrbracket_{\heavy e}, \qquad i \in \Heavy, \\
\hat{\lambda}_{\heavy e} & = \frac{1}{3 k_{\textsc{b}} \Temperature_{e}} \llbracket \mathbf{\Phi}_{\heavy}^{\hat{\lambda}_{\heavy}}, \mathbf{\Phi}_{\heavy}^{\hat{\lambda}_{e}} \rrbracket_{\heavy} = \frac{1}{3 k_{\textsc{b}} \Temperature_{\heavy}} \llbracket \mathbf{\Phi}_{\heavy}^{\hat{\lambda}_{\heavy}}, \mathbf{\Phi}_{e}^{\hat{\lambda}_{e}} \rrbracket_{\heavy e}, \\
\theta_{\heavy e} & = - \frac{1}{3} \llbracket \mathbf{\Phi}_{\heavy}^{\hat{\lambda}_{\heavy}}, \mathbf{\Phi}_{\heavy}^{D_{e}} \rrbracket_{\heavy} = - \frac{1}{3} \frac{\Temperature_{e}}{\Temperature_{\heavy}} \llbracket \mathbf{\Phi}_{\heavy}^{\hat{\lambda}_{\heavy}}, \mathbf{\Phi}_{e}^{D_{e}} \rrbracket_{\heavy e}.
\end{align}
We only prove the first one of those relations. The other ones can be derived by similar arguments. By definition of $\llbracket \cdot \rrbracket_{\heavy}$ and by \eqref{KTPsiiDe}, $D_{ie}$ reads
\begin{align*}
D_{i e} & = \frac{\pressure k_{\textsc{b}} \Temperature_{\heavy}}{3} \llbracket \mathbf{\Phi}_{\heavy}^{D_{i}}, \mathbf{\Phi}_{\heavy}^{D_{e}} \rrbracket_{\heavy} \vphantom{\sum_{k \in \Heavy} \int} \\
 & = \frac{\pressure k_{\textsc{b}} \Temperature_{\heavy}}{3} \sum_{k \in \Heavy} \sum_{\textsc{k} \in \QuantumSpace_{k}} \int f_{k}^{0} \mathbf{\Phi}_{k}^{D_{i}} \cdot \mathbf{\Psi}_{k}^{D_{e}} \, \mathrm{d} \boldsymbol{C}_{k} \\
& = \frac{\pressure k_{\textsc{b}} \Temperature_{\heavy}}{3} \frac{\Mass_{e}}{3} \sum_{k \in \Heavy} \sum_{\textsc{k} \in \QuantumSpace_{k}} \sum_{j \in \Heavy} \sum_{\textsc{j} \in \QuantumSpace_{j}} \density_{j} \left( \int f_{k}^{0} \mathbf{\Phi}_{k}^{D_{i}} \cdot \frac{1}{\pressure_{k}} \left( \delta_{kj} \delta_{\textsc{k} \textsc{j}} - Y^{\heavy}_{k} \frac{a_{j \textsc{j}} e^{- \epsilon_{j \textsc{j}}}}{Q_{j}^{\text{int}}} \right) \boldsymbol{C}_{k} \, \mathrm{d} \boldsymbol{C}_{k} \right) \\
& \qquad \qquad \times \int \Sigma_{\textsc{j} \textsc{j}}^{(1)} (|\boldsymbol{C}_{e}|^{2}) |\boldsymbol{C}_{e}| f_{e}^{0}(\boldsymbol{C}_{e}) ~ \boldsymbol{C}_{e} \cdot \mathbf{\Phi}_{e}^{D_{e}} \, \mathrm{d} \boldsymbol{C}_{e} \vphantom{\sum_{k \in \Heavy} \int} \\
& = \frac{\pressure k_{\textsc{b}} \Temperature_{\heavy}}{3} \frac{\Mass_{e}}{3} \sum_{j \in \Heavy} \sum_{\textsc{j} \in \QuantumSpace_{j}} \density_{j} \left( \sum_{k \in \Heavy} \sum_{\textsc{k} \in \QuantumSpace_{k}} \int f_{k}^{0} \mathbf{\Phi}_{k}^{D_{i}} \cdot \frac{1}{\pressure_{k}} \left( \delta_{kj} \delta_{\textsc{k} \textsc{j}} - Y^{\heavy}_{k} \frac{a_{j \textsc{j}} e^{- \epsilon_{j \textsc{j}}}}{Q_{j}^{\text{int}}} \right) \boldsymbol{C}_{k} \, \mathrm{d} \boldsymbol{C}_{k} \right) \\
& \qquad \qquad \times \int \Sigma_{\textsc{j} \textsc{j}}^{(1)} (|\boldsymbol{C}_{e}|^{2}) |\boldsymbol{C}_{e}| f_{e}^{0}(\boldsymbol{C}_{e}) ~ \boldsymbol{C}_{e} \cdot \mathbf{\Phi}_{e}^{D_{e}} \, \mathrm{d} \boldsymbol{C}_{e}. \vphantom{\sum_{k \in \Heavy} \int}
\end{align*}
Since
\begin{align*}
\sum_{k \in \Heavy} \sum_{\textsc{k} \in \QuantumSpace_{k}} \frac{Y_{k}}{\pressure_{k}} \frac{a_{j \textsc{j}} e^{- \epsilon_{j \textsc{j}}}}{Q_{j}^{\text{int}}}   \int f_{k}^{0} \mathbf{\Phi}_{k}^{D_{i}} \cdot \boldsymbol{C}_{k} \, \mathrm{d} \boldsymbol{C}_{k} & = \frac{1}{\rho k_{\textsc{b}} \Temperature_{\heavy}} \frac{a_{j \textsc{j}} e^{- \epsilon_{j \textsc{j}}}}{Q_{j}^{\text{int}}} \sum_{k \in \Heavy} \sum_{\textsc{k} \in \QuantumSpace_{k}} \int f_{k}^{0} \mathbf{\Phi}_{k}^{D_{i}} \cdot \Mass_{k} \boldsymbol{C}_{k} \, \mathrm{d} \boldsymbol{C}_{k} \\
& = \frac{1}{\rho k_{\textsc{b}} \Temperature_{\heavy}} \frac{a_{j \textsc{j}} e^{- \epsilon_{j \textsc{j}}}}{Q_{j}^{\text{int}}} \llangle \mathbf{\Phi}_{\heavy}^{D_{i}} f_{\heavy}^{0} , \Mass_{\heavy} \boldsymbol{C}_{\heavy} \rrangle_{\heavy},
\end{align*}
and since $\llangle \mathbf{\Phi}_{\heavy}^{D_{i}} f_{\heavy}^{0} , \Mass_{\heavy} \boldsymbol{C}_{\heavy} \rrangle_{\heavy} = 0$ by \eqref{KTPhiOrtho}, the $i^{\text{th}}$-heavy-species electron diffusion coefficient reads finally
\[
D_{ie} = \frac{\pressure}{3} \frac{\Mass_{e}}{3} \sum_{j \in \Heavy} \sum_{\textsc{j} \in \QuantumSpace_{j}} \int f_{j}^{0} \mathbf{\Phi}_{j}^{D_{i}} \cdot \boldsymbol{C}_{j} \, \mathrm{d} \boldsymbol{C}_{j} \int \Sigma_{\textsc{j} \textsc{j}}^{(1)} (|\boldsymbol{C}_{e}|^{2}) |\boldsymbol{C}_{e}| f_{e}^{0}(\boldsymbol{C}_{e}) ~ \boldsymbol{C}_{e} \cdot \mathbf{\Phi}_{e}^{D_{e}} \, \mathrm{d} \boldsymbol{C}_{e},
\]
which completes the proof.

\subsection{Properties of the heavy-species transport coefficients}

The matrix $D_{\heavy} = \left( D_{ij} \right)_{i, j \in \Heavy}$ is symmetric, i.e.,
\begin{equation}
D_{ij} = D_{ji}, \qquad i \in \Heavy, j \in \Heavy.
\label{KTDhSymmetric}
\end{equation}
Indeed, $D_{i j} = \tfrac{1}{3} \pressure k_{\textsc{b}} \Temperature_{\heavy} \llbracket \mathbf{\Phi}_{\heavy}^{D_{i}}, \mathbf{\Phi}_{\heavy}^{D_{j}} \rrbracket_{\heavy}$ and the bracket $\llbracket \cdot \rrbracket_{\heavy}$ is symmetric \eqref{KTHeavyBracketSymmetry}.
$D_{\heavy}$ is also positive semi-definite, i.e., for any $U = \left( U_{i} \right)_{i \in \Heavy}$:
\begin{equation}
U^{\textsc{t}} D_{\heavy} U \geq 0,
\label{KTDhPositiveSemiDefinite}
\end{equation}
and its kernel is the space spanned by the vector $Y_{\heavy} = \left( Y_{i} \right)_{i \in \Heavy}$, i.e.,
\begin{equation}
N(D_{\heavy}) = \mathbb{R} Y_{\heavy}. \label{KTDhKernel}
\end{equation}
Indeed, since the bracket $\llbracket \cdot \rrbracket_{\heavy}$ is bilinear by definition and positive semi-definite by \eqref{KTHeavyBracketPositivity}:
\begin{eqnarray*}
U^{\textsc{t}} D_{\heavy} U & = & \sum_{i \in \Heavy} \sum_{j \in \Heavy} D_{i j} U_{i} U_{j} \\
 & = & \frac{\pressure k_{\textsc{b}} \Temperature_{\heavy}}{3} \sum_{i \in \Heavy} \sum_{j \in \Heavy} \llbracket \mathbf{\Phi}_{\heavy}^{D_{i}}, \mathbf{\Phi}_{\heavy}^{D_{j}} \rrbracket_{\heavy} U_{i} U_{j} \\
 & = & \frac{\pressure k_{\textsc{b}} \Temperature_{\heavy}}{3} \llbracket \sum_{i \in \Heavy} \mathbf{\Phi}_{\heavy}^{D_{i}} U_{i}, \sum_{j \in \Heavy} \mathbf{\Phi}_{\heavy}^{D_{j}} U_{j} \rrbracket_{\heavy} \\
 & \geq & 0, \vphantom{ \frac{\pressure k_{\textsc{b}} \Temperature_{\heavy}}{3} \llbracket \sum_{i \in \Heavy} \rrbracket_{\heavy}}
\end{eqnarray*}
and the $\mathbf{\Phi}_{\heavy}^{D_{i}}$, $i \in \Heavy$, are orthogonal to $\Invariants_{\heavy}$ with respect to the scalar product $\llangle \cdot \rrangle_{\heavy}$.

Furthermore, from the definition \eqref{KTPsiiDj} of $\mathbf{\Psi}_{\heavy}^{D_{i}}$, $i \in \Heavy$
\begin{equation}
\sum_{i \in \Heavy} Y_{i} \mathbf{\Psi}_{\heavy}^{D_{i}}= 0,
\end{equation} 
so that from \eqref{KTDij}-\eqref{KTThetaie}
\begin{align}
\sum_{i \in \Heavy} Y_{i} D_{i j} & = 0, \qquad j \in \Heavy,
\label{KTOrthogYhDhh} \\
\sum_{i \in \Heavy} Y_{i} D_{i e} & = 0,
\label{KTOrthogYhDhe}
\end{align}
and as well
\begin{align}
\sum_{i \in \Heavy} Y_{i} \theta_{i \heavy} & = 0,
\label{KTOrthogYhThetahh} \\
\sum_{i \in \Heavy} Y_{i} \theta_{ie} & = 0.
\label{KTOrthogYhThetahe}
\end{align}

Finally, from expressions \eqref{KTEtah}, \eqref{KTKappah} and \eqref{KTLambdahathh}, respectively, the shear viscosity $\eta_{\heavy}$, the volume viscosity $\kappa_{\heavy}$, and the heavy-species self-partial-thermal-conductivity $\hat{\lambda}_{\heavy \heavy}$, have the same sign as $\llbracket \mathbf{\Phi}_{\heavy}^{\eta_{\heavy}}, \mathbf{\Phi}_{\heavy}^{\eta_{\heavy}} \rrbracket_{\heavy}$, $\llbracket \phi_{\heavy}^{\kappa_{\heavy}}, \phi_{\heavy}^{\kappa_{\heavy}} \rrbracket_{\heavy}$, and $\llbracket \mathbf{\Phi}_{\heavy}^{\hat{\lambda}_{\heavy}}, \mathbf{\Phi}_{\heavy}^{\hat{\lambda}_{\heavy}} \rrbracket_{\heavy}$, respectively. For $\mu = \eta_{\heavy}$, $\kappa_{\heavy}$ and $\hat{\lambda}_{\heavy}$, it is readily seen that $\llbracket \phi_{\heavy}^{\mu}, \phi_{\heavy}^{\mu} \rrbracket_{\heavy} \geq 0$ since the bracket product is positive semi-definite. Moreover, if $\llbracket \phi_{\heavy}^{\mu}, \phi_{\heavy}^{\mu} \rrbracket_{\heavy} = 0$, then $\phi_{\heavy}^{\mu}$ must be in $\Invariants_{\heavy}$ by \eqref{KTHeavyBracketKernel}. Because of the constraints \eqref{KTPhiOrtho}, $\phi_{\heavy}^{\mu}$ must also be orthogonal to $\Invariants_{\heavy}$, so that $\phi_{\heavy}^{\mu} = 0$. Now, from expressions \eqref{KTPsiiEtah}, \eqref{KTPsiiKappah} and \eqref{KTPsiiLambdahath}, the expansion coefficients $\mathbf{\Psi}_{\heavy}^{\eta_{\heavy}}$ and $\mathbf{\Psi}_{\heavy}^{\hat{\lambda}_{\heavy}}$ cannot vanish, while the expansion coefficient $\Psi_{\heavy}^{\kappa_{\heavy}}$ can vanish if there are no internal energy levels. By linearity the same is true for the perturbation coefficients $\mathbf{\Phi}_{\heavy}^{\eta_{\heavy}}$, $\mathbf{\Phi}_{\heavy}^{\hat{\lambda}_{\heavy}}$, and $\phi_{\heavy}^{\kappa_{\heavy}}$, so that finally \cite{Giovangigli}
\begin{align}
\eta_{\heavy} & > 0, \\
\kappa_{\heavy} & \geq 0, \\
\hat{\lambda}_{\heavy \heavy} & > 0.
\end{align}

%From equations \eqref{KTOrthogYhDhh}-\eqref{KTOrthogYhThetahe}, and from expression \eqref{KTHeavyDiffusionVelocities} for the heavy-species diffusion velocities, the sum of the heavy-species diffusion fluxes is seen to vanish:
%\begin{equation}
%\sum_{i \in \Heavy} Y_{i} \boldsymbol{\MeanV}_{i} = 0.
%\label{KTHeavyMassConservationDiffusion}
%\end{equation}

\subsection{Electron Kolesnikov transport coefficients}

The resolution of the linearized Boltzmann equation for the second-order electron perturbation function $\phi_{e}^{2}$ yields first-order electron transport fluxes and associated transport coefficients. These transport fluxes should not be confused with Burnett transport coefficients \cite{GrailleMaginMassot2009} \cite{FerzigerKaper}, since one retrieves the first-order Chapman-Enskog expansion for multicomponent mixtures in the limit $\Temperature_{e} = \Temperature_{\heavy}$ \cite{GiovangigliGrailleMaginMassot}.

In the case of a weakly magnetized plasma, i.e., when $b = 0$, equation \eqref{KTSecondOrderElectronPerturbationEquation} for the second-order electron perturbation function $\phi_{e}^{2}$ becomes
\begin{equation}
\mathcal{F}_{e}(\phi_{e}^{2}) = \Psi_{e}^{2},
\label{KTSecondOrderElectronPerturbationEquationWeakB}
\end{equation}
where
\begin{align}
\Psi_{e}^{2} = & - \Streaming_{e}^{0}(\ln{f_{e}^{0}}) - \frac{1}{f_{e}^{0}} \Streaming_{e}^{-1}(\phi_{e}f_{e}^{0}) + \frac{1}{f_{e}^{0}} \Scattering_{ee}(\phi_{e} f_{e}^{0},\phi_{e}f_{e}^{0}) \\
{} & + \sum_{j \in \Heavy} \frac{1}{f_{e}^{0}} \left[ \Scattering_{ej}^{0}(\phi_{e}f_{e}^{0},\phi_{j}f_{j}^{0}) + \Scattering_{ej}^{1}(f_{e}^{0},\phi_{j}f_{j}^{0}) + \Scattering_{ej}^{2}(f_{e}^{0},f_{j}^{0}) \right], \nonumber
\end{align}
under the constraints
\begin{equation}
\llangle \phi_{e}^{2} f_{e}^{0}, \psi_{e}^{l} \rrangle_{e} = 0, \quad l \in \left\lbrace e, n^{s}+4 \right\rbrace.
\end{equation}
After some lengthy calculations, the right-member $\Psi_{e}^{2}$ of equation \eqref{KTSecondOrderElectronPerturbationEquationWeakB} can be expanded in the form
\begin{align}
\Psi_{e}^{2} = & - \mathbf{\Psi}_{e}^{\eta_{e}} : \boldsymbol{\partial_{x}} \boldsymbol{\Meanv}_{\heavy} - \delta_{b0} \mathbf{\Psi}_{e}^{D_{e}} \cdot (- \density_{e} q_{e} \boldsymbol{\Meanv}_{\heavy} \wedge \boldsymbol{B}) \vphantom{\sum_{j \in \Heavy} \frac{1}{k_{\textsc{b}} \Temperature_{e}}} \\
 & - \delta_{b0} \frac{q_{e}}{\Mass_{e}} \mathbf{\Phi}_{e}^{D_{e}} \cdot \left[ (\boldsymbol{\partial_{x}} \pressure_{e} - \density_{e} q_{e} \boldsymbol{E}) \wedge \boldsymbol{B} \right] - \delta_{b0} \frac{q_{e}}{\Mass_{e}} \mathbf{\Phi}_{e}^{\hat{\lambda}_{e}} \cdot \left[ \boldsymbol{\partial_{x}} \left( \frac{1}{k_{\textsc{b}} \Temperature_{e}}\right) \wedge \boldsymbol{B} \right] \nonumber \vphantom{\sum_{j \in \Heavy} \frac{1}{k_{\textsc{b}} \Temperature_{e}}} \\
 & - \frac{1}{3} \mathbf{\Psi}_{e}^{\kappa_{\heavy} D_{e}} \cdot (\boldsymbol{\partial_{x}} \cdot \boldsymbol{\Meanv}_{\heavy}) ~ (\boldsymbol{\partial_{x}} \pressure_{e} - \density_{e} q_{e} \boldsymbol{E}) - \frac{1}{3} \mathbf{\Psi}_{e}^{\kappa_{\heavy} \hat{\lambda}_{e}} \cdot (\boldsymbol{\partial_{x}} \cdot \boldsymbol{\Meanv}_{\heavy}) ~ \boldsymbol{\partial_{x}} \left( \frac{1}{k_{\textsc{b}} \Temperature_{e}}\right) \nonumber \vphantom{\sum_{j \in \Heavy} \frac{1}{k_{\textsc{b}} \Temperature_{e}}} \\
 & - \mathbf{\Psi}_{e}^{\Theta D_{e}} \cdot (\Temperature_{e} - \Temperature_{\heavy}) ~ (\boldsymbol{\partial_{x}} \pressure_{e} - \density_{e} q_{e} \boldsymbol{E}) - \mathbf{\Psi}_{e}^{\Theta \hat{\lambda}_{e}} \cdot (\Temperature_{e} - \Temperature_{\heavy}) ~ \boldsymbol{\partial_{x}} \left( \frac{1}{k_{\textsc{b}} \Temperature_{e}}\right) \nonumber \vphantom{\sum_{j \in \Heavy} \frac{1}{k_{\textsc{b}} \Temperature_{e}}} \\
 & - \sum_{j \in \Heavy} \mathbf{\Psi}_{e}^{D_{j}} \cdot (\boldsymbol{\partial_{x}} \pressure_{j} - \density_{j} q_{j} \boldsymbol{E}) - \mathbf{\Psi}_{e}^{D_{e}^{2}} \cdot (\boldsymbol{\partial_{x}} \pressure_{e} - \density_{e} q_{e} \boldsymbol{E}) \nonumber \vphantom{\sum_{j \in \Heavy} \frac{1}{k_{\textsc{b}} \Temperature_{e}}} \\
 & - \mathbf{\Psi}_{e}^{\hat{\lambda}_{\heavy}} \cdot \boldsymbol{\partial_{x}} \left( \frac{1}{k_{\textsc{b}} \Temperature_{\heavy}}\right) - \mathbf{\Psi}_{e}^{\hat{\lambda}_{e}^{2}} \cdot \boldsymbol{\partial_{x}} \left( \frac{1}{k_{\textsc{b}} \Temperature_{e}}\right) - \tilde{\Psi}_{e}^{2}, \nonumber \vphantom{\sum_{j \in \Heavy} \frac{1}{k_{\textsc{b}} \Temperature_{e}}}
\end{align}
where $\tilde{\Psi}_{e}^{2}$ is a scalar function of $\boldsymbol{C}_{e} \cdot \boldsymbol{C}_{e}$. Other expansion coefficients are given by
\begin{align}
\mathbf{\Psi}_{e}^{\eta_{e}} & = \frac{\Mass_{e}}{k_{\textsc{b}} \Temperature_{e}} \left[ \boldsymbol{C}_{e} \otimes \boldsymbol{C}_{e} - \frac{1}{3} \boldsymbol{C}_{e} \cdot \boldsymbol{C}_{e} \, \mathbb{I} \right] \vphantom{\sum_{j \in \Heavy}} \\
\mathbf{\Psi}_{e}^{D_{i}} & = \sum_{j \in \Heavy} \sum_{\textsc{j} \in \QuantumSpace_{j}} \frac{1}{3} \frac{\Mass_{e}}{k_{\textsc{b}} \Temperature_{e}} |\boldsymbol{C}_{e}| \Sigma_{\textsc{j} \textsc{j}}^{(1)}(|\boldsymbol{C}_{e}|^{2}) \left( \int \mathbf{\Phi}_{j}^{D_{i}} \cdot \boldsymbol{C}_{j} ~ f_{j}^{0} \, \mathrm{d} \boldsymbol{C}_{j} \right) \boldsymbol{C}_{e} \\
\mathbf{\Psi}_{e}^{D_{e}^{2}} & = \sum_{j \in \Heavy} \sum_{\textsc{j} \in \QuantumSpace_{j}} \frac{1}{3} \frac{\Mass_{e}}{k_{\textsc{b}} \Temperature_{e}} |\boldsymbol{C}_{e}| \Sigma_{\textsc{j} \textsc{j}}^{(1)}(|\boldsymbol{C}_{e}|^{2}) \left( \int \mathbf{\Phi}_{j}^{D_{e}} \cdot \boldsymbol{C}_{j} ~ f_{j}^{0} \, \mathrm{d} \boldsymbol{C}_{j} \right) \boldsymbol{C}_{e} \\
\mathbf{\Psi}_{e}^{\hat{\lambda}_{\heavy}} & = \sum_{j \in \Heavy} \sum_{\textsc{j} \in \QuantumSpace_{j}} \frac{1}{3} \frac{\Mass_{e}}{k_{\textsc{b}} \Temperature_{e}} |\boldsymbol{C}_{e}| \Sigma_{\textsc{j} \textsc{j}}^{(1)}(|\boldsymbol{C}_{e}|^{2}) \left( \int \mathbf{\Phi}_{j}^{\hat{\lambda}_{\heavy}} \cdot \boldsymbol{C}_{j} ~ f_{j}^{0} \, \mathrm{d} \boldsymbol{C}_{j} \right) \boldsymbol{C}_{e} \\
\mathbf{\Psi}_{e}^{\hat{\lambda}_{e}^{2}} & = \sum_{j \in \Heavy} \sum_{\textsc{j} \in \QuantumSpace_{j}} \frac{1}{3} \frac{\Mass_{e}}{k_{\textsc{b}} \Temperature_{e}} |\boldsymbol{C}_{e}| \Sigma_{\textsc{j} \textsc{j}}^{(1)}(|\boldsymbol{C}_{e}|^{2}) \left( \int \mathbf{\Phi}_{j}^{\hat{\lambda}_{e}} \cdot \boldsymbol{C}_{j} ~ f_{j}^{0} \, \mathrm{d} \boldsymbol{C}_{j} \right) \boldsymbol{C}_{e},
\end{align}
and
\begin{align}
\mathbf{\Psi}_{e}^{\kappa_{\heavy} D_{e}} & = \sum_{j \in \Heavy} \sum_{\textsc{j} \in \QuantumSpace_{j}} \left( \int \phi_{j}^{\kappa_{\heavy}} f_{j}^{0} \, \mathrm{d} \boldsymbol{C}_{j} \right) |\boldsymbol{C}_{e}| \Sigma_{\textsc{j} \textsc{j}}^{(1)}(|\boldsymbol{C}_{e}|^{2}) \mathbf{\Phi}_{e}^{D_{e}} \\
\mathbf{\Psi}_{e}^{\kappa_{\heavy} \hat{\lambda}_{e}} & = \sum_{j \in \Heavy} \sum_{\textsc{j} \in \QuantumSpace_{j}} \left( \int \phi_{j}^{\kappa_{\heavy}} f_{j}^{0} \, \mathrm{d} \boldsymbol{C}_{j} \right) |\boldsymbol{C}_{e}| \Sigma_{\textsc{j} \textsc{j}}^{(1)}(|\boldsymbol{C}_{e}|^{2}) \mathbf{\Phi}_{e}^{\hat{\lambda}_{e}} \\
\mathbf{\Psi}_{e}^{\Theta D_{e}} & = \sum_{j \in \Heavy} \sum_{\textsc{j} \in \QuantumSpace_{j}} \left( \int \phi_{j}^{\Theta} f_{j}^{0} \, \mathrm{d} \boldsymbol{C}_{j} \right) |\boldsymbol{C}_{e}| \Sigma_{\textsc{j} \textsc{j}}^{(1)}(|\boldsymbol{C}_{e}|^{2}) \mathbf{\Phi}_{e}^{D_{e}} \\
\mathbf{\Psi}_{e}^{\Theta \hat{\lambda}_{e}} & = \sum_{j \in \Heavy} \sum_{\textsc{j} \in \QuantumSpace_{j}} \left( \int \phi_{j}^{\Theta} f_{j}^{0} \, \mathrm{d} \boldsymbol{C}_{j} \right) |\boldsymbol{C}_{e}| \Sigma_{\textsc{j} \textsc{j}}^{(1)}(|\boldsymbol{C}_{e}|^{2}) \mathbf{\Phi}_{e}^{\hat{\lambda}_{e}}.
\end{align}
Thanks to the linearity of the linearized collision operator $\mathcal{F}_{e}$, the following similar expansion holds for $\phi_{e}^{2}$:
\begin{align}
\phi_{e}^{2} = & - \mathbf{\Phi}_{e}^{\eta_{e}} : \boldsymbol{\partial_{x}} \boldsymbol{\Meanv}_{\heavy} - \delta_{b0} \mathbf{\Phi}_{e}^{D_{e}} \cdot (- \density_{e} q_{e} \boldsymbol{\Meanv}_{\heavy} \wedge \boldsymbol{B}) \vphantom{\sum_{j \in \Heavy} \frac{1}{k_{\textsc{b}} \Temperature_{e}}} \\
 & - \delta_{b0} \frac{q_{e}}{\Mass_{e}} \mathbf{\Xi}_{e}^{D_{e}} \cdot \left[ (\boldsymbol{\partial_{x}} \pressure_{e} - \density_{e} q_{e} \boldsymbol{E}) \wedge \boldsymbol{B} \right] - \delta_{b0} \frac{q_{e}}{\Mass_{e}} \mathbf{\Xi}_{e}^{\hat{\lambda}_{e}} \cdot \left[ \boldsymbol{\partial_{x}} \left( \frac{1}{k_{\textsc{b}} \Temperature_{e}}\right) \wedge \boldsymbol{B} \right] \nonumber \vphantom{\sum_{j \in \Heavy} \frac{1}{k_{\textsc{b}} \Temperature_{e}}} \\
 & - \frac{1}{3} \mathbf{\Phi}_{e}^{\kappa_{\heavy} D_{e}} \cdot (\boldsymbol{\partial_{x}} \cdot \boldsymbol{\Meanv}_{\heavy}) ~ (\boldsymbol{\partial_{x}} \pressure_{e} - \density_{e} q_{e} \boldsymbol{E}) - \frac{1}{3} \mathbf{\Phi}_{e}^{\kappa_{\heavy} \hat{\lambda}_{e}} \cdot (\boldsymbol{\partial_{x}} \cdot \boldsymbol{\Meanv}_{\heavy}) ~ \boldsymbol{\partial_{x}} \left( \frac{1}{k_{\textsc{b}} \Temperature_{e}}\right) \nonumber \vphantom{\sum_{j \in \Heavy} \frac{1}{k_{\textsc{b}} \Temperature_{e}}} \\
 & - \mathbf{\Phi}_{e}^{\Theta D_{e}} \cdot (\Temperature_{e} - \Temperature_{\heavy}) ~ (\boldsymbol{\partial_{x}} \pressure_{e} - \density_{e} q_{e} \boldsymbol{E}) - \mathbf{\Phi}_{e}^{\Theta \hat{\lambda}_{e}} \cdot (\Temperature_{e} - \Temperature_{\heavy}) ~ \boldsymbol{\partial_{x}} \left( \frac{1}{k_{\textsc{b}} \Temperature_{e}}\right) \nonumber \vphantom{\sum_{j \in \Heavy} \frac{1}{k_{\textsc{b}} \Temperature_{e}}} \\
 & - \sum_{j \in \Heavy} \mathbf{\Phi}_{e}^{D_{j}} \cdot (\boldsymbol{\partial_{x}} \pressure_{j} - \density_{j} q_{j} \boldsymbol{E}) - \mathbf{\Phi}_{e}^{D_{e}^{2}} \cdot (\boldsymbol{\partial_{x}} \pressure_{e} - \density_{e} q_{e} \boldsymbol{E}) \nonumber \vphantom{\sum_{j \in \Heavy} \frac{1}{k_{\textsc{b}} \Temperature_{e}}} \\
 & - \mathbf{\Phi}_{e}^{\hat{\lambda}_{\heavy}} \cdot \boldsymbol{\partial_{x}} \left( \frac{1}{k_{\textsc{b}} \Temperature_{\heavy}}\right) - \mathbf{\Phi}_{e}^{\hat{\lambda}_{e}^{2}} \cdot \boldsymbol{\partial_{x}} \left( \frac{1}{k_{\textsc{b}} \Temperature_{e}}\right) - \tilde{\phi}_{e}^{2}. \nonumber \vphantom{\sum_{j \in \Heavy} \frac{1}{k_{\textsc{b}} \Temperature_{e}}}
\end{align}
where for each value of $\mu = \eta_{e}, D_{e}$, $\kappa_{\heavy} D_{e}$, $\kappa_{\heavy} \hat{\lambda}_{e}$, $\Theta D_{e}$, $\Theta \hat{\lambda}_{e}$, $D_{j}, j \in \Heavy$, $D_{e}^{2}$, $\hat{\lambda}_{\heavy}$, $\hat{\lambda}_{e}^{2}$, the function $\phi_{e}^{\mu}$ is solution to
\begin{align}
& \mathcal{F}_{e}(\phi_{e}^{\mu}) = \Psi_{e}^{\mu} \\
&  \llangle f_{e}^{0} \phi_{e}^{\mu}, \psi_{e}^{l} \rrangle_{e} = 0, \qquad l \in \left\lbrace e, n^{s}+4 \right\rbrace,
\end{align}
while for each value of $\mu = D_{e}, \hat{\lambda}_{e}$, the function $\mathbf{\Xi}_{e}^{\mu}$ is solution to
\begin{align}
& \mathcal{F}_{e}(\mathbf{\Xi}_{e}^{\mu}) = \mathbf{\Phi}_{e}^{\mu} \\
&  \llangle f_{e}^{0} \mathbf{\Xi}_{e}^{\mu}, \psi_{e}^{l} \rrangle_{e} = 0, \qquad l \in \left\lbrace e, n^{s}+4 \right\rbrace.
\end{align}
Furthermore, because of the isotropy of $\mathcal{F}_{e}$, $\tilde{\phi}_{e}^{2}$ is a scalar function of $\boldsymbol{C}_{e} \cdot \boldsymbol{C}_{e}$, and each $\phi_{e}^{\mu}$, respectively $\mathbf{\Xi}_{e}^{\mu}$, is of the same tensorial type \cite{GraillePhD} as $\Psi_{e}^{\mu}$, respectively $\mathbf{\Phi}_{e}^{\mu}$.

The first-order electron self-diffusion coefficient, the first-order electron electron-temperature thermal diffusion coefficient, and the first-order electron self-partial-thermal-conductivity are given by
\begin{align}
D_{ee}^{1} & = \frac{\pressure k_{\textsc{b}} \Temperature_{e}}{3} \llbracket \mathbf{\Phi}_{\heavy}^{D_{e}}, \mathbf{\Phi}_{e}^{D_{e}}  \rrbracket_{\heavy e},
\label{KTDee1} \\
\theta_{ee}^{1} & = - \frac{1}{3} \llbracket \mathbf{\Phi}_{\heavy}^{\hat{\lambda}_{e}}, \mathbf{\Phi}_{e}^{D_{e}}  \rrbracket_{\heavy e} = - \frac{1}{3} \llbracket \mathbf{\Phi}_{\heavy}^{D_{e}}, \mathbf{\Phi}_{e}^{\hat{\lambda}_{e}}  \rrbracket_{\heavy e},
\label{KTThetaee1} \\
\hat{\lambda}_{ee}^{1} & = \frac{1}{3 k_{\textsc{b}} \Temperature_{e}} \llbracket \mathbf{\Phi}_{\heavy}^{\hat{\lambda}_{e}}, \mathbf{\Phi}_{e}^{\hat{\lambda}_{e}}  \rrbracket_{\heavy e}.
\label{KTLambdahatee1}
\end{align}
We also introduce the additional transport coefficients
\begin{align}
D_{ee}^{\kappa_{\heavy}} & = \frac{\pressure k_{\textsc{b}} \Temperature_{e}}{9} ~ \llbracket \phi_{\heavy}^{\kappa_{\heavy}}, \mathbf{\Phi}_{e}^{D_{e}}, \mathbf{\Phi}_{e}^{D_{e}}  \rrbracket_{\heavy e e},
\label{KTDeeKappah} \\
\theta_{ee}^{\kappa_{\heavy}} & = - \frac{1}{9} ~ \llbracket \phi_{\heavy}^{\kappa_{\heavy}}, \mathbf{\Phi}_{e}^{\hat{\lambda}_{e}}, \mathbf{\Phi}_{e}^{D_{e}}  \rrbracket_{\heavy e e},
\label{KTThetaeeKappah} \\
\hat{\lambda}_{ee}^{\kappa_{\heavy}} & = \frac{1}{9 k_{\textsc{b}} \Temperature_{e}} \llbracket \phi_{\heavy}^{\kappa_{\heavy}}, \mathbf{\Phi}_{e}^{\hat{\lambda}_{e}}, \mathbf{\Phi}_{e}^{\hat{\lambda}_{e}}  \rrbracket_{\heavy e e},
\label{KTLambdahateeKappah} \\
D_{ee}^{\Theta} & = \frac{\pressure k_{\textsc{b}} \Temperature_{e}}{3} ~ \llbracket \phi_{\heavy}^{\Theta}, \mathbf{\Phi}_{e}^{D_{e}}, \mathbf{\Phi}_{e}^{D_{e}}  \rrbracket_{\heavy e e},
\label{KTDeeTheta} \\
\theta_{ee}^{\Theta} & = - \frac{1}{3} ~ \llbracket \phi_{\heavy}^{\Theta}, \mathbf{\Phi}_{e}^{\hat{\lambda}_{e}}, \mathbf{\Phi}_{e}^{D_{e}}  \rrbracket_{\heavy e e},
\label{KTThetaeeTheta} \\
\hat{\lambda}_{ee}^{\Theta} & = \frac{1}{3 k_{\textsc{b}} \Temperature_{e}} \llbracket \phi_{\heavy}^{\Theta}, \mathbf{\Phi}_{e}^{\hat{\lambda}_{e}}, \mathbf{\Phi}_{e}^{\hat{\lambda}_{e}} \rrbracket_{\heavy e e},
\label{KTLambdahateeTheta}
\end{align}
and the magnetic transport coefficients \cite{GiovangigliGraille2009} \cite{GrailleMaginMassot2009}
\begin{align}
D_{ee}^{\odot} & = - \frac{\pressure k_{\textsc{b}} \Temperature_{e}}{3} (( \mathbf{\Phi}_{e}^{D_{e}}, \mathbf{\Phi}_{e}^{D_{e}} ))_{e}, \\
\theta_{ee}^{\odot} & = \frac{1}{3} (( \mathbf{\Phi}_{e}^{D_{e}}, \mathbf{\Phi}_{e}^{\hat{\lambda}_{e}} ))_{e}, \\
\hat{\lambda}_{ee}^{\odot} & = - \frac{1}{3 k_{\textsc{b}} \Temperature_{e}} (( \mathbf{\Phi}_{e}^{\hat{\lambda}_{e}}, \mathbf{\Phi}_{e}^{\hat{\lambda}_{e}} ))_{e},
\end{align}
where we have introduced the brackets
\begin{align}
{} & (( \boldsymbol{\xi}_{e}, \boldsymbol{\zeta}_{e} ))_{e} = \frac{q_{e} |\boldsymbol{B}|}{\Mass_{e}} \int f_{e}^{0} ~ \boldsymbol{\xi}_{e} \odot \boldsymbol{\zeta}_{e} \, \mathrm{d} \boldsymbol{C}_{e}, \\
{} & \llbracket \xi_{\heavy}, \boldsymbol{\zeta}_{e}, \boldsymbol{\delta}_{e} \rrbracket_{\heavy e e} = \sum_{j \in \Heavy} \sum_{\textsc{j} \in \QuantumSpace_{j}} \left( \int \xi_{j} f_{j}^{0} \, \mathrm{d} \boldsymbol{C}_{j} \right) \int f_{e}^{0} | \boldsymbol{C}_{e}| \Sigma_{\textsc{j} \textsc{j}}^{(1)} (|\boldsymbol{C}_{e}|^{2}) ~ \boldsymbol{\zeta}_{e} \odot \boldsymbol{\delta}_{e} \, \mathrm{d} \boldsymbol{C}_{e}.
\end{align}
The following expressions are then derived for the first-order electron diffusion velocity
\begin{align}
\boldsymbol{\MeanV}_{e}^{1} = & - D_{ee}^{1} \boldsymbol{\hat{d}}_{e} - \sum_{i \in \Heavy} D_{ei} \boldsymbol{\hat{d}}_{i} - \theta_{ee}^{1} \boldsymbol{\partial_{x}} \ln{\Temperature_{e}} - \theta_{e \heavy} \boldsymbol{\partial_{x}} \ln{\Temperature_{\heavy}}
\label{KTElectronSecondOrderDiffusionVelocity} \\
& \vphantom{\sum_{i \in \Heavy}} - \big[ D_{ee}^{\kappa_{\heavy}} ~ (\boldsymbol{\partial_{x}} \cdot \boldsymbol{\Meanv}_{\heavy}) + D_{ee}^{\Theta} ~ (\Temperature_{e} - \Temperature_{\heavy}) \big] ~ \boldsymbol{\hat{d}}_{e} - \big[ \theta_{ee}^{\kappa_{\heavy}} ~ (\boldsymbol{\partial_{x}} \cdot \boldsymbol{\Meanv}_{\heavy}) + \theta_{ee}^{\Theta} ~ (\Temperature_{e} - \Temperature_{\heavy}) \big] ~ \boldsymbol{\partial_{x}} \ln{\Temperature_{e}} \nonumber \\
 & - \delta_{b0} \frac{\density_{e} q_{e} |\boldsymbol{B}|}{\pressure} D_{ee}^{0} ~ \mathscrbf{B} \wedge \boldsymbol{\Meanv}_{\heavy} - \delta_{b0} D_{ee}^{\odot} ~ \mathscrbf{B} \wedge \boldsymbol{\hat{d}}_{e} - \delta_{b0} \theta_{ee}^{\odot} ~ \mathscrbf{B} \wedge \boldsymbol{\partial_{x}} \ln{\Temperature_{e}},
 \nonumber
\end{align}
and heat flux
\begin{align}
\boldsymbol{\HeatFlux}_{e}^{1} = & - \pressure \theta_{ee}^{1} \boldsymbol{\hat{d}}_{e} - \pressure \sum_{i \in \Heavy} \theta_{ei} \boldsymbol{\hat{d}}_{i} - \hat{\lambda}_{ee}^{1} \boldsymbol{\partial_{x}} \ln{\Temperature_{e}} - \hat{\lambda}_{e \heavy} \boldsymbol{\partial_{x}} \ln{\Temperature_{\heavy}} + \density_{e} \Big( \frac{5}{2}k_{\textsc{b}} \Temperature_{e} \Big) \boldsymbol{\MeanV}_{e}^{1}
\label{KTElectronSecondOrderHeatFlux} \\
 & \vphantom{\sum_{i \in \Heavy}} - \pressure \big[ \theta_{ee}^{\kappa_{\heavy}} ~ (\boldsymbol{\partial_{x}} \cdot \boldsymbol{\Meanv}_{\heavy}) + \theta_{ee}^{\Theta} ~ (\Temperature_{e} - \Temperature_{\heavy}) \big] ~ \boldsymbol{\hat{d}}_{e} - \big[ \hat{\lambda}_{ee}^{\kappa_{\heavy}} ~ (\boldsymbol{\partial_{x}} \cdot \boldsymbol{\Meanv}_{\heavy}) + \hat{\lambda}_{ee}^{\Theta} ~ (\Temperature_{e} - \Temperature_{\heavy}) \big] ~ \boldsymbol{\partial_{x}} \ln{\Temperature_{e}}
 \nonumber \\
 & \vphantom{\sum_{i \in \Heavy}} - \delta_{b0} \density_{e} q_{e} |\boldsymbol{B}| \theta_{ee}^{0} ~ \mathscrbf{B} \wedge \boldsymbol{\Meanv}_{\heavy} - \delta_{b0} \pressure \theta_{ee}^{\odot} ~ \mathscrbf{B} \wedge \boldsymbol{\hat{d}}_{e} - \delta_{b0} \hat{\lambda}_{ee}^{\odot} ~ \mathscrbf{B} \wedge \boldsymbol{\partial_{x}} \ln{\Temperature_{e}}, \nonumber
\end{align}
where $\mathscrbf{B}$ denotes the direction of the magnetic field vector
\begin{equation}
\mathscrbf{B} = \frac{1}{|\boldsymbol{B}|} \boldsymbol{B}.
\end{equation}
Except for the last three terms associated with the magnetic field effect on the electrons, all the terms in expressions \eqref{KTElectronSecondOrderDiffusionVelocity} and \eqref{KTElectronSecondOrderHeatFlux} for $\boldsymbol{\MeanV}_{e}^{1}$ and $\boldsymbol{\HeatFlux}_{e}^{1}$, respectively, are due to the coupling between electrons and heavy species. They may thus be referred to as the \og electron Kolesnikov fluxes \fg{} \cite{Kolesnikov1974} \cite{GrailleMaginMassot2009}.

The terms in the first lines of both \eqref{KTElectronSecondOrderDiffusionVelocity} and \eqref{KTElectronSecondOrderHeatFlux} were already present in the monoatomic case \cite{GrailleMaginMassot2009}, though the corresponding transport coefficients \eqref{KTDie}, \eqref{KTThetaie}, \eqref{KTLambdahathe}, \eqref{KTThetahe}, and \eqref{KTDee1}-\eqref{KTLambdahatee1} are different. The terms proportional to $(\boldsymbol{\partial_{x}} \cdot \boldsymbol{\Meanv}_{\heavy})$ are new and specific to the polyatomic case, since the volume viscosity $\kappa_{\heavy}$ vanishes in the monoatomic limit. The terms proportional to $(\Temperature_{e} - \Temperature_{\heavy})$ are also new, because the corresponding transport coefficients vanish in the monoatomic limit \cite{GrailleMaginMassot2009}. This is because the term
\begin{equation}
\int \phi_{j}^{\Theta} f_{j}^{0} \, \mathrm{d} \boldsymbol{C}_{j}
\label{KTIntPhijThetafj0}
\end{equation}
appearing in expressions \eqref{KTDeeTheta}-\eqref{KTLambdahateeTheta} for $D_{ee}^{\Theta}$, $\theta_{ee}^{\Theta}$ and $\hat{\lambda}_{ee}^{\Theta}$, respectively, becomes equal to $\llangle \phi_{\heavy}^{\Theta} f_{\heavy}^{0}, \psi_{\heavy}^{j} \rrangle_{\heavy}$ in the monoatomic limit and thus vanishes, since the number of particles in the $j^{\text{th}}$ species is a collisional invariant of the scattering operator. Conversely, in the polyatomic case, the term \eqref{KTIntPhijThetafj0} does not necessarily vanish. Indeed, the number of molecules of the $j^{\text{th}}$ species in the $\textsc{j}^{\text{th}}$ internal energy state is not conserved in general in scattering collisions, and the elastic collision cross-section $\CrossSection_{je}^{\textsc{j} \textsc{j}}$ may depend on the energy level $\textsc{j}$. Physically speaking, the transport fluxes appearing in the second lines of \eqref{KTElectronSecondOrderDiffusionVelocity} and \eqref{KTElectronSecondOrderHeatFlux} are associated with the nonuniform diffusion of electrons with respect to the different internal energy states of each heavy species. The last lines of \eqref{KTElectronSecondOrderDiffusionVelocity} and \eqref{KTElectronSecondOrderHeatFlux}, respectively, embed the magnetic field induced electron transport fluxes, and have the same structure in the polyatomic and monoatomic cases, for the last two terms were overlooked in the case $b=0$ in \cite{GrailleMaginMassot2009}.

Finally, we obtain the following expression for the first-order energy exchange term due to scattering collisions:
\begin{align}
\Delta E_{\heavy e}^{1,\text{scatt}} = {} & \sum_{i \in \Heavy} \sum_{\textsc{i} \in \QuantumSpace_{i}} \density_{i} \boldsymbol{\MeanV}_{i \textsc{i}} \cdot \boldsymbol{\Force}_{ie}^{I,0} \\
 & + \sum_{i \in \Heavy} \sum_{\textsc{i} \in \QuantumSpace_{i}} \frac{2}{3} \frac{\Mass_{e}}{\Mass_{i}} \nu_{ie}^{\textsc{i} \textsc{i}} \int f_{i}^{0} \phi_{i} \Big( \frac{3}{2} k_{\textsc{b}} \Temperature_{e} - \frac{1}{2} \Mass_{i} \boldsymbol{C}_{i} \cdot \boldsymbol{C}_{i} \Big) \, \mathrm{d} \boldsymbol{C}_{i} \nonumber \\
 & + \sum_{i \in \Heavy} \sum\limits_{\substack{\textsc{i}, \textsc{i}' \in \QuantumSpace_{i}}} \nu_{ie}^{\textsc{i} \textsc{i}'} \Delta E_{\textsc{i} \textsc{i}'} \int f_{i}^{0} \phi_{i} \, \mathrm{d} \boldsymbol{C}_{i}. \nonumber 
\end{align}
One may prefer the alternative formulation
\begin{align}
\Delta E_{\heavy e}^{1,\text{scatt}} = {} & \pressure \sum_{j \in \Heavy} D_{je} \boldsymbol{\hat{d}}_{j} \cdot \boldsymbol{\hat{d}}_{e} + \pressure \sum_{j \in \Heavy} \theta_{je} \boldsymbol{\hat{d}}_{j} \cdot \boldsymbol{\partial_{x}} \ln{\Temperature_{e}} \\
 & + \pressure D_{ee}^{1} \boldsymbol{\hat{d}}_{e} \cdot \boldsymbol{\hat{d}}_{e} + \pressure \theta_{ee}^{1} \boldsymbol{\hat{d}}_{e} \cdot \boldsymbol{\partial_{x}} \ln{\Temperature_{e}} \nonumber \vphantom{\sum_{j \in \Heavy} \int} \\
 & +\pressure \theta_{\heavy e} \boldsymbol{\partial_{x}} \ln{\Temperature_{\heavy}} \cdot \boldsymbol{\hat{d}}_{e} + \hat{\lambda}_{\heavy e} \boldsymbol{\partial_{x}} \ln{\Temperature_{\heavy}} \cdot \boldsymbol{\partial_{x}} \ln{\Temperature_{e}} \nonumber \vphantom{\sum_{j \in \Heavy} \int} \\
 & + \pressure \theta_{ee}^{1} \boldsymbol{\partial_{x}} \ln{\Temperature_{e}} \cdot \boldsymbol{\hat{d}}_{e} + \hat{\lambda}_{ee}^{1} \boldsymbol{\partial_{x}} \ln{\Temperature_{e}} \cdot \boldsymbol{\partial_{x}} \ln{\Temperature_{e}} \nonumber \vphantom{\sum_{j \in \Heavy} \int} \\
 & + k_{\textsc{b}} \Temperature_{\heavy}^{2} \llangle f_{\heavy}^{0} \phi_{\heavy}, \Psi_{\heavy}^{\Theta} \rrangle_{\heavy} \nonumber \vphantom{\sum_{j \in \Heavy} \int} \\
 & + k_{\textsc{b}} (\Temperature_{e} - \Temperature_{\heavy}) \sum_{i \in \Heavy} \sum_{\textsc{i} \in \QuantumSpace_{i}} \frac{\Mass_{e}}{\Mass_{i}} \nu_{ie}^{\textsc{i} \textsc{i}} \int f_{i}^{0} \phi_{i} \, \mathrm{d} \boldsymbol{C}_{i} \nonumber \\*
 & + \sum_{i \in \Heavy} \sum\limits_{\substack{\textsc{i}, \textsc{i}' \in \QuantumSpace_{i}}} \Big( 1 - \frac{\Temperature_{\heavy}}{\Temperature_{e}} g_{\textsc{i} \textsc{i}'} \Big) \nu_{ie}^{\textsc{i} \textsc{i}'} \Delta E_{\textsc{i} \textsc{i}'} \int f_{i}^{0} \phi_{i} \, \mathrm{d} \boldsymbol{C}_{i}, \nonumber
\end{align}
where $\int f_{i}^{0} \phi_{i} \, \mathrm{d} \boldsymbol{C}_{i}$ can be further expanded in the form
\begin{align}
\int f_{i}^{0} \phi_{i} \, \mathrm{d} \boldsymbol{C}_{i} = & - (\boldsymbol{\partial_{x}} \cdot \boldsymbol{\Meanv}_{\heavy}) \int f_{i}^{0} \phi_{i}^{\kappa_{\heavy}} \, \mathrm{d} \boldsymbol{C}_{i} \\
 & - (\Temperature_{e} - \Temperature_{\heavy}) \int f_{i}^{0} \phi_{i}^{\Theta} \, \mathrm{d} \boldsymbol{C}_{i}. \nonumber
\end{align}

\section{Fluid Equations}
\label{SecKTConsEq}
In this section, we summarize the macroscopic equations obtained for multicomponent plasmas in the Navier-Stokes regime. The fluid equations \eqref{KTNavierStokesMassElectron}, \eqref{KTNavierStokesEnergyElectron}, \eqref{KTNavierStokesMassHeavy}, \eqref{KTNavierStokesMomentumHeavy} and \eqref{KTNavierStokesEnergyHeavy} and the transport fluxes derived from the Chapman-Enskog expansion are redimensionalized. This is equivalent to setting $\varepsilon = 1$ in the full-dimension equations where $\varepsilon$ is taken as a formal parameter.

\subsection{Conservation of mass, momentum, and energy}

The fluid equations \eqref{KTNavierStokesMassElectron}, \eqref{KTNavierStokesEnergyElectron}, \eqref{KTNavierStokesMassHeavy}, \eqref{KTNavierStokesMomentumHeavy} and \eqref{KTNavierStokesEnergyHeavy} are rewritten in the form
\begin{gather}
\partial_{t} \rho_{e} + \boldsymbol{\partial_{x}} \cdot \left( \rho_{e} \boldsymbol{\Meanv}_{\heavy} + \rho_{e} \boldsymbol{\MeanV}_{e} \right) = \Mass_{e} \mathfrak{w}_{e}, \label{KTElectronMass} \\
\partial_{t} \Energy_{e} + \boldsymbol{\partial_{x}} \cdot (\Energy_{e} \boldsymbol{\Meanv}_{\heavy}) = - \, \pressure_{e} \, \boldsymbol{\partial_{x}} \cdot \boldsymbol{\Meanv}_{\heavy} - \boldsymbol{\partial_{x}} \cdot \boldsymbol{\HeatFlux}_{e} + \Delta E_{e \heavy} + \boldsymbol{\Current}_{e} \cdot \boldsymbol{E}' + \delta_{b0} \, \boldsymbol{\Current}_{e}^{0} \cdot (\boldsymbol{\Meanv}_{\heavy} \wedge \boldsymbol{B}), \label{KTElectronEnergy}
\end{gather}
for electrons, and
\begin{gather}
\partial_{t} \rho_{i} + \boldsymbol{\partial_{x}} \cdot (\rho_{i} \boldsymbol{\Meanv}_{\heavy} + \rho_{i} \boldsymbol{\MeanV}_{i}) = \Mass_{i} \mathfrak{w}_{i}, \qquad i \in \Heavy,
 \label{KTHeavyMassDensities} \\
\partial_{t} \left( \rho_{\heavy} \boldsymbol{\Meanv}_{\heavy} \right) + \boldsymbol{\partial_{x}} \cdot \left(  \rho_{\heavy} \boldsymbol{\Meanv}_{\heavy} \otimes \boldsymbol{\Meanv}_{\heavy} + \pressure \, \mathbb{I} \right) = - \, \boldsymbol{\partial_{x}} \cdot \boldsymbol{\Pi}_{\heavy} + \density q \boldsymbol{E} + \boldsymbol{\current} \wedge \boldsymbol{B},
 \label{KTMomentum} \\
\partial_{t} \Energy_{\heavy} + \boldsymbol{\partial_{x}} \cdot \left( \Energy_{\heavy} \boldsymbol{\Meanv}_{\heavy} \right) = - \, \pressure_{\heavy} \, \boldsymbol{\partial_{x}} \cdot \boldsymbol{\Meanv}_{\heavy} - \boldsymbol{\partial_{x}} \boldsymbol{\Meanv}_{\heavy} : \boldsymbol{\Pi}_{\heavy} - \boldsymbol{\partial_{x}} \cdot \boldsymbol{\HeatFlux}_{\heavy} + \Delta E_{\heavy e} + \boldsymbol{\Current}_{\heavy} \cdot \boldsymbol{E}',
 \label{KTHeavyEnergy} 
\end{gather}
for the heavy species.

The electron diffusion velocity in the heavy-species reference frame is given by
\begin{equation}
\boldsymbol{\MeanV}_{e} = \boldsymbol{\MeanV}_{e}^{0} + \boldsymbol{\MeanV}_{e}^{1},
\end{equation}
the electron heat flux in the heavy-species reference frame by
\begin{equation}
\boldsymbol{\HeatFlux}_{e} = \boldsymbol{\HeatFlux}_{e}^{0} + \boldsymbol{\HeatFlux}_{e}^{1},
\end{equation}
and the electron conduction current density in the heavy-species reference frame by
\begin{equation}
\boldsymbol{\Current}_{e} = \boldsymbol{\Current}_{e}^{0} + \boldsymbol{\Current}_{e}^{1} = \density_{e} q_{e} \boldsymbol{\MeanV}_{e}.
\end{equation}
We also introduce the heat flux
\begin{equation}
\boldsymbol{\HeatFlux} = \boldsymbol{\HeatFlux}_{e} + \boldsymbol{\HeatFlux}_{\heavy},
\end{equation}
the conduction current of the mixture
\begin{equation}
\boldsymbol{\Current} = \boldsymbol{\Current}_{e} + \boldsymbol{\Current}_{\heavy} = \density_{e} q_{e} \boldsymbol{\MeanV}_{e} + \sum_{j \in \Heavy} \density_{j} q_{j} \boldsymbol{\MeanV}_{j},
\end{equation}
and the energy exchange terms
\begin{align}
\Delta E_{e \heavy} & = \Delta E_{e \heavy}^{0} + \Delta E_{e \heavy}^{1},
\label{KTDeltaEeh} \\
\Delta E_{\heavy e} & = \Delta E_{\heavy e}^{0} + \Delta E_{\heavy e}^{1},
\label{KTDeltaEhe}
\end{align}
which satisfy the reciprocity relation
\begin{equation}
\Delta E_{\heavy e} = - \Delta E_{e \heavy}.
\end{equation}
We also restate expressions for the zeroth-order and first-order current densities of the mixture in the inertial reference frame
\begin{align}
\boldsymbol{\current}^{0} & = \density_{\heavy} q_{\heavy} \boldsymbol{\Meanv}_{\heavy} + \density_{e} q_{e} ( \boldsymbol{\Meanv}_{\heavy} + \boldsymbol{\MeanV}_{e}^{0} ), \\
\boldsymbol{\current}^{1} & = \density_{\heavy} q_{\heavy} \boldsymbol{\Meanv}_{\heavy} + \sum_{j \in \Heavy} \density_{j} q_{j} \boldsymbol{\MeanV}_{j} + \density_{e} q_{e} ( \boldsymbol{\Meanv}_{\heavy} + \boldsymbol{\MeanV}_{e}^{0} + \boldsymbol{\MeanV}_{e}^{1} ),
\end{align}
and denote by
\begin{equation}
\boldsymbol{\current} = \delta_{b0} \boldsymbol{\current}^{0} + \delta_{b1} \boldsymbol{\current}^{1}
\end{equation}
the current density of the mixture in the inertial reference frame.

From equations \eqref{KTOrthogYhDhh}-\eqref{KTOrthogYhThetahe}, the heavy species diffusion velocities satisfy the mass conservation constraint
\begin{equation}
\sum_{i \in \Heavy} \rho_{i} \boldsymbol{\MeanV}_{i} = 0,
\label{KTMassConservationDiffusion}
\end{equation}
so that summing equation \eqref{KTHeavyMassDensities} over $i \in \Heavy$, the equation for conservation of the heavy-species mass is obtained
\begin{equation}
\partial_{t} \rho_{\heavy} + \boldsymbol{\partial_{x}} \cdot (\rho_{\heavy} \boldsymbol{\Meanv}_{\heavy}) = \sum_{i \in \Heavy} \Mass_{i} \mathfrak{w}_{i}.
\label{KTHeavyMass}
\end{equation}
Since the total mass is conserved in reactive collisions, the conservation constraint
\begin{equation}
\Mass_{e} \mathfrak{w}_{e} +  \sum_{i \in \Heavy} \Mass_{i} \mathfrak{w}_{i} = 0
\end{equation}
is satisfied. Thus, the total mass conservation equation is obtained by summing \eqref{KTElectronMass} and \eqref{KTHeavyMass}, and reads
\begin{equation}
\partial_{t} \rho + \boldsymbol{\partial_{x}} \cdot (\rho \boldsymbol{\Meanv}) = 0,
\label{KTMass}
\end{equation}
where
\begin{equation}
\rho = \rho_{e} + \rho_{\heavy}
\label{KTConservationEquationsRho}
\end{equation}
is the mass density of the mixture, and the mixture-averaged velocity $\boldsymbol{\Meanv}$ is given by
\begin{equation}
\rho \boldsymbol{\Meanv} = \rho_{e} \boldsymbol{\Meanv}_{e} + \rho_{\heavy} \boldsymbol{\Meanv}_{\heavy} = \rho \boldsymbol{\Meanv}_{\heavy} + \rho_{e} \boldsymbol{\MeanV}_{e},
\label{KTConservationEquationsMeanv}
\end{equation}
where the mean electron velocity in the inertial reference frame reads
\begin{equation}
\boldsymbol{\Meanv}_{e} = \boldsymbol{\Meanv}_{\heavy} + \boldsymbol{\MeanV}_{e}.
\end{equation}

The electron momentum relation  \eqref{KTFirstElectronMomentumRelation} is also rewritten in the form
\begin{equation}
\boldsymbol{\partial_{x}} \pressure_{e} = \density_{e} q_{e} \boldsymbol{E} + \boldsymbol{\current}_{e}^{0} \wedge \boldsymbol{B} + \delta_{b1} \, \boldsymbol{\Current}_{e}^{1} \wedge \boldsymbol{B} + \boldsymbol{\Force}_{e \heavy},
\label{KTElectronMomentumRelation}
\end{equation}
where $\boldsymbol{\Force}_{e \heavy} = - \boldsymbol{\Force}_{\heavy e}$ is the average force exerted on electrons by the heavy species:
\begin{equation}
\boldsymbol{\Force}_{e \heavy} = \boldsymbol{\Force}_{e \heavy}^{0} + \boldsymbol{\Force}_{e \heavy}^{1}.
\label{KTFeh}
\end{equation}

Finally, summing equations \eqref{KTElectronEnergy} and \eqref{KTHeavyEnergy}, we obtain the following equation for conservation of the total internal energy $\Energy = \Energy_{e} + \Energy_{\heavy}$:
\begin{equation}
\partial_{t} \Energy + \boldsymbol{\partial_{x}} \cdot \left( \Energy \boldsymbol{\Meanv}_{\heavy} \right) = - \, \pressure \, \boldsymbol{\partial_{x}} \cdot \boldsymbol{\Meanv}_{\heavy} - \boldsymbol{\partial_{x}} \boldsymbol{\Meanv}_{\heavy} : \boldsymbol{\Pi}_{\heavy} - \boldsymbol{\partial_{x}} \cdot \boldsymbol{\HeatFlux} + \boldsymbol{\Current} \cdot \boldsymbol{E}' + \delta_{b0} \, \boldsymbol{\Current}_{e}^{0} \cdot ( \boldsymbol{\Meanv}_{\heavy} \wedge \boldsymbol{B}).
\end{equation}

\subsection{Transport fluxes}

The electron diffusion velocity is obtained from \eqref{KTFirstOrderElectronDiffusionVelocity} and \eqref{KTElectronSecondOrderDiffusionVelocity} in the form
\begin{align}
\boldsymbol{\MeanV}_{e} = & - D_{ee} \boldsymbol{\hat{d}}_{e} - \sum_{i \in \Heavy} D_{ei} \boldsymbol{\hat{d}}_{i} - \theta_{ee} \boldsymbol{\partial_{x}} \ln{\Temperature_{e}} - \theta_{e \heavy} \boldsymbol{\partial_{x}} \ln{\Temperature_{\heavy}}
\label{KTDiffusionVelocityElectron} \\
& \vphantom{\sum_{i \in \Heavy}} - \big[ D_{ee}^{\kappa_{\heavy}} ~ (\boldsymbol{\partial_{x}} \cdot \boldsymbol{\Meanv}_{\heavy}) + D_{ee}^{\Theta} ~ (\Temperature_{e} - \Temperature_{\heavy}) \big] ~ \boldsymbol{\hat{d}}_{e} - \big[ \theta_{ee}^{\kappa_{\heavy}} ~ (\boldsymbol{\partial_{x}} \cdot \boldsymbol{\Meanv}_{\heavy}) + \theta_{ee}^{\Theta} ~ (\Temperature_{e} - \Temperature_{\heavy}) \big] ~ \boldsymbol{\partial_{x}} \ln{\Temperature_{e}} \nonumber \\
 & - \delta_{b0} \frac{\density_{e} q_{e} |\boldsymbol{B}|}{\pressure} D_{ee}^{0} ~ \mathscrbf{B} \wedge \boldsymbol{\Meanv}_{\heavy} - \delta_{b0} D_{ee}^{\odot} ~ \mathscrbf{B} \wedge \boldsymbol{\hat{d}}_{e} - \delta_{b0} \theta_{ee}^{\odot} ~ \mathscrbf{B} \wedge \boldsymbol{\partial_{x}} \ln{\Temperature_{e}},
 \nonumber
\end{align}
where the electron self-diffusion coefficient and the electron electron-temperature thermal diffusion coefficient read
\begin{align}
D_{ee} & = D_{ee}^{0} + D_{ee}^{1},
\label{KTDee} \\
\theta_{ee} & = \theta_{ee}^{0} + \theta_{ee}^{1},
\label{KTThetaee}
\end{align}
and the unconstrained diffusion driving force read
\begin{align}
\boldsymbol{\hat{d}}_{e} & = \frac{1}{\pressure} (\boldsymbol{\partial_{x}} \pressure_{e} - \density_{e} q_{e} \boldsymbol{E}), \\
\boldsymbol{\hat{d}}_{i} & = \frac{1}{\pressure} (\boldsymbol{\partial_{x}} \pressure_{i} - \density_{i} q_{i} \boldsymbol{E}), \quad i \in \Heavy.
\end{align}
Similarly, the electron self-partial-thermal-conductivity is given by
\begin{equation}
\hat{\lambda}_{ee} = \hat{\lambda}_{ee}^{0} + \hat{\lambda}_{ee}^{1},
\label{KTLambdahatee}
\end{equation}
and from \eqref{KTFirstOrderElectronHeatFlux} and \eqref{KTElectronSecondOrderHeatFlux} the electron heat flux thus reads
\begin{align}
\boldsymbol{\HeatFlux}_{e} = & - \pressure \theta_{ee} \boldsymbol{\hat{d}}_{e} - \pressure \sum_{i \in \Heavy} \theta_{ei} \boldsymbol{\hat{d}}_{i} - \hat{\lambda}_{ee} \boldsymbol{\partial_{x}} \ln{\Temperature_{e}} - \hat{\lambda}_{e \heavy} \boldsymbol{\partial_{x}} \ln{\Temperature_{\heavy}} + \density_{e} \Big( \frac{5}{2}k_{\textsc{b}} \Temperature_{e} \Big) \boldsymbol{\MeanV}_{e}
\label{KTHeatFluxElectron} \\
 & \vphantom{\sum_{i \in \Heavy}} - \pressure \big[ \theta_{ee}^{\kappa_{\heavy}} ~ (\boldsymbol{\partial_{x}} \cdot \boldsymbol{\Meanv}_{\heavy}) + \theta_{ee}^{\Theta} ~ (\Temperature_{e} - \Temperature_{\heavy}) \big] ~ \boldsymbol{\hat{d}}_{e} - \big[ \hat{\lambda}_{ee}^{\kappa_{\heavy}} ~ (\boldsymbol{\partial_{x}} \cdot \boldsymbol{\Meanv}_{\heavy}) + \hat{\lambda}_{ee}^{\Theta} ~ (\Temperature_{e} - \Temperature_{\heavy}) \big] ~ \boldsymbol{\partial_{x}} \ln{\Temperature_{e}}
 \nonumber \\
 & \vphantom{\sum_{i \in \Heavy}} - \delta_{b0} \density_{e} q_{e} |\boldsymbol{B}| \theta_{ee}^{0} ~ \mathscrbf{B} \wedge \boldsymbol{\Meanv}_{\heavy} - \delta_{b0} \pressure \theta_{ee}^{\odot} ~ \mathscrbf{B} \wedge \boldsymbol{\hat{d}}_{e} - \delta_{b0} \hat{\lambda}_{ee}^{\odot} ~ \mathscrbf{B} \wedge \boldsymbol{\partial_{x}} \ln{\Temperature_{e}}. \nonumber
\end{align}
The heavy-species diffusion velocities, viscous tensor and heat flux where stated in \eqref{KTHeavyDiffusionVelocities}, \eqref{KTViscousTensor} and \eqref{KTHeavyHeatFlux}, respectively. We recall here their expressions
\begin{align}
\boldsymbol{\MeanV}_{i} & = - \sum_{j \in \Heavy} D_{i j} \boldsymbol{\hat{d}}_{j} - D_{ie} \boldsymbol{\hat{d}}_{e} - \theta_{i \heavy} \boldsymbol{\partial_{x}} \ln{\Temperature_{\heavy}} - \theta_{i e} \boldsymbol{\partial_{x}} \ln{\Temperature_{e}}, \quad i \in \Heavy,
\label{KTDiffusionVelocitiesHeavy} \\
\boldsymbol{\Pi}_{\heavy} & = - \eta_{\heavy} \, \Big( \boldsymbol{\partial_{x}} \boldsymbol{\Meanv}_{\heavy} + (\boldsymbol{\partial_{x}} \boldsymbol{\Meanv}_{\heavy})^{\textsc{t}} - \frac{2}{3} (\boldsymbol{\partial_{x}} \cdot \boldsymbol{\Meanv}_{\heavy}) \, \mathbb{I} \Big) - \kappa_{\heavy} (\boldsymbol{\partial_{x}} \cdot \boldsymbol{\Meanv}_{\heavy}) \, \mathbb{I} - \zeta (\Temperature_{e}-\Temperature_{\heavy}) \, \mathbb{I}, \vphantom{\sum_{j \in \Heavy}}
\label{KTViscousTensorHeavy} \\
\boldsymbol{\HeatFlux}_{\heavy} & = - \pressure \sum_{j \in \Heavy} \theta_{\heavy j} \boldsymbol{\hat{d}}_{j} - \pressure \theta_{\heavy e} \boldsymbol{\hat{d}}_{e} - \hat{\lambda}_{\heavy \heavy} \boldsymbol{\partial_{x}} \ln{\Temperature_{\heavy}} - \hat{\lambda}_{\heavy e} \boldsymbol{\partial_{x}} \ln{\Temperature_{e}} + \sum_{j \in \Heavy} \big( \frac{5}{2}k_{\textsc{b}} \Temperature_{\heavy} + \overline{E}_{j} \big) \density_{j} \boldsymbol{\MeanV}_{j}.
\label{KTHeatFluxHeavy}
\end{align}

\section{Center-of-Mass Reference Frame}

The conservation equations may be rewritten in the center-of-mass reference frame. From the definition \eqref{KTConservationEquationsMeanv} of $\boldsymbol{\Meanv}$, the heavy-species velocity reads
\begin{equation}
\boldsymbol{\Meanv}_{\heavy} = \boldsymbol{\Meanv} - Y_{e} \boldsymbol{\MeanV}_{e},
\label{KTvhv}
\end{equation}
where $Y_{e} = \rho_{e} / \rho$ is the electron mass fraction. Equations \eqref{KTElectronMass} and \eqref{KTHeavyMassDensities} expressing the mass conservation of the respective species thus read in the center-of-mass reference frame
\begin{align}
\partial_{t} \rho_{e} + \boldsymbol{\partial_{x}} \cdot \big( \rho_{e} \boldsymbol{\Meanv} + \rho_{e} \boldsymbol{\MeanV}_{e}^{\boldsymbol{\Meanv}} \big) &  = \Mass_{e} \mathfrak{w}_{e},
 \label{KTElectronMassInertialOriginal} \\
\partial_{t} \rho_{i} + \boldsymbol{\partial_{x}} \cdot \big( \rho_{i} \boldsymbol{\Meanv} + \rho_{i} \boldsymbol{\MeanV}_{i}^{\boldsymbol{\Meanv}} \big) & = \Mass_{i} \mathfrak{w}_{i}, \qquad i \in \Heavy.
 \label{KTHeavyMassDensitiesInertialOriginal}
\end{align}
where $\boldsymbol{\MeanV}_{e}^{\boldsymbol{\Meanv}}$ and $\boldsymbol{\MeanV}_{i}^{\boldsymbol{\Meanv}}$, $i \in \Heavy$, are the species diffusion velocities in the center-of-mass reference frame:
\begin{align}
\boldsymbol{\MeanV}_{e}^{\boldsymbol{\Meanv}} & = (1- Y_{e}) \boldsymbol{\MeanV}_{e}
\label{KTElectronDiffusionVelocityInertial} \\
\boldsymbol{\MeanV}_{i}^{\boldsymbol{\Meanv}} & = \boldsymbol{\MeanV}_{i} - Y_{e} \boldsymbol{\MeanV}_{e}, \quad i \in \Heavy.
\label{KTHeavyDiffusionVelocitiesInertial}
\end{align}

The macroscopic equations \eqref{KTElectronMass}-\eqref{KTHeavyEnergy} read in the center-of-mass reference frame
\begin{gather}
\partial_{t} \rho_{e} + \boldsymbol{\partial_{x}} \cdot \big( \rho_{e} \boldsymbol{\Meanv} + \rho_{e} (1- Y_{e}) \boldsymbol{\MeanV}_{e} \big) = \Mass_{e} \mathfrak{w}_{e}, \label{KTElectronMassInertialExact} \\
\partial_{t} \Energy_{e} + \boldsymbol{\partial_{x}} \cdot \big( \Energy_{e} (\boldsymbol{\Meanv} - Y_{e} \boldsymbol{\MeanV}_{e}) \big) = - \, \pressure_{e} \, \boldsymbol{\partial_{x}} \cdot (\boldsymbol{\Meanv} - Y_{e} \boldsymbol{\MeanV}_{e}) - \boldsymbol{\partial_{x}} \cdot \boldsymbol{\HeatFlux}_{e} + \Delta E_{e \heavy} + \boldsymbol{\Current}_{e} \cdot \boldsymbol{E}' 
 \label{KTElectronEnergyInertialExact} \\
 + \, \delta_{b0} \, \boldsymbol{\Current}_{e}^{0} \cdot \big( (\boldsymbol{\Meanv} - Y_{e} \boldsymbol{\MeanV}_{e}) \wedge \boldsymbol{B} \big),
 \nonumber
\end{gather}
for electrons, and
\begin{gather}
\partial_{t} \rho_{i} + \boldsymbol{\partial_{x}} \cdot \big( \rho_{i} \boldsymbol{\Meanv} + \rho_{i} (\boldsymbol{\MeanV}_{i} - Y_{e} \boldsymbol{\MeanV}_{e}) \big) = \Mass_{i} \mathfrak{w}_{i}, \qquad i \in \Heavy,
 \label{KTHeavyMassInertialExact} \\
\partial_{t} \big( \rho (1- Y_{e}) (\boldsymbol{\Meanv} - Y_{e} \boldsymbol{\MeanV}_{e}) \big) + \boldsymbol{\partial_{x}} \cdot \big(  \rho (1- Y_{e}) (\boldsymbol{\Meanv} - Y_{e} \boldsymbol{\MeanV}_{e}) \otimes (\boldsymbol{\Meanv} - Y_{e} \boldsymbol{\MeanV}_{e}) + \pressure \, \mathbb{I} \big)
 \label{KTMomentumInertialExact} \\
 = - \, \boldsymbol{\partial_{x}} \cdot \boldsymbol{\Pi}_{\heavy} + \density q \boldsymbol{E} + \boldsymbol{\current} \wedge \boldsymbol{B},
 \nonumber \\
\partial_{t} \Energy_{\heavy} + \boldsymbol{\partial_{x}} \cdot \big( \Energy_{\heavy} (\boldsymbol{\Meanv} - Y_{e} \boldsymbol{\MeanV}_{e}) \big) = - \pressure_{\heavy} \, \boldsymbol{\partial_{x}} \cdot (\boldsymbol{\Meanv} - Y_{e} \boldsymbol{\MeanV}_{e}) - \boldsymbol{\partial_{x}} (\boldsymbol{\Meanv} - Y_{e} \boldsymbol{\MeanV}_{e}) : \boldsymbol{\Pi}_{\heavy}
 \label{KTHeavyEnergyInertialExact} \\
 - \, \boldsymbol{\partial_{x}} \cdot \boldsymbol{\HeatFlux}_{\heavy} + \Delta E_{\heavy e} + \boldsymbol{\Current}_{\heavy} \cdot \boldsymbol{E}',
 \nonumber
\end{gather}
for the heavy species.

Given that $Y_{e}$ is of order $\varepsilon^{2}$, the latter model is equivalent at first-order in $\varepsilon$ to the following one:
\begin{gather}
\partial_{t} \rho_{e} + \boldsymbol{\partial_{x}} \cdot \big( \rho_{e} \boldsymbol{\Meanv} + \rho_{e} \boldsymbol{\MeanV}_{e}^{\boldsymbol{\Meanv}} \big) = \Mass_{e} \mathfrak{w}_{e}, \label{KTElectronMassInertial} \\
\partial_{t} \Energy_{e} + \boldsymbol{\partial_{x}} \cdot ( \Energy_{e} \boldsymbol{\Meanv}) = - \, \pressure_{e} \, \boldsymbol{\partial_{x}} \cdot \boldsymbol{\Meanv} - \boldsymbol{\partial_{x}} \cdot \boldsymbol{\HeatFlux}_{e}^{\boldsymbol{\Meanv}} + \Delta E_{e \heavy} + \boldsymbol{\Current}_{e}^{\boldsymbol{\Meanv}} \cdot \boldsymbol{E}' + \, \delta_{b0} \, \boldsymbol{\Current}_{e}^{0} \cdot (\boldsymbol{\Meanv} \wedge \boldsymbol{B} ),
 \label{KTElectronEnergyInertial} \\
\partial_{t} \rho_{i} + \boldsymbol{\partial_{x}} \cdot \big( \rho_{i} \boldsymbol{\Meanv} + \rho_{i} \boldsymbol{\MeanV}_{i}^{\boldsymbol{\Meanv}} \big) = \Mass_{i} \mathfrak{w}_{i}, \qquad i \in \Heavy,
 \label{KTHeavyMassInertial} \\
\partial_{t} ( \rho \boldsymbol{\Meanv} ) + \boldsymbol{\partial_{x}} \cdot (  \rho \boldsymbol{\Meanv} \otimes \boldsymbol{\Meanv} + \pressure \, \mathbb{I} ) = - \, \boldsymbol{\partial_{x}} \cdot \boldsymbol{\Pi} + \density q \boldsymbol{E} + \boldsymbol{\current} \wedge \boldsymbol{B},
 \label{KTMomentumInertial} \\
\partial_{t} \Energy_{\heavy} + \boldsymbol{\partial_{x}} \cdot ( \Energy_{\heavy} \boldsymbol{\Meanv}) = - \pressure_{\heavy} \, \boldsymbol{\partial_{x}} \cdot \boldsymbol{\Meanv} - \boldsymbol{\partial_{x}} \boldsymbol{\Meanv} : \boldsymbol{\Pi} - \, \boldsymbol{\partial_{x}} \cdot \boldsymbol{\HeatFlux}_{\heavy}^{\boldsymbol{\Meanv}} + \Delta E_{\heavy e} + \boldsymbol{\Current}_{\heavy} \cdot \boldsymbol{E}',
 \label{KTHeavyEnergyInertial}
\end{gather}
where the species diffusion velocities in the center-of-mass reference frame have been taken such as to satisfy the mass conservation constraint
\begin{equation}
\sum_{k \in \Species} \rho_{k} \boldsymbol{\MeanV}_{k}^{\boldsymbol{\Meanv}} = \rho_{e} \boldsymbol{\MeanV}_{e}^{\boldsymbol{\Meanv}} + \sum_{i \in \Heavy} \rho_{i} \boldsymbol{\MeanV}_{i}^{\boldsymbol{\Meanv}} = 0.
\end{equation}
The viscous tensor may be written in the form
\begin{equation}
\boldsymbol{\Pi} = - \eta_{\heavy} \, \Big( \boldsymbol{\partial_{x}} \boldsymbol{\Meanv} + (\boldsymbol{\partial_{x}} \boldsymbol{\Meanv})^{\textsc{t}} - \frac{2}{3} (\boldsymbol{\partial_{x}} \cdot \boldsymbol{\Meanv}) \, \mathbb{I} \Big) - \kappa_{\heavy} (\boldsymbol{\partial_{x}} \cdot \boldsymbol{\Meanv}) \, \mathbb{I} - \zeta (\Temperature_{e}-\Temperature_{\heavy}) \, \mathbb{I},
\label{KTViscousTensorInertial}
\end{equation}
the electron diffusion velocity and heat flux in the center-of-mass reference frame in the form
\begin{align}
\boldsymbol{\MeanV}_{e}^{\boldsymbol{\Meanv}} = & - D_{ee}^{\boldsymbol{\Meanv}} \boldsymbol{\hat{d}}_{e} - \sum_{i \in \Heavy} D_{ei}^{\boldsymbol{\Meanv}} \boldsymbol{\hat{d}}_{i} - \theta_{ee}^{\boldsymbol{\Meanv}} \boldsymbol{\partial_{x}} \ln{\Temperature_{e}} - \theta_{e \heavy}^{\boldsymbol{\Meanv}} \boldsymbol{\partial_{x}} \ln{\Temperature_{\heavy}}
\label{KTDiffusionVelocityElectronInertial} \\
& \vphantom{\sum_{i \in \Heavy}} - (1-Y_{e}) \big[ D_{ee}^{\kappa_{\heavy}} ~ (\boldsymbol{\partial_{x}} \cdot \boldsymbol{\Meanv}) + D_{ee}^{\Theta} ~ (\Temperature_{e} - \Temperature_{\heavy}) \big] ~ \boldsymbol{\hat{d}}_{e} 
\nonumber \\
 & \vphantom{\sum_{i \in \Heavy}} - (1-Y_{e}) \big[ \theta_{ee}^{\kappa_{\heavy}} ~ (\boldsymbol{\partial_{x}} \cdot \boldsymbol{\Meanv}) + \theta_{ee}^{\Theta} ~ (\Temperature_{e} - \Temperature_{\heavy}) \big] ~ \boldsymbol{\partial_{x}} \ln{\Temperature_{e}} \nonumber \\
 & \vphantom{\sum_{i \in \Heavy}} - \delta_{b0} (1-Y_{e}) \frac{\density_{e} q_{e} |\boldsymbol{B}|}{\pressure} D_{ee}^{0} ~ \mathscrbf{B} \wedge \boldsymbol{\Meanv} - \delta_{b0} (1-Y_{e}) D_{ee}^{\odot} ~ \mathscrbf{B} \wedge \boldsymbol{\hat{d}}_{e}
 \nonumber \\
 & \vphantom{\sum_{i \in \Heavy}} - \delta_{b0} (1-Y_{e}) \theta_{ee}^{\odot} ~ \mathscrbf{B} \wedge \boldsymbol{\partial_{x}} \ln{\Temperature_{e}},
 \nonumber \\
\boldsymbol{\HeatFlux}_{e}^{\boldsymbol{\Meanv}} = & - \pressure \theta_{ee}^{\boldsymbol{\Meanv}} \boldsymbol{\hat{d}}_{e} - \pressure \sum_{i \in \Heavy} \theta_{ei}^{\boldsymbol{\Meanv}} \boldsymbol{\hat{d}}_{i} - \hat{\lambda}_{ee} \boldsymbol{\partial_{x}} \ln{\Temperature_{e}} - \hat{\lambda}_{e \heavy} \boldsymbol{\partial_{x}} \ln{\Temperature_{\heavy}} + \density_{e} \Big( \frac{5}{2}k_{\textsc{b}} \Temperature_{e} \Big) \boldsymbol{\MeanV}_{e}^{\boldsymbol{\Meanv}}
\label{KTHeatFluxElectronInertial} \\
 & \vphantom{\sum_{i \in \Heavy}} - \pressure \big[ \theta_{ee}^{\kappa_{\heavy}} ~ (\boldsymbol{\partial_{x}} \cdot \boldsymbol{\Meanv}) + \theta_{ee}^{\Theta} ~ (\Temperature_{e} - \Temperature_{\heavy}) \big] ~ \boldsymbol{\hat{d}}_{e} - \big[ \hat{\lambda}_{ee}^{\kappa_{\heavy}} ~ (\boldsymbol{\partial_{x}} \cdot \boldsymbol{\Meanv}) + \hat{\lambda}_{ee}^{\Theta} ~ (\Temperature_{e} - \Temperature_{\heavy}) \big] ~ \boldsymbol{\partial_{x}} \ln{\Temperature_{e}}
 \nonumber \\
 & \vphantom{\sum_{i \in \Heavy}} - \delta_{b0} \density_{e} q_{e} |\boldsymbol{B}| \theta_{ee}^{0} ~ \mathscrbf{B} \wedge \boldsymbol{\Meanv} - \delta_{b0} \pressure \theta_{ee}^{\odot} ~ \mathscrbf{B} \wedge \boldsymbol{\hat{d}}_{e} - \delta_{b0} \hat{\lambda}_{ee}^{\odot} ~ \mathscrbf{B} \wedge \boldsymbol{\partial_{x}} \ln{\Temperature_{e}}, \nonumber
\end{align}
respectively, and the heavy-species diffusion velocities and heat flux in the form
\begin{align}
\boldsymbol{\MeanV}_{i}^{\boldsymbol{\Meanv}} & = - \sum_{j \in \Heavy} D_{i j}^{\boldsymbol{\Meanv}} \boldsymbol{\hat{d}}_{j} - D_{ie}^{\boldsymbol{\Meanv}} \boldsymbol{\hat{d}}_{e} - \theta_{i \heavy}^{\boldsymbol{\Meanv}} \boldsymbol{\partial_{x}} \ln{\Temperature_{\heavy}} - \theta_{i e}^{\boldsymbol{\Meanv}} \boldsymbol{\partial_{x}} \ln{\Temperature_{e}},
\label{KTDiffusionVelocitiesHeavyInertial} \\
& \vphantom{\sum_{i \in \Heavy}} +Y_{e} \big[ D_{ee}^{\kappa_{\heavy}} ~ (\boldsymbol{\partial_{x}} \cdot \boldsymbol{\Meanv}) + D_{ee}^{\Theta} ~ (\Temperature_{e} - \Temperature_{\heavy}) \big] ~ \boldsymbol{\hat{d}}_{e} + Y_{e} \big[ \theta_{ee}^{\kappa_{\heavy}} ~ (\boldsymbol{\partial_{x}} \cdot \boldsymbol{\Meanv}) + \theta_{ee}^{\Theta} ~ (\Temperature_{e} - \Temperature_{\heavy}) \big] ~ \boldsymbol{\partial_{x}} \ln{\Temperature_{e}} \nonumber \\
 & \vphantom{\sum_{i \in \Heavy}} + \delta_{b0} Y_{e} \frac{\density_{e} q_{e} |\boldsymbol{B}|}{\pressure} D_{ee}^{0} ~ \mathscrbf{B} \wedge \boldsymbol{\Meanv} + \delta_{b0} Y_{e} D_{ee}^{\odot} ~ \mathscrbf{B} \wedge \boldsymbol{\hat{d}}_{e} + \delta_{b0} Y_{e} \theta_{ee}^{\odot} ~ \mathscrbf{B} \wedge \boldsymbol{\partial_{x}} \ln{\Temperature_{e}}, \quad i \in \Heavy,
 \nonumber \\
\boldsymbol{\HeatFlux}_{\heavy}^{\boldsymbol{\Meanv}} & = - \pressure \sum_{j \in \Heavy} \theta_{\heavy j}^{\boldsymbol{\Meanv}} \boldsymbol{\hat{d}}_{j} - \pressure \theta_{\heavy e}^{\boldsymbol{\Meanv}} \boldsymbol{\hat{d}}_{e} - \hat{\lambda}_{\heavy \heavy} \boldsymbol{\partial_{x}} \ln{\Temperature_{\heavy}} - \hat{\lambda}_{\heavy e} \boldsymbol{\partial_{x}} \ln{\Temperature_{e}} + \sum_{j \in \Heavy} \big( \frac{5}{2}k_{\textsc{b}} \Temperature_{\heavy} + \overline{E}_{j} \big) \density_{j} \boldsymbol{\MeanV}_{j}^{\boldsymbol{\Meanv}}.
\label{KTHeatFluxHeavyInertial}
\end{align}
The electron conduction current density in the center-of-mass reference frame may be written in the form
\begin{equation}
\boldsymbol{J}_{e}^{\boldsymbol{\Meanv}} = \density_{e} q_{e} \boldsymbol{\MeanV}_{e}^{\boldsymbol{\Meanv}},
\end{equation}
and the current density in the form
\begin{equation}
\boldsymbol{\current} = \delta_{b0} \boldsymbol{\current}^{0} + \delta_{b1} \boldsymbol{\current}^{1},
\end{equation}
where
\begin{align}
\boldsymbol{\current}^{0} & = \density q \boldsymbol{\Meanv} + \density_{e} q_{e} \boldsymbol{\MeanV}_{e}^{0}, \\
\boldsymbol{\current}^{1} & = \density q \boldsymbol{\Meanv} + \density_{e} q_{e} \boldsymbol{\MeanV}_{e}^{\boldsymbol{\Meanv}} + \sum_{i \in \Heavy} \density_{i} q_{i} \boldsymbol{\MeanV}_{i}^{\boldsymbol{\Meanv}}.
\end{align}
The diffusion coefficients $D_{ee}^{\boldsymbol{\Meanv}}$, $D_{ei}^{\boldsymbol{\Meanv}}$, $i \in \Heavy$, $D_{ie}^{\boldsymbol{\Meanv}}$, $i \in \Heavy$, and $D_{ij}^{\boldsymbol{\Meanv}}$, $i,j \in \Heavy$, may be expressed in the form
\begin{align}
D_{ee}^{\boldsymbol{\Meanv}} & = (1-Y_{e}) D_{ee},
\label{KTDeeInertial} \\
D_{ei}^{\boldsymbol{\Meanv}} & = (1-Y_{e}) D_{ei}, \quad i \in \Heavy,
\label{KTDeiInertial} \\
D_{i e}^{\boldsymbol{\Meanv}} & = D_{i e} - Y_{e} D_{ee}, \quad i \in \Heavy,
\label{KTDieInertial} \\
D_{ij}^{\boldsymbol{\Meanv}} & = D_{ij} - Y_{e} D_{ej}, \quad i,j \in \Heavy,
\label{KTDijInertial}
\end{align}
and the thermal diffusion coefficients  $\theta_{ee}^{\boldsymbol{\Meanv}}$, $\theta_{e \heavy}^{\boldsymbol{\Meanv}}$, $\theta_{ie}^{\boldsymbol{\Meanv}}$, $i \in \Heavy$, $\theta_{i \heavy}^{\boldsymbol{\Meanv}}$, $i \in \Heavy$, may be expressed in the form
\begin{align}
\theta_{ee}^{\boldsymbol{\Meanv}} & = (1-Y_{e}) \theta_{ee}, \\
\theta_{e \heavy}^{\boldsymbol{\Meanv}} & = \theta_{\heavy e}^{\boldsymbol{\Meanv}} = (1-Y_{e}) \theta_{e \heavy}, \\
\theta_{ie}^{\boldsymbol{\Meanv}} & = \theta_{ei}^{\boldsymbol{\Meanv}} = \theta_{ie} - Y_{e} \theta_{ee}, \quad i \in \Heavy, \\
\theta_{i \heavy}^{\boldsymbol{\Meanv}} & = \theta_{\heavy i}^{\boldsymbol{\Meanv}} = \theta_{i \heavy} - Y_{e} \theta_{e \heavy}, \quad i \in \Heavy.
\end{align}
However, one can see from expressions \eqref{KTDeeInertial}-\eqref{KTDijInertial} that the diffusion matrix associated with the center-of-mass reference frame $(D_{kl}^{\boldsymbol{\Meanv}})_{k,l \in \Species}$ is not symmetric, and neither is the heavy-species diffusion matrix $(D_{ij}^{\boldsymbol{\Meanv}})_{i,j \in \Heavy}$. This confirms that the center-of-mass reference frame is not adapted to the study of non-thermal multicomponent plasmas \cite{GrailleMaginMassot2009}.

\section{Conclusion}

We have derived from the kinetic theory a unified multicomponent fluid model for non-thermal, partially ionized, polyatomic, chemically reactive plasmas. We have applied the classical Chapman-Enskog procedure, upon expanding the species distribution functions in powers of $\varepsilon$. The ratio $\varepsilon$ of electron to heavy-species characteristic masses was taken proportional to the Knudsen number. For the scaling adopted here, the equilibrium distribution functions are shown to be Maxwellian, with a different temperature for electrons and heavy species. We retrieve the zeroth-order and first-order drift-diffusion equations for electrons, while the macroscopic equations for the heavy species are the Euler equations at order zero, and the Navier-Stokes-Fourier equations at order one. Those equations involve transport fluxes, which have been expressed in terms of macroscopic variable gradients and source terms, by means of transport coefficients.

The characteristic cross-section for inelastic scattering between electrons and heavy species was taken two orders of magnitude lower than other relevant scattering cross-sections: ${\CrossSection_{\heavy}^{\text{in},0} = \varepsilon^{2} \CrossSection^{0}}$. We have shown that, other things being equal, this assumption is necessary to ensure that the electron and heavy-species respective temperatures, $\Temperature_{e}$ and $\Temperature_{\heavy}$, are distinct. In a future study, an alternative scaling will be investigated where the assumption over the inelastic scattering cross-section between electrons and heavy species is relaxed. For such a new scaling, we expect some of the internal energy states, or internal modes, of the heavy species to thermalize at $\Temperature_{e}$, while the others will thermalize at $\Temperature_{\heavy}$. The Chapman-Enskog expansion will thus require a splitting between the internal energy modes of the heavy-species. Typically, we may assume that vibrational modes are at equilibrium between them at $\Temperature_{\text{vib}} = \Temperature_{e}$, while rotational and translational modes thermalize at $\Temperature_{\text{rot}} = \Temperature_{\heavy}$ \cite{NagnibedaKustova}.

Expressions of transport fluxes for polyatomic plasmas have been derived in the weakly-magnetized case. The structure of the macroscopic \og Navier-Stokes type \fg {} equations and associated transport fluxes are similar to the monoatomic case treated in \cite{GrailleMaginMassot2009}, although expressions of transport coefficients now involve summations over the internal energy states of the heavy species. Additional terms in the electron second-order transport fluxes have been derived, associated with the interaction between the thermal non-equilibrium and the volume viscosity with the electron diffusion driving force and the electron temperature gradient. In the weakly-magnetized regime, the electron diffusion velocity and heat flux involve transverse driving forces. In a future study, the strongly-magnetized case will be investigated. In this regime, the electron transport fluxes exhibit an anisotropic behaviour with respect to the direction of the magnetic field \cite{GiovangigliGraille2009} \cite{GiovangigliGrailleMaginMassot}.

The derivation should also be completed with the investigation of the mathematical structure of the macroscopic equations and of the transport linear systems obtained. In particular, the entropy conservation equation must be derived, and the sign of the entropy production rate must be ascertained. Indeed, a positive entropy structure is desirable for numerical stability \cite{Bobylev1982} \cite{Torrilhon2016}. Besides, the numerical computation of the transport coefficients derived above, in particular of the electron zeroth-order and first-order diffusion coefficients, should be carried out and compared to both experimental and numerical values, when accessible \cite{ErnGiovangigli} \cite{GiovangigliGrailleMaginMassot}. Such a calculation will require data for the various collision integrals involved.

However, the set of equations derived in this paper, and associated expressions for transport fluxes, are a sound basis for the numerical modeling of non-thermal plasmas. Indeed, the first-order Chapman-Enskog expansion for multicomponent gas mixtures is retrieved in the limit where $\Temperature_{e} = \Temperature_{\heavy}$ \cite{GiovangigliGrailleMaginMassot}. Also, the present model encompasses almost all existing fluid models for low-temperature plasmas.

\section*{Acknowledgements}
This work has been supported by the Region Ile-de-France in the framework of DIM Nano-K, the nanoscience competence center of Paris Region.

{\small
\bibliographystyle{plain}
\bibliography{KTheory}
}

\end{document}